\documentclass[twocolumn]{aastex62}

\usepackage{color}
\usepackage{framed}
\usepackage{natbib}
\usepackage{textcomp}
\usepackage{amsmath}

\newcommand{\ms}{\ensuremath{\mathrm{m\,s}^{-1}}}

\newcommand{\degree}{\ensuremath{\mathrm{^{\circ}}}}
\newcommand{\logrphk}{\ensuremath{\log R^\prime_{\rm HK}}}
\newcommand{\vsini}{\ensuremath{v\sin i}}
\newcommand{\msun}{\ensuremath{\rm M_\sun}}
\newcommand{\mearth}{\ensuremath{\rm M_{\oplus}}}
\newcommand{\rearth}{\ensuremath{\rm R_{\oplus}}}
\newcommand{\msini}{\ensuremath{\rm m \sin i}}

\newcommand{\rsun}{\ensuremath{\rm R_\sun}}

\newcommand{\teff}{T$_{\rm eff}$}
\newcommand{\Teff}{T$_{\rm eff}$}
\newcommand{\kepler}{\emph{Kepler}}
\newcommand{\TESS}{\emph{TESS}}

\newcommand{\sysonebPerunc}{11.72$\pm$0.001}
\newcommand{\sysonebKunc}{5.64$\pm$0.55} 
\newcommand{\sysonebEccunc}{0.14$\pm$0.07}  
\newcommand{\sysonebOmegaunc}{254.64$\pm$109.93}
\newcommand{\sysonebMounc}{100.78$\pm$92.27}
\newcommand{\sysonebMsiniunc}{16.46$\pm$1.66}
\newcommand{\sysonebAunc}{0.092$\pm$0.0008}

\newcommand{\sysonebPer}{11.72}
\newcommand{\sysonebK}{5.64} 
\newcommand{\sysonebEcc}{0.14}

\newcommand{\sysonebMsini}{16.46}
\newcommand{\sysonebA}{0.092}

\newcommand{\sysonecPerunc}{29.180$\pm$0.012}
\newcommand{\sysonecKunc}{3.54$\pm$0.90}
\newcommand{\sysonecEccunc}{0.31$\pm$0.15}
\newcommand{\sysonecOmegaunc}{225$\pm$47}
\newcommand{\sysonecMounc}{214$\pm$47}

\newcommand{\sysonecPer}{29.18}

\newcommand{\systwobPerunc}{35.45$\pm$0.011}
\newcommand{\systwobKunc}{2.28$\pm$0.20}
\newcommand{\systwobEccunc}{0.09$\pm$0.06} 
\newcommand{\systwobOmegaunc}{220.95$\pm$125.17}
\newcommand{\systwobMounc}{48.72$\pm$9.69}
\newcommand{\systwobMsiniunc}{10.26$\pm$0.99}
\newcommand{\systwobAunc}{0.198$\pm$0.0004}

\newcommand{\systwobPer}{35.45}

\newcommand{\systwobMsini}{10.26}

\newcommand{\systwocPerunc}{154.43$\pm$0.44}
\newcommand{\systwocKunc}{1.29$\pm$0.22}
\newcommand{\systwocEccunc}{0.12$\pm$0.08} 
\newcommand{\systwocOmegaunc}{198.12$\pm$96.34}
\newcommand{\systwocMounc}{48.72$\pm$9.69}
\newcommand{\systwocMsiniunc}{$9.44\pm1.63$}
\newcommand{\systwocAunc}{$0.528\pm0.010$}

\newcommand{\systwocPer}{154.43}

\newcommand{\systwocMsini}{9.44}

\newcommand{\systwodPerunc}{7767.35$\pm$1464.31}
\newcommand{\systwodKunc}{1.90$\pm$0.35}
\newcommand{\systwodEccunc}{0.19$\pm$0.12}
\newcommand{\systwodOmegaunc}{95.32$\pm$82.22}
\newcommand{\systwodMounc}{113.61$\pm$9310.53}

\newcommand{\systwodPer}{7767.35}

\newcommand{\systhreebPer}{15.53}
\newcommand{\systhreebK}{3.004}
\newcommand{\systhreebEcc}{0.050}

\newcommand{\systhreebMsini}{6.624}
\newcommand{\systhreebA}{0.091}

\newcommand{\systhreebPerunc}{15.530$\pm$0.0011}
\newcommand{\systhreebKunc}{3.004$\pm$0.180}
\newcommand{\systhreebEccunc}{0.050$\pm$0.030}
\newcommand{\systhreebOmegaunc}{197.062$\pm$153.673}
\newcommand{\systhreebMounc}{223.884$\pm$27.596}
\newcommand{\systhreebMsiniunc}{6.624$\pm$0.432}
\newcommand{\systhreebAunc}{0.091$\pm$0.001}

\newcommand{\sysfourbPer}{105.91}
\newcommand{\sysfourbK}{2.696}
\newcommand{\sysfourbEcc}{0.101}

\newcommand{\sysfourbMsini}{12.214}
\newcommand{\sysfourbA}{0.343}

\newcommand{\sysfourbPerunc}{105.911$\pm$0.109}
\newcommand{\sysfourbKunc}{2.696$\pm$0.224}
\newcommand{\sysfourbEccunc}{0.101$\pm$0.053}
\newcommand{\sysfourbOmegaunc}{257.433$\pm$51.256}
\newcommand{\sysfourbMounc}{123.594$\pm$50.897}
\newcommand{\sysfourbMsiniunc}{12.214$\pm$1.05}
\newcommand{\sysfourbAunc}{0.343$\pm$0.004}

\submitjournal{The Astronomical Journal}
\accepted{October 13th, 2020}

\shortauthors{Burt et al. 2020}

\begin{document}
\title{A collage of small planets from the Lick Carnegie Exoplanet Survey : \\ Exploring the super-Earth and sub-Neptune mass regime \footnote{This paper includes data gathered with the 6.5 meter Magellan Telescopes located at Las Campanas Observatory, Chile.}}

\correspondingauthor{Jennifer A. Burt}
\email{jennifer.burt@jpl.nasa.gov}

\author[0000-0002-0040-6815]{Jennifer Burt}
\affil{Jet Propulsion Laboratory, California Institute of Technology, 4800 Oak Grove drive, Pasadena, CA, 91109, USA}

\author[0000-0001-6039-0555]{Fabo Feng}
\affil{Tsung-Dao Lee Institute, Shanghai Jiao Tong University, 800 Dongchuan Road, Shanghai 200240, People's Republic of China}
\affil{Department of Astronomy, School of Physics and Astronomy, Shanghai Jiao Tong University, 800 Dongchuan Road, Shanghai 200240, People's Republic of China}

\author[0000-0002-6153-3076]{Bradford Holden}
\affil{UCO/Lick Observatory, Department of Astronomy and Astrophysics, University of California at Santa Cruz, Santa Cruz, CA, 95064, USA}

\author[0000-0003-2008-1488]{Eric~E.~Mamajek}
\affiliation{Jet Propulsion Laboratory, California Institute of Technology, 4800 Oak Grove Drive, Pasadena, CA, 91109, USA}

\author[0000-0003-0918-7484]{Chelsea~ X.~Huang}
\affiliation{Department of Physics and Kavli Institute for Astrophysics and Space Research, Massachusetts Institute of Technology, Cambridge, MA, 02139, USA}
\affiliation{Juan Carlos Torres Fellow}

\author[0000-0003-3938-3099]{Mickey M. Rosenthal}
\affiliation{Department of Astronomy and Astrophysics, University of California, Santa Cruz, CA, 95064, USA}

\author[0000-0002-7846-6981]{Songhu Wang}
\affil{Department of Astronomy, Indiana University, Bloomington, IN, 47405, USA}

\author[0000-0003-1305-3761]{R. Paul Butler}
\affil{Earth and Planets Laboratory, Carnegie Institution for Science, 5241 Broad Branch Road, NW, Washington, DC, 20015, USA}

\author[0000-0001-7177-7456]{Steven S. Vogt}
\affil{UCO/Lick Observatory, Department of Astronomy and Astrophysics, University of California at Santa Cruz, Santa Cruz, CA, 95064, USA}

\author[0000-0002-3253-2621]{Gregory Laughlin}
\affil{Department of Astronomy, Yale University, New Haven, CT, 06511, USA}

\author{Gregory W. Henry}
\affil{Center of Excellence in Information Systems, Tennessee State University, Nashville, TN, 37209, USA}

\author{Johanna K. Teske}
\affil{Earth and Planets Laboratory, Carnegie Institution for Science, 5241 Broad Branch Road, NW, Washington, DC, 20015, USA}
\affil{NASA Hubble Fellow}

\author{Sharon X. Wang}
\affil{Department of Astronomy, Tsinghua University, Beijing 100084, People’s Republic of China}

\author[0000-0002-5226-787X]{Jeffrey D. Crane}
\affil{Observatories of the Carnegie Institution for Science, 813 Santa Barbara St., Pasadena, CA, 91101, USA}

\author[0000-0002-8681-6136]{Steve A. Shectman}
\affil{Observatories of the Carnegie Institution for Science, 813 Santa Barbara St., Pasadena, CA, 91101, USA}

\begin{abstract}

Analysis of new precision radial velocity (RV) measurements from the Lick Automated Planet Finder (APF) and Keck HIRES have yielded the discovery of three new exoplanet candidates orbiting two nearby K dwarfs not previously reported to have companions (HD 190007 and HD 216520. 
We also report new velocities from both the APF and the Planet Finder Spectrograph (PFS) for the previously reported planet host stars GJ 686 and HD 180617 and update the corresponding exoplanet orbital models. 
Of the newly discovered planets, HD~190007~b has a period of \sysonebPer\ days, an RV semi-amplitude of K = \sysonebKunc\ \ms, a minimum mass of \sysonebMsiniunc\ \mearth, and orbits the slightly metal-rich, active K4 dwarf star HD~190007 (d = 12.7 pc).  
HD~216520~b has an orbital period of \systwobPer\ days, an RV semi-amplitude of K = \systwobKunc\ \ms, and a minimum mass of \systwobMsiniunc\ \mearth, while HD~216520~c has an orbital period of P = \systwocPer\ days, an RV semi-amplitude of K = \systwocKunc\ \ms, and a minimum mass of \systwocMsiniunc\ \mearth. Both of these planets orbit the slightly metal-poor, inactive K0 dwarf star HD~216520 (d = 19.6 pc). 
We find that our updated best fit models for HD 180617 b and GJ 686 b are in good agreement with the previously published results. For HD~180617~b we obtain an orbital period of \sysfourbPer\ days, an RV semi-amplitude of K = \sysfourbKunc\ \ms, and a minimum mass of \sysfourbMsiniunc\ \mearth. For GJ~686~b we find the orbital period to be \systhreebPer\ days, the RV semi-amplitude to be K = \systhreebKunc\ \ms, and the minimum mass to be \systhreebMsiniunc\ \mearth.
Using an injection-recovery exercise, we find that HD~190007~b and HD~216520~b are unlikely to have additional planets with masses and orbital periods within a factor of two, in marked contrast to $\sim$85\% of planets in this mass and period range found with Kepler.
\end{abstract}


\section{Introduction} \label{sec:intro}

The use of precision ground-based Doppler spectrometers enabled the first generation of exoplanet detections, producing 47 exoplanet detections before the first exoplanet was discovered to transit in 1999 \citep{Henry1999b,Henry2000}. Since then an additional 773 planets have been discovered using such instruments, bringing the total number of radial velocity (RV) detected planets to 820 as of September  2020\footnote{https://exoplanetarchive.ipac.caltech.edu}. 

More recently, numerous precision RV instruments have been spending more time pursuing targeted mass measurements of transiting planets discovered by space-based missions like {\it K2} and \TESS. Such planets, for which we can obtain both mass and radius measurements, are valuable additions to the field as they enable studies of planetary composition and evolution \citep[see e.g.]{Burt2020, Christiansen2017, Teske2018}. However more traditional Doppler surveys, which focus on repeated observations of nearby stars without any {\it a priori} knowledge of whether or not they host planets, are still a key component in our efforts to understand the Galaxy's exoplanet population. Because RV detections do not require an exoplanet to transit its host star, an alignment that happens with only a few percent probability for even promising short period planets \citep{Winn2010}, this method is sensitive to a wider range of orbital configurations. As such, many of our constraints on the characteristics and occurrence rates of long-period gas and ice giant planets come from RV surveys \citep{Rowan2016, Bryan2019}. And unlike transit observations, which detect planets during their fleeting inferior conjunctions, RV observations permit the planetary signal to build in the data over the entire span of observing time so they do not suffer as badly from short-term weather closures or varied observing times/locations. Examples of such surveys include the Lick Carnegie Exoplanet Survey \citep[LCES,][ hereafter B17]{Butler2017}, the California Planet Search \citep[CPS,][]{Howard2010}, the HARPS search for southern extra-solar planets \citep{Pepe2004}, and the CARMENES search for exoplanets around M dwarfs \citep{Reiners2018}. 

While they are less susceptible to short-term disruptions, RV surveys are more observationally expensive than transit surveys, requiring both long exposure times to reach the required high S/N levels on high-resolution spectrometers and long observing baselines to properly fill out a planet's RV phase curve. This combination means that surveys often require months or years of data on a star before any potential planet signals are interpreted as robust detections. Additionally, RV surveys can often begin by selecting stars that seem promising based on their V magnitudes, effective temperatures (T$_{\rm eff}$), rotational velocities, and \logrphk\, activity indicators, only to learn months or years later that the stars are actually too active to allow for the detection of Keplerian signals that are only a few \ms in amplitude. When such stars are ultimately dropped from a survey, however, there is often no public record of the discontinuation of RV monitoring, nor is there a write up of {\it why} the star was excluded. This can result in future surveys observing that same target again, and spending many nights of telescope time only to realize once again that the star is not well-suited to RV science. Dropping targets also has strong, detrimental impacts on efforts to infer RV planet population statistics and makes it difficult to draw robust conclusions about planet occurrence rates from most RV surveys.

In 2017 the LCES team released a compendium of all the Keck HIRES RV observations acquired since the advent of the survey in 1994, resulting in 64,949 radial velocities spread over 1,624 stars in B17. The majority of stars included in the survey have V magnitudes between 6 and 10, with a median value of V = 8. Stars brighter than 5th magnitude were handled by the Hamilton spectrograph on the Shane 3m telescope at Lick Observatory. Observing times were limited to roughly 10 minutes per target per night in order to accommodate a relatively large target list which led to a faint magnitude limit of about 14th magnitude. The stellar spectral types covered by the survey run from about F5V to M6V. Stars hotter than F5V were generally avoided due to the decrease in RV information content \citep{Bouchy2001, BeattyGaudi2015} while stars cooler than M6V were generally past the faint end of the survey's magnitude limit. For new or ongoing planet search surveys, the ability to sift through two decades of archival data and use it for either initial target selection or in combination with recent RV observations to search for potential Keplerian signals (or trends that could indicate notable stellar activity cycles which can last many years) is a powerful tool.

In this work, we make use of the LCES RV catalog as the backbone for two new planet detections and updated fits for two recently published planets. Two of these stars are among the 357 found in B17 to have significant signals that are of constant period and phase, but not coincident in period and/or phase with stellar activity indices in the original catalog publication. The other two display significant RV periodicities only after the HIRES velocities have been combined with RV data from other observatories. We begin in \S \ref{sec:methods} with an overview of the specifics of each step of our analysis methods, as they are shared between the four stellar data sets. Then we detail the characteristics of exoplanet host stars and present a compilation of their stellar parameters in \S \ref{sec:HostStars}. Next, we cover each star and its associated planet detection individually in \S 4-7 before concluding in \S \ref{sec:Discussion} with a discussion about how these planets compare to the multi-planet systems discovered in large part by the \kepler\ mission. 

\section{Overview of data and methodology}\label{sec:methods}

Each star is analyzed using the same methodology and series of steps, which we summarize here. The initial data product for each target is a set of time series radial velocity measurements and the corresponding spectral activity indicators from various PRV instruments which are monitored using the interactive SYSTEMIC Console \citep{Meschiari2009} over the months/years that the stars spend on our survey target lists. When the RV data sets show signs of a significant signal, we then work to determine whether or not the signal could have been caused by our observing approach or by the host star itself via chromospheric activity and/or stellar rotation \citep[see, e.g.,][]{Rajpaul2016}. To this end we first compute the spectral window function of the RV data sets to ensure the signal is not due to our observational cadence. We then gather, when available, long baseline photometric measurements to try and measure the stars' rotation periods via seasonal changes in stellar brightness. We also derive S- and H-index activity indicators (described in \S \ref{sec:activityindicators}) from the RV spectra to map the stars' chromospheric activity signatures. And finally we assess the consistency of any signals over time by computing a moving periodogram with a moving average noise model (MA(1) model, described in \S \ref{sec:bayesian}) for those signals whose orbital phases are well-covered by the RV data. If none of these analyses produce signals that can explain the signature seen in the RV data, we then test whether the signal originally noted with SYSTEMIC is ``significant'' under a full Bayesian treatment as seen in \citet{Feng2019}. If so, we assume the signal is caused by a planet in orbit around the host star, and fit a Keplerian model to determine the exact parameters of the planet and its orbit. Additional details on each of these steps follow below.

\subsection{Radial velocities}

The data sets presented in this work consist of unbinned radial velocity observations taken with seven different instruments; the Levy spectrometer \citep[on the 2.4m Automated Planet Finder telescope,][]{Vogt2014}, the High Resolution spectrometer \citep[HIRES, on the 10m Keck I telescope,][]{Vogt1994}, the Planet Finder spectrometer \citep[PFS, on the 6.5m Magellan Clay telescope,][]{Crane2006, Crane2008, Crane2010}, the High Accuracy Radial Velocity Planet Searcher \citep[HARPS, on the ESO 3.6m telescope,][]{Mayor2003}, the High Accuracy Radial velocity Planet Searcher for the Northern hemisphere \citep[HARPS-N, on the 3.58m Telescopio Nazionale Galileo,][]{Cosentino2012}, the Spectrographe pour l'Observation des PH\'{e}nom\`{e}nes des Int\'{e}rieurs stellaires et des Exoplan\`{e}tes \citep[SOPHIE, on the 1.93m reflector telescope at the Haute-Provence Observatory,][]{Bouchy2011}, and the CARMENES spectrometer \citep[on the 3.5 m telescope at the Calar Alto Observatory,][]{Quirrenbach2018}. Samples of the velocities used in our analyses are presented in the appendix, while the full data sets will be made available as machine readable tables.

The APF, HIRES, and PFS RV values are all measured by placing a cell of gaseous I$_2$ in the converging beam of each telescope. This imprints the 5000-6200\AA\ region of incoming stellar spectra with a dense forest of I$_2$ lines that act as a wavelength calibrator and provide a proxy for the point spread function (PSF) of each spectrometer. To ensure a constant I$_2$ column density over multiple decades, the cells are held at a constant temperature of 50.0 $\pm$ 0.1\degree C. The instruments have typical spectral resolutions of 90,000, 60,000 and 80,000 for the APF, HIRES and PFS, respectively. While only the 5000-6200\AA\ spectral region is used for measuring radial velocities, the instruments produce spectra from 3700-7700\AA\ for the APF, 3700-8000\AA\ for HIRES, and 3900-6700\AA\ for PFS.

Once the iodine region of the spectrum has been extracted, it is split into 2\AA\ chunks. Each chunk is analyzed using the spectral synthesis technique described in \citet{Butler1996}, which deconvolves the stellar spectrum from the I$_2$ absorption lines and produces an independent measure of the wavelength, instrument PSF, and Doppler shift. The  final Doppler velocity from a given observation is the weighted mean of the velocities of all the individual chunks ($\sim$700 for the APF and HIRES, and $\sim$800 for PFS). The final internal uncertainty of each velocity is the standard deviation of all 700 chunk velocities about that mean. 

In contrast, HARPS, HARPS-N, and SOPHIE make use of multiple observing fibers, one of which is placed on the stellar target while the other is pointed at a Th-Ar calibration lamp to provide a simultaneous wavelength reference. The two HARPS instruments operate in the same region of wavelength space, from 3800-6900\AA, and have the same peak resolving power of $\sim$115,000 \citep{Pepe2002, Cosentino2012}. SOPHIE covers the same wavelength range as the two HARPS instruments, but has a slightly lower resolution of $\sim$75,000.

The HARPS-TERRA velocities presented here are measured using the approach laid out in \citet{Anglada-Escude12a}, using data obtained from the HARPS instrument described above. Specifically, the spectra for these observations are downloaded from the ESO archive and then each observation is decomposed into 1) a high signal-to-noise template, 2) the RV shift for that observation, and 3) a multiplicative background set of polynomials to account for flux variations. TERRA first derives approximate RVs measured against an observed spectrum and then improves the template by co-adding all observed spectra and recomputing the RVs. The spectra are co-added via a weighted least-squares regression with a cubic B-spline.

For the HARPS-N velocities the data reduction and spectral extraction were carried out using the Data Reduction Software (DRS v3.7). Once an observation is complete a 2-D spectra is optimally extracted from the resulting FITS file. The spectrum is cross-correlated with a numerical mask corresponding to the appropriate spectral type (F0, G2, K0, K5, or M4), and the resulting cross-correlation function (CCF) is fit with a Gaussian curve to a produce radial velocity measurement \citep{Baranne1996, Pepe2002} and calibrated to determine the RV photon-noise uncertainty $\sigma_{RV}$. The SOPHIE radial velocities are calculated using a similar data-reduction pipeline, which is adapted from the HARPS DRS software.

The CARMENES spectrometer consists of two cross-dispersed echelles and also employs multiple observing fibers in order to obtain simultaneous stellar and wavelength calibration data. The first spans 5200-9600\AA\ in the visible (VIS) at a resolution of 94,600, while the second covers 9600-17100 \AA\ in the near infrared (NIR) at a resolution of 80,400 \citep{Quirrenbach2014}. CARMENES makes use of the spectrum radial velocity analyser (SERVAL) software package to produce radial velocity measurements \citep{Zechmeister2018}. SERVAL is based upon the least-squares fitting approach described above and again computes precision RV measurements using a least-squares matching of each observed spectrum to a high signal-to-noise ratio template derived from the same observations. 

\subsection{Photometry and stellar rotation}

To search for evidence of a given star's stellar rotational period, we have acquired high-precision, long-baseline photometric data of HD~190007, GJ~686, and HD 180617 taken with the T12 0.8m and T4 0.75m APTs at Fairborn Observatory in the Stromgren $b$ \& $y$ pass bands. The two-color observations have been combined to produce a $\Delta (b+y)/2$ joint-filter time series, which improves measurement precision. Program stars on these telescopes have their observations interlaced with three nearby comparison stars in the sequence: dark, A, B, C, D, A, SKYA, B, SKYB, C, SKYC, D, SKYD, A, B, C, D, where A, B, and C are the comparison stars and D is the program star. Integration times are 20-30 seconds (depending on stellar brightness) on the 0.75 m APT, where the Stromgren $b$ \& $y$ observations are made sequentially, and 40 seconds on the 0.80 m APT, where the two bands are measured simultaneously  \citep{Henry1999a}. The photometric data for each target and the resulting conclusions drawn from the data set is described in each star's respective section of the paper. 

\subsection{Bayesian search and orbital fitting}
\label{sec:bayesian}
The significance of signals initially identified with SYSTEMIC are assessed by calculating likelihood ratios and Bayes factors based on the Bayesian information criterion \citep[e.g.,][]{liddle2007,feng2016} and by applying a noise model that accounts for the most important sources of variability in a radial velocity time-series \citep{feng2016}, i.e. Keplerian signals, an unknown amount of white noise, different unknown levels of correlated noise and potential linear correlations between velocities and available spectroscopic activity proxies. The model has been discussed in great detail in e.g. \citet{feng2016,feng2017,diaz2018}.

The model for a given star's RV solution is made up of a combination of signal and noise components, where the signal for N$_{p}$ planets in the k$^{th}$ RV data set is given by Equation 1: 
\begin{equation}
\hat{v}^{k}_{s}(t_{j}) = \sum_{i=1}^{N_{p}} K_{i}[\sin(\omega_{i} + \nu_{i}(t_{j})) + e_{i}\cos\omega_{i}] + \gamma_{k}+\dot{\gamma}_{k}t_{j}
\end{equation}
where K$_{i}$ is the RV semi-amplitude of stellar variation induced by the i$^{th}$ planet's gravitational pull on the host star, $\nu_{i}$(t$_{j}$) is the true anomaly derived from the planet's orbital period P$_{i}$, eccentricity e$_{i}$ and the reference mean anomaly M$_{0}$ after solving Kepler's equation. Unlike \citet{Feng2019}, we do not include a linear trend in the model as in some cases we find evidence of signals with periods approaching the RV data time span. We therefore replace the linear trend by an offset to avoid introducing degeneracies between potential long-period Keplerian signals and a linear trend term in the model. 

We assess each RV dataset to determine whether it is best represented by a white noise model (a constant jitter is used to fit excess noise), or a moving average model \citep[MA(1),][]{Tuomi2013}. In order to determine the order, q, of the MA model  we calculate the maximum likelihood for a MA model using the Levenberg-Marquardt (LM) optimization algorithm. We calculate ln(BF) for MA(q+1) and MA(q). If ln(BF) \textless\ 5, we select MA(q). If ln(BF) \textgreater\ 5, we select MA(q+1) and keep increasing the order of MA model until the model with the highest order passing the ln(BF) \textgreater\ 5 criterion is found. Additional details on this model determination process can be found in \citet{Feng2019}.

Once the appropriate model has been determined, we search for periodic signals in the data by applying an adaptive Metropolis algorithm \cite{Haario2006} combined with a parallel tempering scheme designed by \citet{Feng2019}. This Markov-chain Monte Carlo technique has been applied to similar data sets \citep[e.g.][]{Butler2017,Vogt2017,Tuomi2018} and found to identify significant signals reliably with a low sensitivity to false positives \citep{Dumusque2017}. A signal is considered to be strong if the logarithmic Bayes factor is larger than 3 (i.e. ln(BF) \textgreater\ 3) or equivalently its Bayesian information criterion (BIC) is larger than 6 \citep{KassRaftery1995}. A signal is considered to be very strong or significant if ln(BF) \textgreater\ 5. The exact number of free parameters needed for the RV fit is unknown, however, as a circular Keplerian model has three free parameters while an eccentric Keplerian orbit requires five. We therefore define  k = 3 and k = 5 as the boundaries for the real BIC value, which are represented in the fit summaries as ln(BF$_{3}$) and ln(BF$_{5}$), respectively. Since we are using a sinusoidal model in the calculation of the BFPs, we use $\ln({\rm BF}_3) = 5$ as the initial threshold for declaring a signal to be significant. Although a more conservative criterion of $\ln(\rm {BF}) > 6.9$ is proposed by \citet{Nelson2020}, we keep using $\ln(\rm{BF}) > 5$ in order to be consistent with the threshold proposed by \citet{KassRaftery1995} and suggested by \citet{feng2016} through analyses of RV data sets. Nevertheless, the results presented in this work are not sensitive to the choice of $\ln(\rm{BF})$ criteria because all planets satisfy the more conservative $\ln(\rm{BF}) > 6.9$ criterion. The real significance of a signal, however, is determined through posterior sampling combined with the BF thresholds. This analysis is visualized by creating a ln(BF) periodogram (BFP) for each of the individual RV data sets, in addition to a BFP for the combined RV data set. 

In an attempt to identify those signals caused by stellar activity, we calculate BFPs for the activity indices and the observational window function for each data set and perform visual inspections to determine whether they exhibit overlapping periods with potential planetary signals in the RV data sets. To assess the consistency of signals over time, we compute a moving periodogram for those signals whose phase is well covered by the RV data. Since the RVs are typically not measured in a uniform way, the consistency of a true signal may depend on the sampling cadence even if the power is normalized. However, it is easy to identify false positives if inconsistency is found at high cadence epochs with a timescale comparable with or longer than the signal period \citep{Feng2019}.

A full set of white noise, moving average, and auto-regressive BFPs for each target star's RV data sets, activity indicators, and window function is presented in Appendix 1.

\subsection{Stellar activity indicators}
\label{sec:activityindicators}

The derived spectral activity indicators from each of our HIRES, PFS, and APF spectra serve as proxies for chromospheric activity in the visible stellar hemisphere at the moments when the spectra were obtained. The S-index is obtained from measurement of the emission reversal at the cores of the Fraunhofer H and K lines of Ca II located at at 3968 {\AA} and 3934 {\AA}, respectively \citep{Duncan1991}. 

For the APF data, we employ an adaptation of the S-index analysis presented in \citet{Isaacson2010} where, for each star, we identify the observation with the highest SNR level as the ``reference spectrum'' and then compute the redshift of that spectrum by cross correlating with the NSO solar atlas. All other spectra are then shifted into the same reference frame and continuum aligned to that reference spectrum using 10\AA\ continuum regions in order to remove flux differentials arising from different SNRs between the exposures. The flux in the Ca II H \& K bands and their associated continuum bands are measured and recorded, and the final data sets are calibrated against the original Mt. Wilson S-index survey results to allow for comparisons with our Keck and PFS data.

We also report H-index measurements for our APF, PFS, and post-fix (taken after the detector upgrade in August 2004) HIRES spectra. Similarly to the S-index, the H-index quantifies the amount of flux within the H$\alpha$ Balmer line core compared to the local continuum. Details on the prescription used for the HIRES and PFS data sets can be found in \citet{Butler2017}. For the APF, we use the \citet{Gomes2011} prescription, which defines the H-index as the ratio of the flux within $\pm$0.8{\AA} of the H$\alpha$ line at 6562.808 {\AA} to the combined flux of two broader flanking wavelength regions: 6550.87 $\pm$ 5.375 {\AA} and 6580.31 $\pm$ 4.375 {\AA}. The H-index measurements often show peaks at periods of roughly one year due to the presence of shallow telluric lines that can shift into and out of the H$\alpha$ filter if the star's systemic radial velocity places them on the edge of the H$\alpha$ line. We find that attempts to remove these lines change the shape and flux level of the H$\alpha$ line in substantial ways that compromise the search for flux modulations indicative of stellar activity. Therefore we do not remove the tellurics, but rather search for additional signals in the H-index periodograms after removing the one-year cycles.

For the HARPS data sets, which we process through the HARPS TERRA pipeline, we also make use of the Ca II H \& K activity indicators that TERRA measures, in addition to the line bisectors (BIS), full width half max (FWHM), and CCF which are taken from the original HARPS DRS results.

We analyze the resulting activity indicators by computing Bayes factor periodograms for each activity indicator in each RV data set and looking for any well-defined peaks with ln(BF$_{5}$) \textgreater 5 at or near the period of our suspected planets. In cases where we see peaks in these regions, we then compute the correlation coefficients between the RV measurements and the activity indicator measurements to look for evidence that the stellar activy is influencing the star's radial velocities. 

\section{Stellar Parameters} \label{sec:HostStars}

All four stellar hosts described herein are nearby, dwarf stars with spectral types from K0V to M3V. These stars make excellent candidates for traditional iodine cell-based RV spectrometers thanks to their relatively bright V magnitudes and abundance of narrow absorption lines in the visible part of the spectrum \citep{Burt2015}. These stars in particular were selected for inclusion in the long running Lick-Carnegie Exoplanet Survey (LCES) carried out in B17 using Keck HIRES after analysis of early spectra revealed them to be chromospherically inactive and slowly rotating - two other key characteristics for RV candidates. 
Two of the stars (HD 190007 and HD 216520) do not have previously known exoplanets, which often prompts in-depth study of stellar characteristics. Their proximity to Earth and resulting bright magnitudes, however, have lead to their inclusion in a number of different surveys and large stellar characterization efforts (see, e.g., \citealt{Lepine2011}; \citealt{Mann2015}; \citealt{Brewer2016}) that provide us with the parameters for each star presented in Table \ref{table:stellarparams}. The stars have also been examined by \citet{Mishenina2013}, which provides individual elemental abundances (Table \ref{tab:abundances}). For the other two targets with previously published planets (GJ 686 and HD 180617) we take the stellar parameters from \citet{Schweitzer2019}.

\begin{deluxetable*}{l|l|l|l|l}
\tablecaption{\label{table:stellarparams} Stellar parameters}
\tablehead{{Parameter}&{HD~190007}&{HD~216520}&{GJ~686}&{HD~180617}}
\tablecolumns{6}
\startdata
Right Ascension	&	  20 02 47.05$^{a}$  	&	  22 47 31.88$^{a}$  	&	  17 37 53.35$^{a}$ & 19 16 55.26$^{a}$ \\ 	
Declination    	&	  03 19 34.27$^{a}$  	&	  83 41 49.30$^{a}$  	&	  18 35 30.16$^{a}$ & 05 10 08.04$^{a}$ \\ 	
Spectral type  	&	  K4V(k)$^{b}$  	     &	  K0V$^{c}$  	&	  M1.0V$^{d}$  	& M2.5V$^{d}$\\ 	
$\varpi$ (mas) 	&	  78.6238 $\pm$ 0.0617$^{a}$ &	 51.1167 $\pm$ 0.0292$^{a}$ 	& 122.5609 $\pm$ 0.0346$^{a}$ & 169.1590 $\pm$ 0.0520$^{a}$\\	
Distance (pc)  	&	  12.72 $\pm$ 0.01$^{a}$ & 19.56 $\pm$ 0.011$^{a}$ & 8.159 $\pm$ 0.0023$^{a}$ & 5.912 $\pm$ 0.0018$^{a}$\\ 	
Systemic RV (k\ms)&	-30.467 $\pm$ 0.133$^{a}$ &	-18.536 $\pm$ 0.182$^{a}$	&	-10.092 $\pm$ 0.232$^{a}$ & 35.554 $\pm$ 0.158$^{a}$\\
$V$    	& 7.46$^{e}$    &	 7.53$^{e}$     & 9.62$^{e}$	& 9.12$^{e}$\\ 	
$G$    	& 7.0634$^{a}$ 	&	 7.2790$^{a}$ 	& 8.7390$^{a}$ 	& 8.0976$^{a}$\\	
$M_V$  	& 6.94 $\pm$ 0.01$^{a,e}$	&	 6.08 $\pm$ 0.02$^{a,e}$ &	10.04 $\pm$ 0.03$^{a,e}$ & 10.26 $\pm$ 0.005$^{a,e}$\\ 	
$M_G$  	& 6.54$^{a,e}$  &	5.82$^{a,e}$	&	9.18$^{a,e}$	&	  9.24$^{a,e}$\\	
$B-V$  	& 1.128 $\pm$ 0.015$^{e}$ 	&	 0.867 $\pm$ 0.010$^{e}$ 	&	 1.530 $\pm$ 0.015$^{e}$ & 1.464 $\pm$ 0.005$^{e}$\\ 	
$Bp-Rp$	& 1.3534$^{a}$  &	 1.0496$^{a}$   &	 2.1173$^{a}$  & 2.3816$^{a}$\\	
$K_s$  	& 4.796 $\pm$ 0.017$^{f}$   &  5.449 $\pm$ 0.021$^{f}$	&	 5.572 $\pm$ 0.020$^{g}$  &	4.673 $\pm$ 0.020$^{g}$\\ 
Mass (\msun) & 0.77 $\pm$ 0.02$^{tw}$ & 0.82 $\pm$ 0.04$^{tw}$ & 0.426 $\pm$ 0.017$^{d}$ & 0.484 $\pm$ 0.019$^{d}$ \\ 	
Radius (\rsun)  & 0.79 $\pm$ 0.039$^{i}$ & 0.760 $\pm$ 0.007$^{tw}$ & 0.427 $\pm$ 0.013$^{d}$ & 0.481 $\pm$ 0.014$^{d}$ \\ 	
log Luminosity ($L/L_\sun$) & -0.68 $\pm$ 0.01$^{j}$ & -0.452 $\pm$ 0.004$^{tw}$ & -1.53 $\pm$ 0.011$^{d}$ & -1.49 $\pm$ 0.0052$^{d}$ \\ 	
$[\mathrm{Fe/H}]$ 	&	  0.16 $\pm$ 0.05$^{k}$  	&	  -0.16$^{h}$  	&	  -0.23 $\pm$ 0.16$^{d}$  	& -0.04 $\pm$ 0.16$^{d}$ \\ 	
T$_{eff}$ ($K$)   	&	  4610\,$\pm$\,20$^{tw}$	&	  5103 $\pm$ 20$^{tw}$  	&	  3656 $\pm$ 51$^{d}$  	& 3534 $\pm$ 51$^{d}$ \\ 	
$\log(g)$ (cm s$^{-2}$)  	&	  4.58 $\pm$ 0.02$^{l}$  	&	  4.54 $\pm$ 0.028$^{h}$  	& 4.87 $\pm$ 0.07$^{d}$ & 4.90 $\pm$ 0.07$^{d}$\\ 	
P$_{rot}$ (days)  	&	 28.626$\pm$0.046$^{[tw]}$  	&	unknown	&	38.732 $\pm$ 0.286$^{[tw]}$	&	  50.60 $\pm$ 0.41$^{[tw]}$\\ 	
v sini (km s$^{-1}$)  	&	  2.55$^{m}$  	&	  0.2 $\pm$ 0.5$^{h}$	&	  2.49$^{n}$  	&	  \textless 2 $^{p}$\\ 	
RV data  &	 APF, HIRES &	 APF, HIRES	& APF, HIRES, PFS &	 APF, HARPS\\ 	
      	 &              &	 	        & HARPS, HARPS-N  & HIRES\\
      	 &              &	 	        & SOPHIE, CARMENES        &	 CARMENES\\ 
\enddata
\tablecomments{
$^{a}$\citet{GaiaDR2}, 
$^{b}$\citet{Gray2006}, 
$^{c}$\citet{Gray2003}, 
$^{d}$\citet{Schweitzer2019}, 
$^{e}$\citet{ESA1997}, 
$^{f}$\citet{Cutri2003}, 
$^{h}$\citet{Brewer2016}, 
$^{i}$\citet{Kervella2019}, 
$^{j}$\citet{Biazzo2007}, 
$^{k}$\citet{Mishenina2013}, 
$^{l}$\citet{Franchini2014}, 
$^{m}$\citet{Mishenina2012},
$^{n}$\citet{Houdebine2016},
$^{p}$\citet{Reiners2018},
$^{[tw]}$this work,
} 
\end{deluxetable*}

\begin{table}
\centering
\caption{Stellar Abundances from \citet{Mishenina2013}}
\begin{tabular}{rcc}
\hline\hline
Species & HD~190007 & HD~216520 \\
\hline \vspace{2pt}
 $\rm{[Fe/H]}$ & 0.16 & -0.17 \\
 $\rm{[O/Fe]}$ & -0.16 & 0.20 \\
 $\rm{[Mg/Fe]}$ & -0.11 & 0.09 \\
 $\rm{[Si/Fe]}$ & 0.10 & 0.01 \\
 $\rm{[Ca/Fe]}$ & 0.13 & 0.02 \\
 $\rm{[Ni/Fe]}$ & 0.03 & -0.07 \\
 $\rm{[Y/Fe]}$ & -0.06 & -0.13 \\
 $\rm{[Zr/Fe]}$ & -0.19 & 0.00 \\
 $\rm{[Ba/Fe]}$ & -0.03 & -0.20 \\
 $\rm{[La/Fe]}$ & -0.18 & -0.12 \\
 $\rm{[Ce/Fe]}$ & -0.20 & 0.00 \\
 $\rm{[Nd/Fe]}$ & -0.16 & 0.07 \\
 $\rm{[Sm/Fe]}$ & -0.15 & 0.08 \\
 $\rm{[Eu/Fe]}$ & -0.04  & 0.09 \\
\hline\hline
\end{tabular}
\label{tab:abundances}
\tablecomments{Based on discussion in \citet{Mishenina2008, Mishenina2013} the abundance uncertainties are on the order of $\pm$0.06 dex for Fe, Mg, and Si, $\pm$0.1 dex for O, Ca, and Ni, and $\pm$0.1-0.15 dex for Y, Zr, Ba, La, Ce, Nd, Sm, and Eu.}
\end{table}


\section{HD~190007}\label{sec:190007}  

HD~190007 (GJ 775, HIP 98698) is a nearby 
(12.714\,$\pm$\,0.010 pc)\footnote{$\varpi$ = 78.6238\,$\pm$\,0.0617 mas \citep{GaiaDR2}, and distance calculated as 1/$\varpi$, including parallax zero point shift from \citet{Lindegren2018}.}, relatively 
bright \citep[V\,=\,7.46;][]{ESA1997}, K4V(k) star \citep[][; see summary of
parameters in Table \ref{table:stellarparams}]{Gray2006} 
The star is a moderately active star showing elevated
chromospheric emission via Ca H \& K
(\logrphk\, = -4.652) \citep[][]{Olspert2018} and coronal X-ray emission (log\,L$_X$/L$_{bol}$ = -4.98) \citep[][]{Hinkel2017}.
HD~190007 was classified as a BY Dra variable and designated V1654 Aql in the General Catalogue of Variable Stars (GCVS) by \citet{Kazarovets2003} based on the observed photometric variability reported by \citet[][; $b$-band amplitude of 0.016 mag]{Lockwood1997}. 
BY Draconis variables are usually K or M-type dwarfs that display quasiperiodic photometric variations on timescales of hours to months with amplitudes from 1-500 mmag that are believed to be driven by surface spots and chromospheric activity \citep[e.g.][]{LopezMorales2006}.
While the eponymous star BY Dra itself is a close binary (P = 6 days), and many BY Dra stars are close binaries, the manifestations of magnetic activity (starspots, strong emission lines) are tied to relatively rapid rotation \citep{Bopp1977}.
The star's level of chromospheric and coronal activity, and observed amplitude of photometric variability \citep[e.g.][]{MamajekHillenbrand2008,Radick1998}, appears to be normal for a mid-K dwarf with its rotation period of $P_{\rm{rot}}$ = 28.626 $\pm$ 0.046 days (Table 1, Section \ref{sec:190007photometry}).
The published \Teff\, estimates mostly cluster between $\sim$4500\,K and 4700K (see Table \ref{tab:mass1}) and we adopt \Teff\, = 4610\,$\pm$\,20\,K. 

Previous analyses of the star show good agreement on HD~190007 being slightly metal-rich
(e.g., 
[Fe/H] = 0.14$\pm$0.06 \citep{Ramirez2012},
0.16$\pm$0.05 \citep{Mishenina2013}), 
similar to the well-studied Hyades cluster \citep[Fe/H $\simeq$ 0.15-0.18][]{Dutra-Ferreira2016,Cummings2017,Liu2016}.
This metallicity enhancement likely explains why the star is slightly above the Main Sequence in B-V vs. M$_{V}$ space (dM$_{V}$ $\sim$ -0.16) \citep{Isaacson2010}.
An upper limit on the ages of thin-disk stars this metal-rich is roughly $<$8 Gyr \citep{Mishenina2013}.

The star has multiple published mass estimates, compiled in Table \ref{tab:mass1}, however we complement these values with some new independent estimates based on luminosity and absolute magnitude. 
The star has metallicity similar to the Hyades cluster,
which has [Fe/H] $\simeq$ +0.18 \citep{Dutra-Ferreira2016} and
an age of $\sim$700\,$\pm$\,100 Myr \citep[][]{Brandt2015,Martin2018,Gossage2018,Lodieu2019}. 
\citet{Torres2019} has found that stars in eclipsing binaries
in the Hyades show reasonable agreement in their mass versus
absolute magnitude trend against the PARSEC evolutionary tracks
\citep{Chen2014} for [Fe/H] = +0.18 and 625-800 Myr isochrones.
Using the \citet{Chen2014} isochrone matched to the Hyades, the absolute
magnitude ($M_V$ = 6.94) corresponds to mass 0.774 \msun.
K dwarfs evolve very slowly, and indeed a 5 Gyr isochrone for the same
chemical composition yields mass 0.756 \msun\, for the same $M_V$. 
Using the mass-luminosity trend for main-sequence stars
from \citet{Eker2018} yields a mass estimate of 0.772 \msun. 
If one fits a quadratic\footnote{log$_{10}$(M/\msun) =  0.4391 - 0.107645\,M$_V$ + 
3.818613e-3\,$M_{V}^{2}$. The fit is appropriate for FGK dwarfs (although anchored
to late A's and early M's) over 1.6 $<$ $M_V$ $<$ 9.0. The RMS scatter is about 5.7\%\,
and likely dominated by the differences in chemical abundances and ages among the
field dwarfs in the \citet{Torres2010} review.} to the absolute
magnitudes vs. log(mass) for the well-characterized FGK dwarf detached binaries compiled 
in the review by \citet{Torres2010}, for $M_V$ = 6.94 one would predict 0.752\,$\pm$\,0.043 \msun.
The distribution of mass estimates is tightly clustered
in Table \ref{tab:mass1} and we simply adopt $M_{\star}$ = 0.77\,$\pm$\,0.02 \msun\, based on the mean and standard
deviation (representing the scatter among multiple methods
to derive the mass). 

\begin{table}
\centering
\caption{\Teff\, and M$_{\star}$ Estimates for HD 190007}
\begin{tabular}{ll}
\hline\hline
\teff\, (K) & Reference \\
\hline \vspace{2pt}
4352\,$\pm$\,70 & \citet{Houdebine2012}\\
4466  & \citet{Katz2011}\\
4541$^{+80}_{-43}$ & \cite{Boeche2016}\\
4568\,$\pm$\,23 & \citet{Houdebine2019}\\
4571\,$\pm$\,117 & \citet{Bai2019}\\
4596\,$\pm$\,40 & \citet{Luck2017}\\
4597\,$\pm$\,8  & \citet{Franchini2014}\\
4599\,$\pm$\,85 & \citet{Ramirez2012}\\
4603\,$\pm$\,91 & \citet{Aguilera-Gomez2018}\\
4611\,$\pm$\,40 & \citet{GonzalezHernandez2009}\\
4616\,$\pm$\,10 & \citet{Stevens2017}\\
4637\,$\pm$\,51 & \citet{Casagrande2006}\\
4640\,$\pm$\,51 & \citet{Casagrande2010}\\
4650   & \citet{Luck2005}\\
4681\,$\pm$\,1 & \cite{Biazzo2007}\\
4709   & \citet{Anders2019}\\
4722   & \citet{Stassun2018}\\
4724\,$\pm$\,7 & \citet{Mishenina2008}\\
4724 & \citet{Frasca2015}\\
\hline
4610\,$\pm$\,20 & adopted \teff \\
\hline\hline
M$_{\star}$ ($\msun$) & Reference \\
\hline \vspace{2pt}
0.80                & \citet{Wright2011}\\
0.785 (0.773-0.801) & \citet{Takeda2007}\\
0.778\,$\pm$\,0.039 & \citet{Kervella2019}\\
0.774               & a\\
0.772               & b\\
0.76                & \citet{Luck2017}\\
0.760\,$\pm$\,0.091 & \citet{Stassun2018}\\
0.752\,$\pm$\,0.043 & c\\
0.751 (0.732-0.801) & \citet{Anders2019}\\
0.73$^{+0.03}_{-0.01}$ & \citet{Ramirez2012}\\
\hline
0.77$\pm$0.02       & adopted M$_{\star}$ \\
\hline\hline
\end{tabular}
\label{tab:mass1}
\tablecomments{
a) from \citet{Chen2014} isochrone matching Hyades ([Fe/H]=0.18, Z=0.0193, log(age/yr) = 8.85), following \citet{Torres2019}. 
b) using luminosity-mass calibration from \citet{Eker2018}.
c) from fit of M$_V$ to mass for FGK dwarfs for detached binary stars compiled by \citet{Torres2010}.
}
\end{table}

Multiple independent age estimates can be made using the star's rotation and activity indicators.
Using the X-ray age calibrations from  \citet{MamajekHillenbrand2008} we estimate an X-ray age estimate of 0.9 Gyr, which is in general agreement with previously published ages from X-rays \citep[1.08 Gyr, ][]{Vican2012}, rotational age-dating \citep[1.5$\pm$0.3 Gyr and 2.0$\pm$0.4 Gyr from ][respectively]{Ramirez2012, Aguilera-Gomez2018}, and high resolution visible spectra \citep[0.88 Gyr,][]{Kospal2009}.
However, recent results from \kepler\ and \textit{K2} studies of young clusters show that the rotational evolution of K dwarfs shows significant stalling between ages of $\sim$0.7 and 1 Gyr, and that a Skumanich-like rotational evolution law is a poor approximation \citep{Curtis2019}.
Comparing the star's combination of color \citep[$B_p - R_p = 1.3534$][]{GaiaDR2} or \Teff, and rotation period, to young clusters and field dwarfs in the Kepler field \citep[][]{Curtis2019}, we find that the star is clearly older than 1 Gyr, and its age is likely typical for field stars in the \kepler\ field. 

Variations in published RVs show evidence of a linear trend on the order of 0.5 k\ms over $\sim$16 years that could be due to a stellar or sub-stellar companion \citep{Soubiran2013}. HD 190007 does not, however, show evidence of a significant tangential velocity anomaly (dV$_{t}$= 13.89$\pm$8.16\ms) nor an abnormally high Gaia DR2 RUWE value (RUWE = 0.89), either of which would lend additional support to the possibility of a binary companion \citep{Kervella2019, Gaia2016}. HD~190007 has not previously been the target of high resolution imaging efforts to look for stellar companions using the VLT (NACO or SPHERE) or Gemini (using GPI, NIRI, or NICI), nor has it been targeted by Robo-AO. Such observations may help clarify the trend seen in the previously published radial velocity data and we encourage such follow up efforts.

Based on its Gaia DR2 results, we find that HD 190007 has a barycentric galactic velocity of U = -22.097$\pm$0.006 k\ms, V = -16.461$\pm$0.006 k\ms, W = 15.442$\pm$0.007 k\ms with $U$ measured towards the Galactic center, $V$ in the direction of  Galactic rotation, and $W$ towards the North Galactic Pole \citep[][]{ESA1997}. Adopting the Local Standard of Rest (LSR) from \citet{Bland-Hawthorn2016} gives U$_{LSR}$ = -12.1 k\ms, V$_{LSR}$ = -5.5 and W$_{LSR}$ = 22.4 k\ms, for a total LSR velocity of 26.1 k\ms. Following the
kinematic selection criteria of \citet{Bensby2014}, we estimate kinematic
membership probabilities of P$_{thin}$ = 98.74\%, P$_{thick}$=1.26\%,
P$_{halo}$ = 0.0003\%, and the probability of membership in the Hercules dynamical stream as P$_{Hercules}$ = 0.00007\%.
\citet{Mackereth2018} calculates that HD 190007's Galactic orbit has 3D pericenter and apocenter radii of 6.998 and 8.103 kpc, respectively, an eccentricity of 0.07306 and a maximum vertical excursion of 0.336 kpc. The star's Banyan $\Sigma$, a Bayesian classification tool that identifies members of young moving groups, yields P$_{field}$=99.9\% independent of whether the star's default RV or Gaia DR2 RV is used - so it does not appear to belong to any known nearby young moving groups \citep{Gagne2018}.

We use the measured abundances for the $\alpha$ elements Mg, Si, and Ca, and the Fe abundance from \citet{Mishenina2013} to estimate the $\alpha$ enrichment [$\alpha$/Fe] \citep[e.g.][]{Bovy2016}. Adopting unweighted means of the abundances of three $\alpha$ elements with respect to iron, we estimate [$\alpha$/Fe] $\simeq$ 0.04. 
Similarly, \citet{Franchini2014} estimate [$\alpha$/Fe] = 0.03\,$\pm$\,0.01. 
These kinematic calculations and abundance ratios suggest that the 
star is a metal-rich thin disk star \citep[e.g.][]{Mishenina2013}. 

HD~190007 will not be surveyed during the primary mission of NASA's \TESS\ satellite, which would have allowed for more detailed characterization of the star's brightness modulations, due to its proximity to the Earth and Moon contamination zone that the \TESS\ observing plan has been shifted to avoid. 

\subsection{HD~190007 Radial velocities}\label{sec:190007rvs}

\begin{figure}
\centering
\includegraphics[width=.48\textwidth]{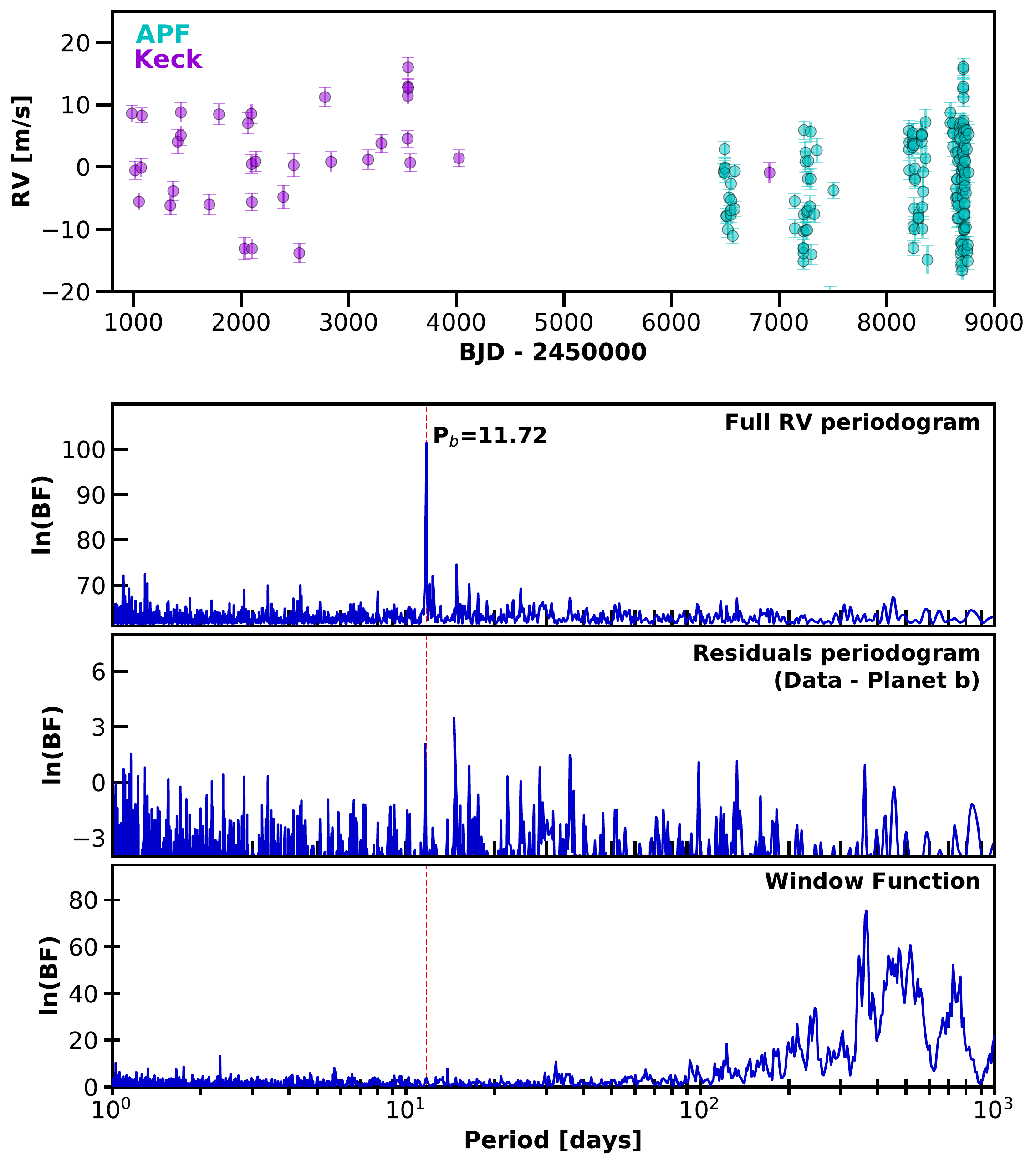}
\caption{\label{fig:HD190007vels} First panel: Unbinned radial velocity measurements of HD~190007 taken with the APF (cyan), and Keck HIRES (purple). Second panel: Bayes factor periodogram of the RV data showing the potential planet peak at 11.72 days. Third panel: Bayes factor periodogram of the RV residuals after the 11.72 day signal has been modeled and removed. Fourth panel: Spectral window function of the combined RV data sets.}
\end{figure}

The RV data set for HD~190007 contains observations from two instruments, HIRES on Keck I and the APF's Levy spectrometer. The Keck data is comprised of 34 unbinned velocities (32 individual epochs) obtained from June 1998 - September 2014 with a mean internal uncertainty of 1.54 \ms. The APF data is comprised of 157 unbinned velocities (91 individual epochs) obtained from July 2013 - October 2019 with a mean internal uncertainty of 1.50 \ms. Analysis of the combined data set reveals a prominent peak at 11.72 days (Figure \ref{fig:HD190007vels}).

To discern whether this signal could be caused by non-planetary sources, we first examine the spectral window function of the combined data set (Figure \ref{fig:HD190007vels}). Notable peaks in the window function - such as the peak at P=1.0d that results from night time observing restrictions - can cause aliases to appear at $f_{alias} = f_{planet} + f_{WF}$  \citep{Dawson2010}. None of the periods that we would expect from this star's combined window function show up prominently in the 11-12 day range of the RV periodogram, and thus we do not suspect observational aliases of masquerading as the 11.72 day signal.

\subsection{HD~190007 Activity indicators}\label{sec:190007activity}
\begin{figure}
\centering
\includegraphics[width=.45 \textwidth]{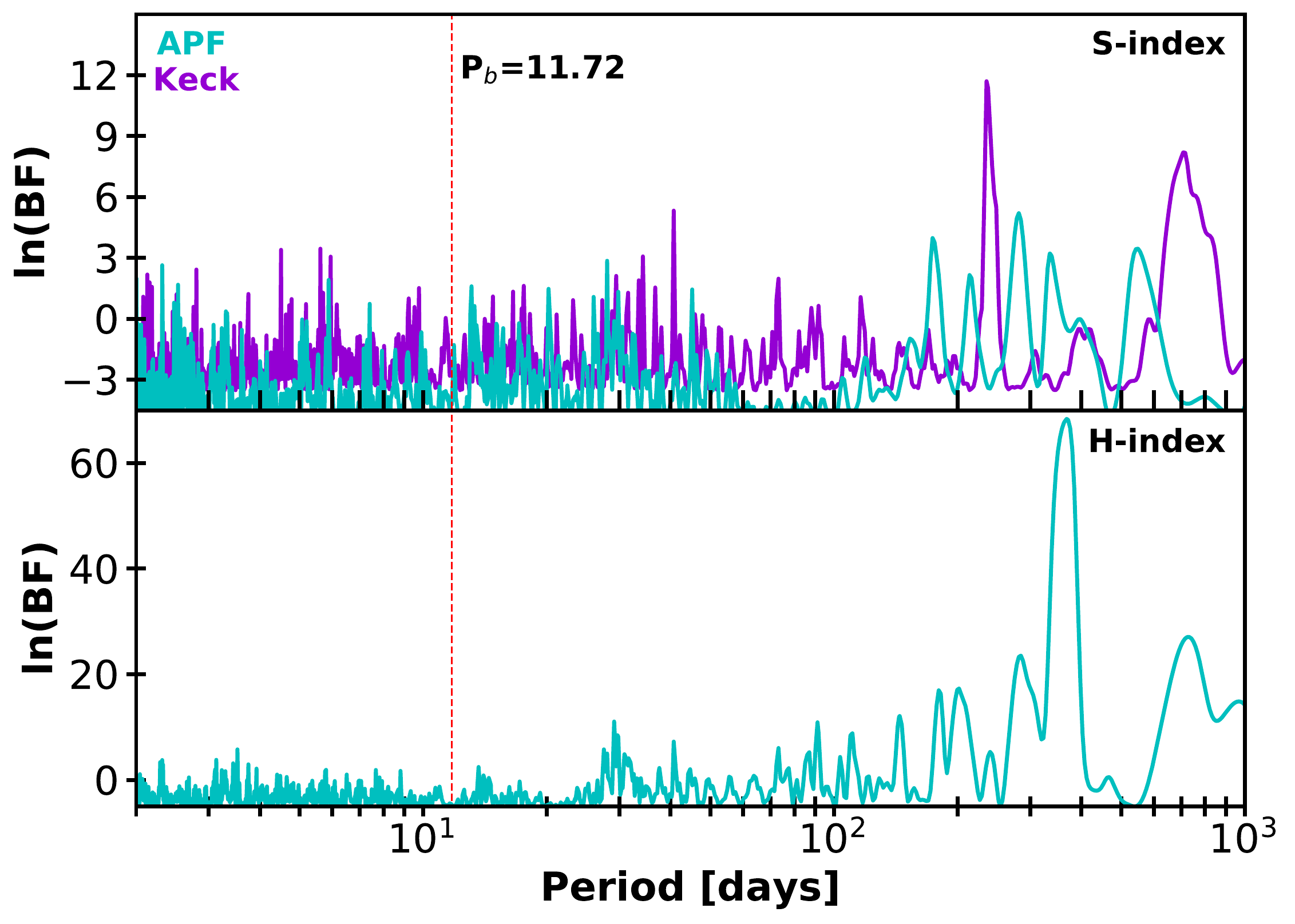}
\caption{\label{fig:HD190007activity_per} Top: Bayes factor periodogram of the S-index values measured from the APF and Keck HIRES data sets in cyan and purple, respectively. Bottom: Same as above, but for the H-index measurements extracted from the APF data. In both panels there are no prominent peaks in the vicinity of the proposed planet period of 11.72 days, but there is a broad activity-based peak at 29.18 days that is likely tied to rotational modulation evident in the APF activity indicators. The Keck HIRES data set contains only 6 observations able to produce H-index measurements, which is insufficient to produce an informative periodogram.}
\end{figure}

Only seven observations of HD~190007 were taken with Keck HIRES after the detector upgrade that expanded HIRES' wavelength range and enabled measurement of the H$\alpha$ line, which is not a large enough data set to compute a meaningful Bayes factor periodogram. None of the periodograms of the three available activity data sets (S- and H-index from the APF, and S-index from HIRES) show peaks at or near the proposed 11.72 day planet period (Figure \ref{fig:HD190007activity_per}). Based on the lack of notable signals at periods matching our potential planet signal, we conclude that the 11.72 day signal is not due to the varieties of stellar activity that produce variance in the emission of the Ca II H\&K and H$\alpha$ spectral features.

\begin{figure}
\includegraphics[width=.48\textwidth]{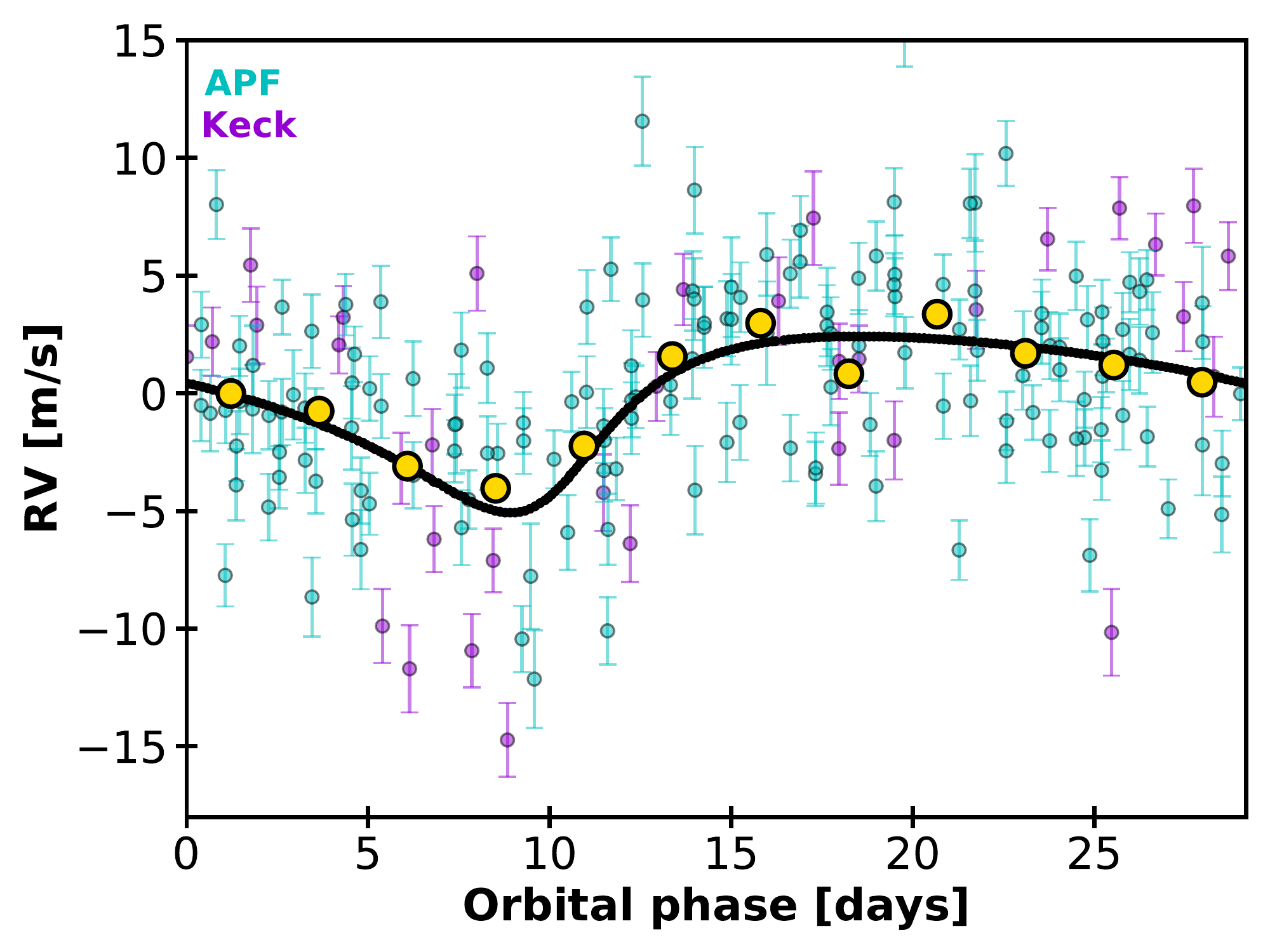}
\caption{\label{fig:HD190007_PhasedActivity} RV observations of HD~190007 phase folded to the best fit period of the suspected rotational activity cycle, P$_{\rm{act}}$= \sysonecPer\ days, with APF observations shown in cyan and Keck HIRES observations shown in purple. The error bars include the excess white noise ``jitter'' from our analysis, and the black solid curve denotes the maximum a posteriori Keplerian model. Yellow points depict the phase-binned RV data. Yellow points depict the phase-binned RV data.}
\end{figure}

The APF S- and H-index measurements show evidence of a relatively broad signal with period, P $\sim$ \sysonecPer\ days which we take to be a sign of rotational modulation due to its close match with the stellar rotation period as identified in Section \ref{sec:190007photometry}. Fitting a Keplerian model to this long term signal produces a best fit period of P=\sysonecPerunc\ days and an RV semi-amplitude of K = \sysonecKunc\ \ms (Figure \ref{fig:HD190007_PhasedActivity}). While we believe the signal to be due to stellar variability and not an additional planet, as the signal does not appear with significant power in the combined RV periodogram, we still included the signal in our overall RV model of the system.

\subsection{HD~190007 Photometry and stellar rotation} \label{sec:190007photometry}
\begin{figure}
\centering
\includegraphics[width=.48 \textwidth]{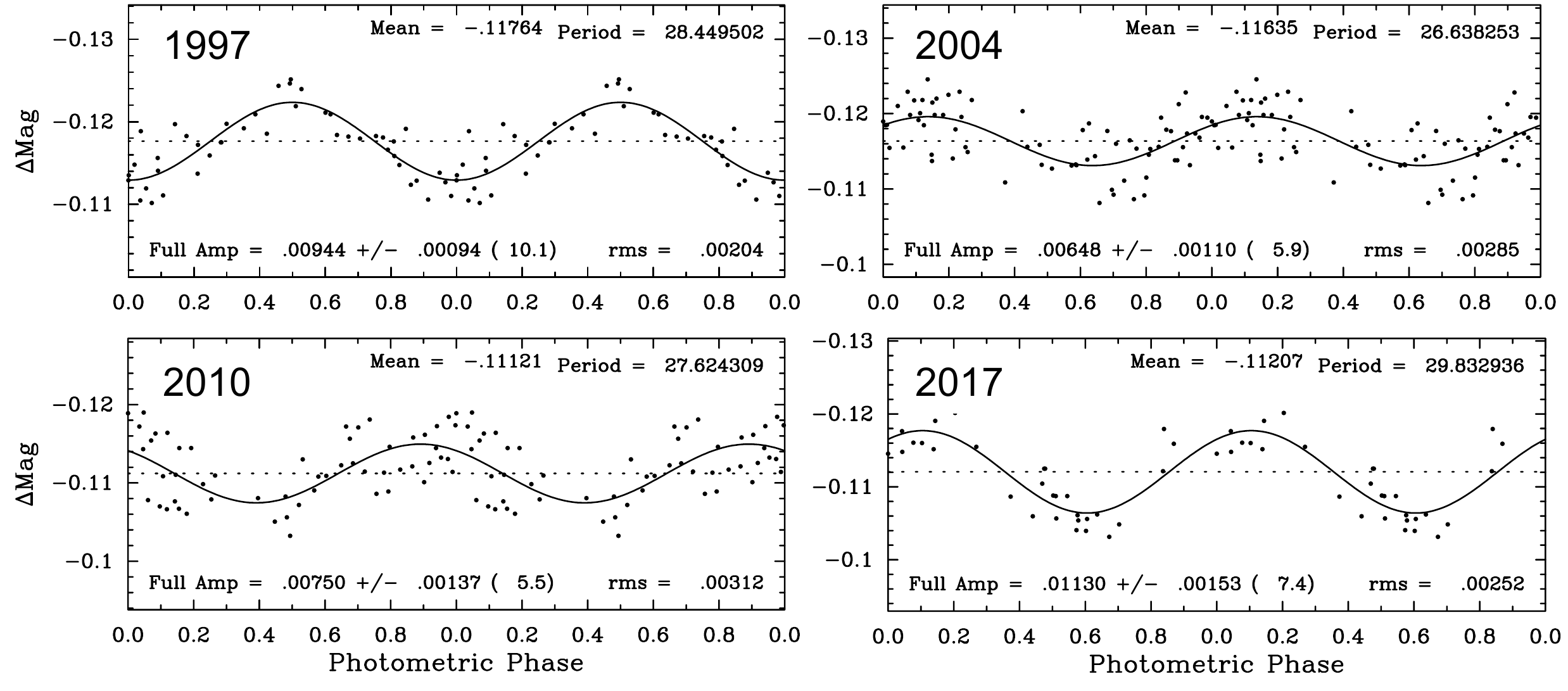}
\caption{\label{fig:HD190007Phot} A selection of four of the 19 seasons of photometry in which HD~190007 displayed a significant rotation signal. Analyzing the combined set of 19 seasons of rotational periods, we derive a weighted-mean rotation period of 28.626$\pm$0.046 days, which is in general agreement with the \sysonecPer\ day activity signal measured using the APF S- and H-index activity indicators.}
\end{figure}

A total of 1092 observations were obtained over 21 observing seasons, spanning 1997 through 2017 using the T4 0.75-m APT at Fairborn Observatory. The comparison stars used in the data analysis are HD~190521, (a K0III star with V = 7.60, B-V=1.15) and HD~187406 (an F3V star with V=7.67, B-V=0.48). The photometry shows that HD~190007 varies from year to year by roughly $\sim$0.016 mag. When searching the data for rotation periods, we find significant signals in all seasons except those covering 1999 and 2001. The star does, however, appeared to be ``double spotted" in the years 2000, 2002, 2006, and 2015. In these cases, the rotation period is twice the observed photometric period. Analyzing the 19 seasons where we measure rotational periods, we derive a weighted- mean rotation period of 28.626 $\pm$ 0.046 days (Figure \ref{fig:HD190007Phot}), which is well separated from the proposed planet periods of 11.72 days but overlaps closely with the $\sim$29 day signal seen in the APF activity indicators. This rotation period is also in agreement with the $P_{\rm{rot}}$ = 27.68 $\pm$ 0.36 day derived in \citet[][]{Olspert2018}.

\subsection{HD~190007 Orbital parameters}\label{sec:190007orbit}

Having concluded that neither our observing scheme, the star's chromospheric activity, nor the star's rotation could be causing the 11.72 day signal we observe in HD~190007's RV periodogram, we move on to testing whether or not the RV data provide enough Bayesian evidence to support the existence of the suspected planet. We apply a statistical model accounting for Keplerian signals and red noise, as well as correlations between radial velocities and activity data (see Section \ref{sec:methods}). 

\begin{figure}
\includegraphics[width=.48\textwidth]{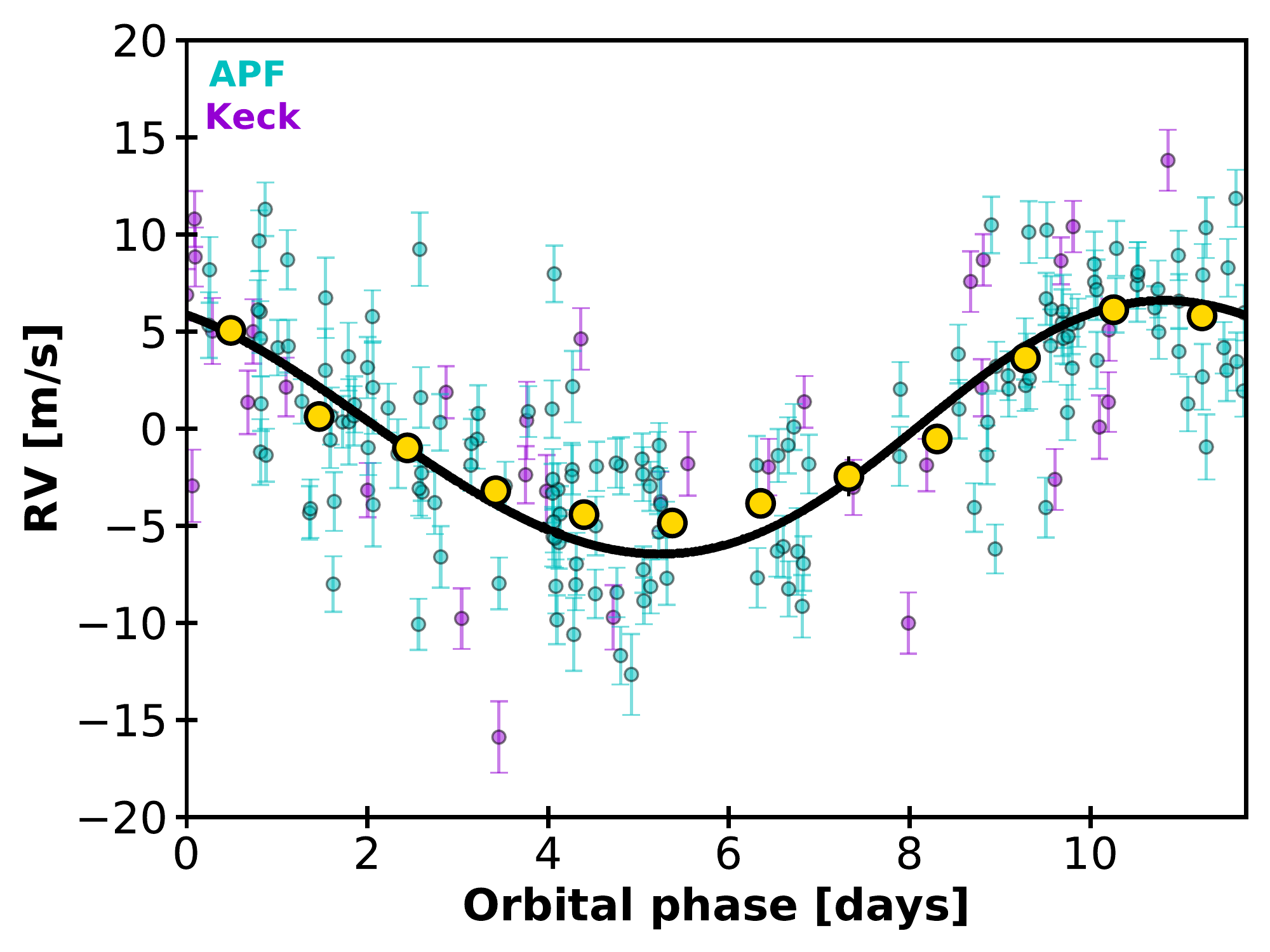}
\caption{\label{fig:HD190007PhasedRVs} RV observations of HD~190007 phase folded to the best fit period of P = 11.72 days, with APF observations in cyan and Keck HIRES observations in purple. The error bars include the excess white noise ``jitter'' from our analysis, and the black solid curve denotes the maximum \emph{a posteriori} Keplerian model. Yellow points depict the phase-binned RV data.}
\end{figure}

The resulting fit to the data reveals a planet with a period of \sysonebPer\ days, a semi-amplitude of K = \sysonebK \ms, and an eccentricity of e = \sysonebEcc\ (Figure \ref{fig:HD190007PhasedRVs}, Table \ref{table:OrbitalParamsTable}). This corresponds to a \sysonebMsini\ \mearth\ planet orbiting \sysonebA\ AU from its host star. The signal is detected at ln(BF$_{5}$) = 29.2, well above the ln(BF$_{5}$) = 5 limit we set for identifying a signal as statistically significant. Thus we promote the 11.72 day signal to being labeled as a newly identified planet candidate, HD~190007~b.

\section{HD~216520}\label{sec:216520}

HD~216520 (HIP 112527) is a V\,=\,7.53, K0V star \citep{Gray2003} located 19.6 parsecs away in the constellation of Cepheus \citep[$\varpi$ = 51.1167\,$\pm$\,0.0292 mas;][]{GaiaDR2}. 
The metallicity of the star has been estimated in several studies, with values ranging between [Fe/H] = -0.22 and -0.14, with median value [Fe/H] = -0.17, i.e. slightly metal poor \citep{Brewer2016,Luck2017,Mishenina2008,Mishenina2012,Mikolaitis2019}.
Effective temperature and mass estimates for HD 216520 are shown in Table \ref{tab:mass2}.
We adopt the recent \teff\, value from \citet{Luck2017}, whose value (5130\,$\pm$\,20\,K) falls near the middle of a tight cluster of recent estimates from high resolution spectroscopy surveys including \citet{Mishenina2008,Bermejo2013,Brewer2016,Luck2017} and \citet{Mikolaitis2019}. 
We calculate the bolometric luminosity using the Virtual Observatory SED Analyzer (VOSA)\footnote{http://svo2.cab.inta-csic.es/theory/main/} from \citep{Bayo2008}.
We use VOSA to query archival UV/Vis/IR photometry from several sources (GALEX, Tycho, Gaia DR2,
2MASS, WISE)\footnote{GALEX: \citet{Bianchi2017}, Tycho:
\citet{ESA1997}, 2MASS: \citet{Cutri2003}, WISE: \citet{Cutri2012}. Gaia
DR2 photometry was omitted as it led to large uncertainty (0.2) in the
derived bolometric magnitude.} and fit synthetic stellar spectra in
order to constrain the star's bolometric flux and independently check
the \Teff\, estimates.
We find a best fit BT-Settl-CIFIST model (constraining log($g$) = 4.5 and solar metallicity) with 5100\,K, which
has bolometric flux $f_{bol}$ = 2.949 $\times$ 10$^{-8}$ ($\pm$1\%) erg\,cm$^{-2}$\,s$^{-1}$ 
(m$_{bol}$ = 7.33\,$\pm$\,0.01 on the IAU 2015 bolometric scale).
Combined with the Gaia DR2 parallax, we estimate the absolute bolometric magnitude to be M$_{bol}$ = 5.87\,$\pm$\,0.01, and the luminosity to be log(L/L$_{\odot}$) = -0.452\,$\pm$\,0.004.
Combining this luminosity estimate with our adopted \teff\, value yields
an estimated radius of 0.760\,$\pm$\,0.007 $R_{\odot}$.
This is similar to recent estimates by \citet{Brewer2016} (0.79\,$\pm$\,0.02 \rsun), \citet{Stassun2019} (0.794\,$\pm$\,0.043 \rsun), and \citet{GaiaDR2} (0.77$^{+0.01}_{-0.02}$ \rsun).
All of these radii estimates are systematically smaller than that
predicted in the JMMC Stellar Diameter Catalog (JSDC) \citep[Version 2;][]{Chelli2016,Bourges2017}, which estimates an angular diameter
of $\theta_{LD}$ = 0.3989\,$\pm$\,0.0093 mas
(and when combined with the Gaia DR2 parallax yields an radii estimate
of 0.839\,$\pm$\,0.020 R$_{\odot}$). 
In Table \ref{tab:mass2} we list several published mass estimates and three new independent ones based on the
the estimates of the star's luminosity and absolute magnitude. 
We adopt a mass of 0.82\,$\pm$\,0.04 \msun\, for HD 216520, which spans the range
of masses estimated using evolutionary tracks and empirical trends. 


Measurements of the star's chromospheric activity through the Ca H \& K S-index and \logrphk\, indices show the star to be relatively inactive, indeed similar to that of the Sun. 
\citet{Isaacson2010} report 124 epochs of Ca H \& K measurements, showing the star's activity ranging from \logrphk\, = -4.860 to -5.006.
These values span a range not too much wider than that observed over typical solar cycles \citep[indeed almost identical to that seen over solar cycle 19;][]{Egeland2018}.
Using the rotation-activity-age relations from \citet{MamajekHillenbrand2008}, the mean \logrphk\, value from \citet{Brewer2016} (-4.93) is consistent with the star's Rossby number being 2.02, with a predicted rotation period of 43 days and an age of $\sim$6.7 Gyr. 
Previous estimates based on chromospheric activity by \citet{Wright2004} and \citet{Isaacson2010} similarly estimated ages and predicted rotation periods of 5.25 Gyr/44.0 day and 5.19 Gyr/42 day, respectively. 
Isochronal age estimates incorporating the star's HR diagram constraints and metallicity and using several different sets of evolutionary tracks by \citet{Luck2017} spanned 4.84\,$\pm$\,2.81 Gyr.


Based on its Gaia DR2 results, we find that HD 216520 has a heliocentric velocity of U = 17.458$\pm$0.007 k\ms, V = -16.905$\pm$0.006 k\ms, W = 8.341$\pm$0.009 k\ms\, with $U$ measured towards the Galactic center, $V$ in the direction of  Galactic rotation, and $W$ towards the North Galactic Pole \citep[][]{ESA1997}.
With respect to the LSR of \citet{Bland-Hawthorn2016}, the velocities
are U$_{LSR}$ = +27.5 k\ms, V$_{LSR}$ = -5.9 k\ms, W$_{LSR}$ = +15.3 k\ms, with overall LSR velocity of 32.0 k\ms. 
Using the kinematic criteria of \citet{Bensby2014}, we estimate kinematic membership probabilities for HD 216520 of for the thin disk, thick disk, halo, and Hercules dynamical stream of: P$_{thin}$ = 99.005\%, P$_{thick}$=1.004\%, 
P$_{halo}$ = 0.0002\%, and P$_{Hercules}$ = 0.000007\%.  
\citet{Mackereth2018} calculate that HD 216520's Galactic orbit has 3D pericenter and apocenter radii of 6.753 and 8.431 kpc, respectively, an eccentricity of 0.11, and a maximum vertical excursion of 0.224 kpc. 
Chemically and kinematically, the star is squarely consistent with being a member of the thin disk, corroborating previous classifications by \citet{Mishenina2013} and \citet{Hinkel2017}. 
The star's lithium abundance, logA(Li) = -0.30 \citep{Mishenina2012}, sets a lower age limit of roughly 0.5 Gyr. Note that the combination of chromospheric activity and metallicity/kinematic constraints show that the isochronal age of 11.1 Gyr as listed by \citet{Brewer2016} (6.9-13.8 Gyr) seems unlikely. 
Based on the available constraints from chromospheric activity measurements, isochronal age estimates, and Galactic kinematic constraints, we adopt an age of 6\,$\pm$\,3 Gyr.

\begin{table}
\centering
\caption{\Teff\, and M$_{\star}$ Estimates for HD 216520}
\begin{tabular}{ll}
\hline\hline
\teff\, (K) & Reference \\
\hline \vspace{2pt}
5082\,$\pm$\,25    & \citet{Brewer2016}\\
5103\,$\pm$\,20    & \citet{Luck2017}\\
5119\,$\pm$\,7.3     & \citet{Mishenina2008}\\
5119\,$\pm$\,50    & \citet{Mikolaitis2019}\\
5140$^{+59}_{-27}$ & \citet{GaiaDR2}\\
5156\,$\pm$\,132   & \citet{Bai2019}\\
5163\,$\pm$\,72    & \citet{Bermejo2013}\\
\hline
5103\,$\pm$\,20 & adopted \teff \\
\hline\hline
M$_{\star}$ ($\msun$) & Reference \\
\hline \vspace{2pt}
0.78  & a\\
0.79\,$\pm$\,0.02 & \citet{Brewer2016}\\
0.808\,$\pm$\,0.040 & \citet{Kervella2019}\\
0.84 & b\\
0.84\,$\pm$\,0.02 & \citet{Luck2017}\\
0.844\,$\pm$\,0.049 & b\\
0.850\,$\pm$0.103 & \citet{Stassun2019}\\
0.900\,$\pm$\,0.045 & \citet{Kervella2019}\\
\hline
0.82\,$\pm$\,0.04 & adopted M$_{\star}$ \\
\hline\hline
\end{tabular}
\label{tab:mass2}
\tablecomments{
a) Using \citet{Chen2014} isochrone for [M/H]=-0.16 for age 6 Gyr.
b) using \citet{Eker2018} mass-luminosity trend for adopted luminosity.
c) from fit of M$_V$ to mass for FGK dwarfs for detached binary stars compiled by \citet{Torres2010}.
}
\end{table}

\subsection{Radial velocities}\label{sec:216520rvs}

The RV data set for HD~216520 contains observations from two instruments, Keck HIRES and the APF's Levy spectrometer. The Keck data is comprised of 504 unbinned velocities (210 individual epochs) obtained from October 2001 - June 2017 with a mean internal uncertainty of 1.26 \ms. The APF data is comprised of 300 unbinned velocities (91 individual epochs) obtained from October 2014 - June 2020 with a mean internal uncertainty of 2.02 \ms.

\begin{figure}
\centering
\includegraphics[width=.48 \textwidth]{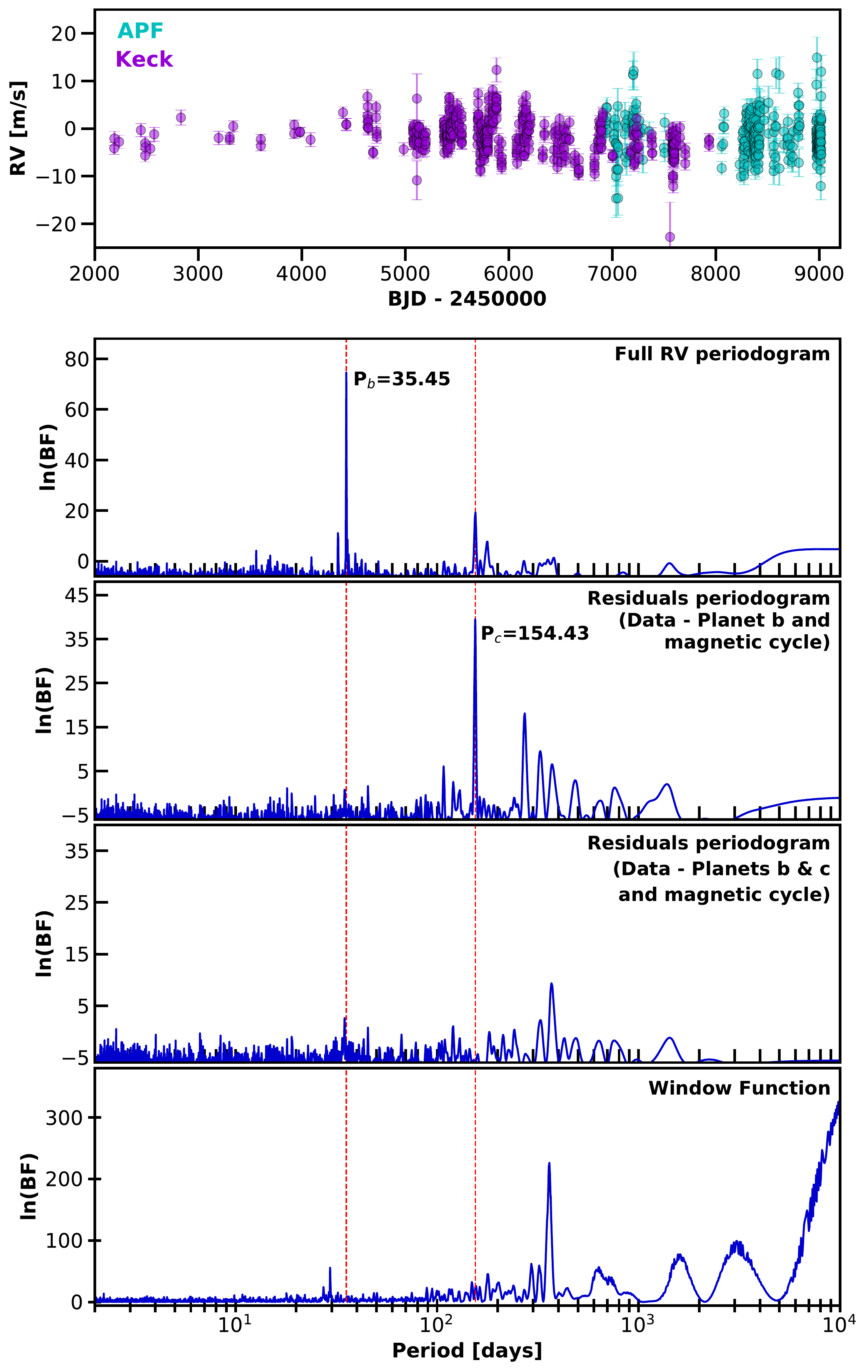}
\caption{\label{fig:HD216520vels} First panel: Unbinned radial velocity measurements of HD~216520 taken with the APF (cyan), and Keck HIRES (purple). Second panel: Bayes factor periodogram of the RV data showing the suspected planet peak at \systwobPer\ days. Third panel: Bayes factor periodogram of the RV residuals after the \systwobPer\ day Keplerian signal and the $\sim$7,700 activity signal have been fit and removed from the data sets. The second suspected planet peak at \systwocPer\ days is clearly visible. Fourth panel: Bayes factor periodogram of the RV residuals after both suspected planets and the activity signal have been modeled and removed. Fifth panel: Spectral window function of the combined RV data sets showing a lack of significant peaks that could cause alias signals at the period observed in the RVs.}
\end{figure}

Combining these data sets, we find two strong, well-defined peaks in the RV periodogram at periods of P = \systwobPer\ and \systwocPer days. When comparing this period with the data sets' combined spectral window function, we find no corresponding alias peaks (Figure \ref{fig:HD216520vels}). 

\subsection{Activity indicators}\label{sec:216520activity}

\begin{figure}
\centering
\includegraphics[width=.45 \textwidth]{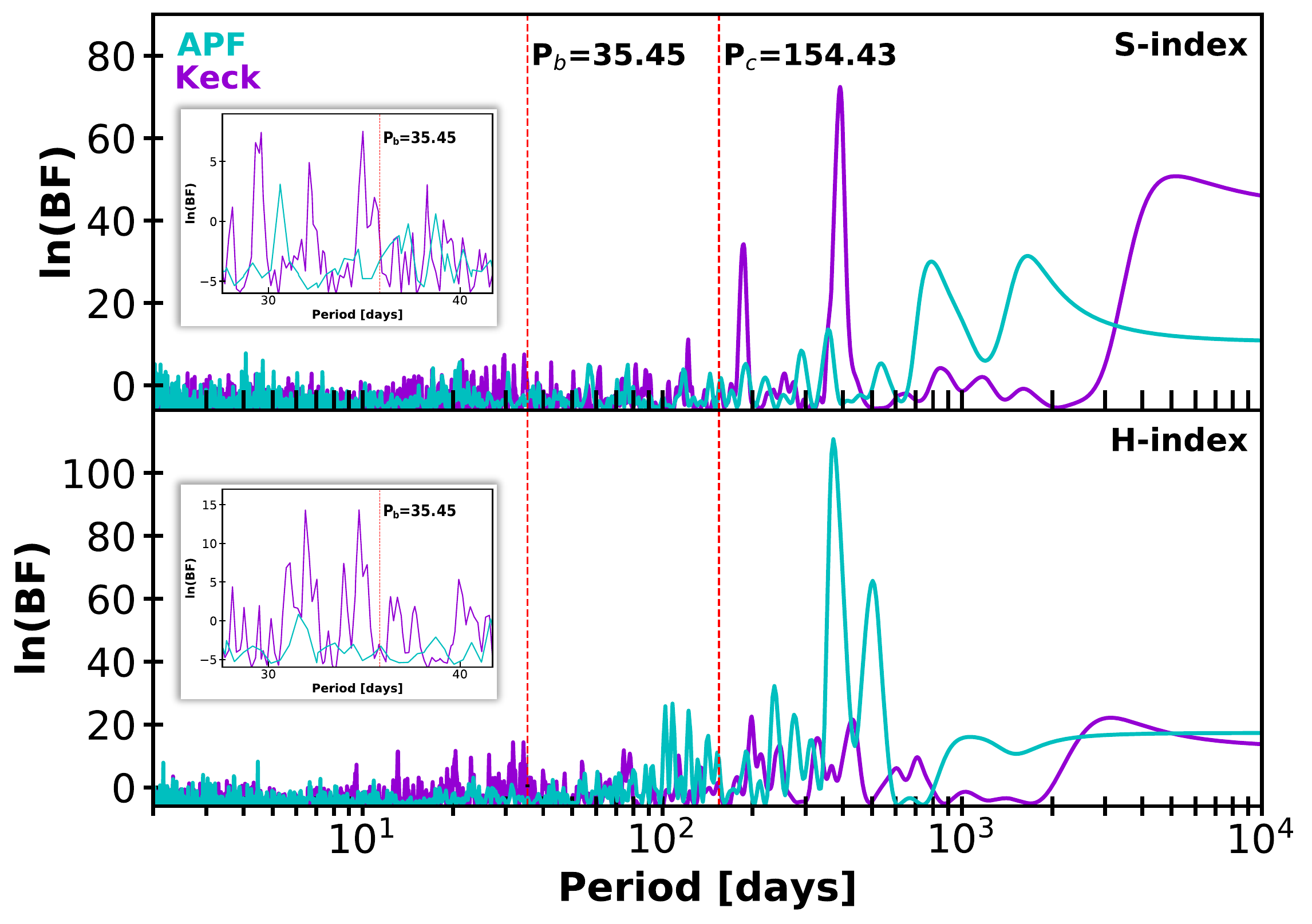}
\caption{\label{fig:HD216520activity_per} Top: Bayes factor periodogram of the S-index values measured from the APF and Keck HIRES data sets for HD~216520 in cyan and purple, respectively. Inset is a zoomed version focusing on the region around the suspected \systwobPer\ day planet signal. Bottom: Same as above, but for the H-index measurements extracted from the APF and Keck HIRES data. The planet period does not overlap with any significant peaks in either the S-index or H-index periodogram.}
\end{figure}

Plotting the Bayes factor periodograms of the S-index values extracted from the APF and HIRES spectra for HD~216520 reveals a  set of peaks in the 20-40 day period range in the Keck HIRES S-index and H-$\alpha$ data (see inset panels in Figure \ref{fig:HD216520activity_per}). Closer inspection shows that none of the HIRES activity peaks overlap directly with the suspected planet signal, and that the APF data does not show significant power in this period range, but given the proximity of the HIRES activity peaks we compare the HIRES RVs to their corresponding S- and H-index values and measure the Pearson correlation coefficients (PCC). The PCC values are 0.12 and 0.22 for the S- and H-index values, respectively, both of which fall below the PCC \textgreater 0.30 that is used as the threshold for identifying a moderate linear relationship. These low PCC values suggest that the stellar variability mapped by the S- and H-index indicators is not driving periodicity within the RV data.

\begin{figure}
\centering
\includegraphics[width=.45 \textwidth]{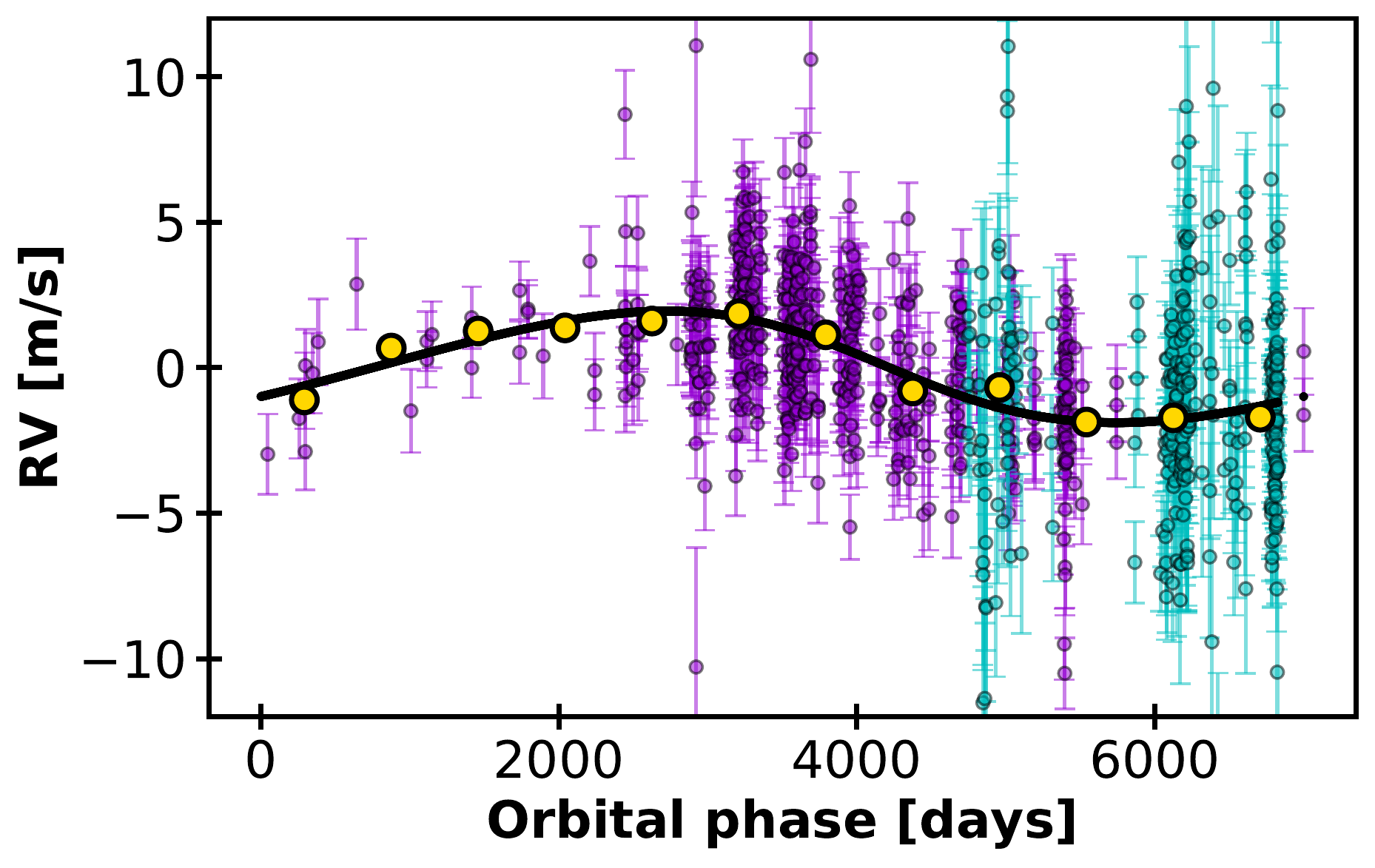}
\caption{\label{fig:HD216520_PhasedActivity} RV  observations of HD 216520 phase folded to the best fit period of the long term activity cycle, P$_{\rm{act}}$= \systwodPer\ days, with APF observations shown in cyan and Keck HIRES observations shown in purple. The error bars include the excess white noise ``jitter'' from our analysis, and the black solid curve denotes the maximum a posteriori Keplerian model. Yellow points depict the phase-binned RV data.}
\end{figure}

Both the HIRES S-index measurements and the combined HIRES and APF RV measurements show evidence of a signal with period, P $\sim$ 7700 days which we take to be a long term magnetic activity cycle (see Figures \ref{fig:HD216520vels}, \ref{fig:HD216520activity_per}). The lack of this signal in the APF S-index periodogram is not surprising as the APF data set spans only 2081 days. Fitting a Keplerian model to this long term signal produces a best fit period of P=\systwodPerunc\ days and an RV semi-amplitude of K = \systwodKunc\ \ms (Figure \ref{fig:HD216520_PhasedActivity}). While we believe the signal to be due to stellar variability and not an additional planet, given the overlap in periodicity seen in the combined RV and HIRES S-index data sets, we still included the signal in our overall RV model of the system.

\subsection{Photometry and stellar rotation}\label{sec:HD216520phot}

HD 216520 was observed by NASA's \TESS\ mission \citep{Ricker2014} during Sectors 18 (UT Nov 3 - 27 2019), 19 (UT Nov 28 - Dec 23 2019), and 20 (UT Dec 24 - Jan 21 2020). The star fell on Camera 3 during all three sectors, and on CCD 3 in sector 18 and CCD 4 in sectors 19 and 20.

\begin{figure}
\centering
\includegraphics[width=.48\textwidth]{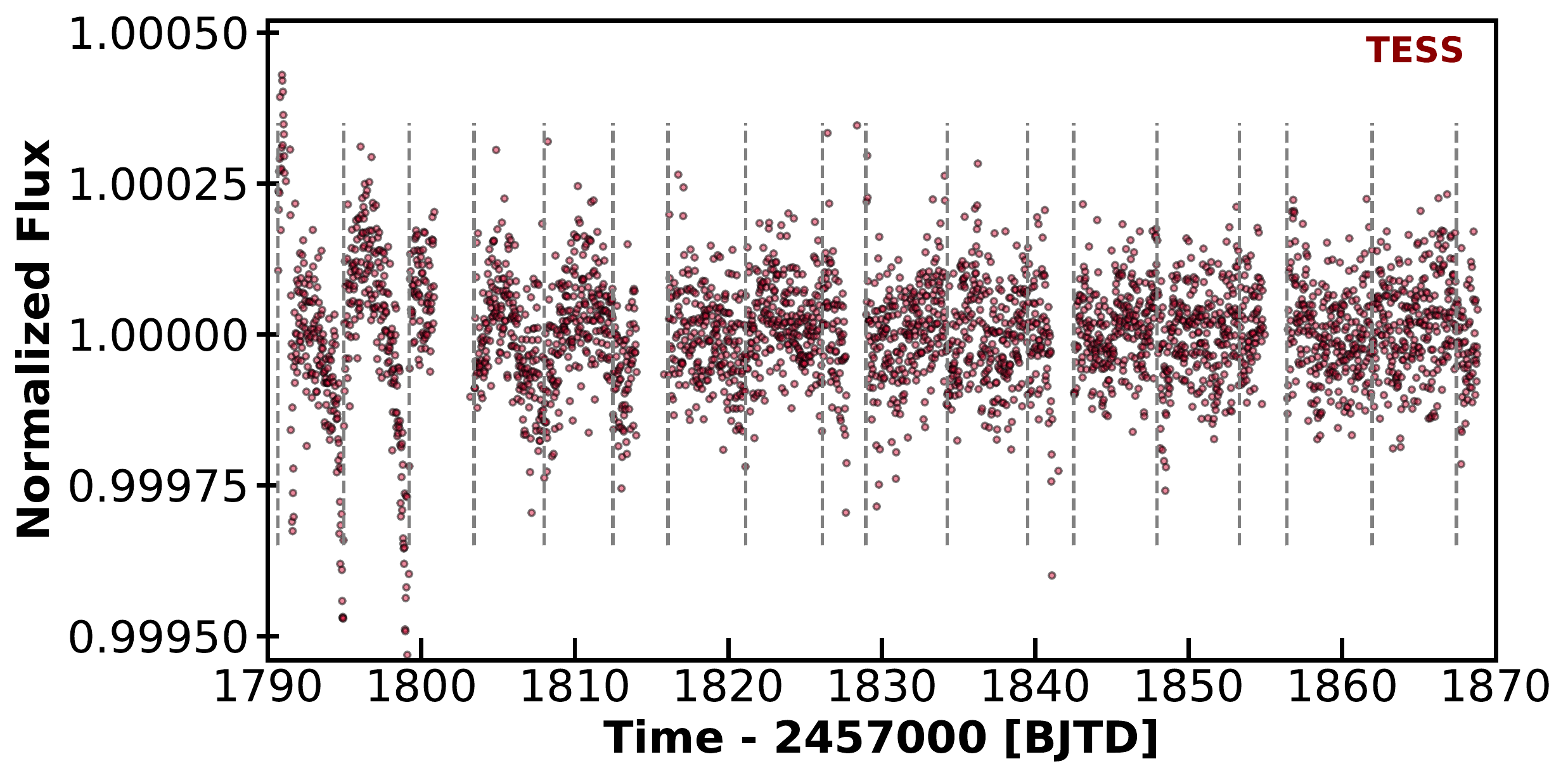}
\caption{\label{fig:HD216520_TESS} Top panel: Three sectors of presearch data conditioned simple aperture photometry (PDCSAP) flux measurements of HD 216520 taken with the \TESS\ spacecraft. The 5 day periodicity seen in the light curve is caused by the spacecraft's regularly scheduled momentum dumps. No signs of stellar periodicity are visible, leaving us unable to discern the star's rotation period from the \TESS\ data.}
\end{figure}

The Science Processing Operations Center (SPOC) data \citep{Jenkins2016} for HD~216520, available at the the Mikulski Archive for Space Telescopes (MAST) website\footnote{https://mast.stsci.edu}, includes the pre-search data conditioned simple aperture photometry (PDCSAP) flux measurements \citep{Smith2012,Stumpe2012,Stumpe2014} shown in Figure \ref{fig:HD216520_TESS}. At the start of each orbit, thermal effects and scattered light can impact the systematic error removal in PDC (see \TESS\, data release note DRN16 and DNR17). Therefore we have used the quality flags provided by SPOC to mask out unreliable segments of the time series. 

While many of the instrumental variations present in the SAP flux are removed in the PDCSAP result, there are still obvious signs of the spacecraft's momentum dumps that occur on a roughly five day period. Beyond the five day periodicity, we do not see any evidence of rotational modulation in the \TESS\ photometry. Given that HD~216520 has an estimated rotation period of 42-44 days based on its chromospheric activity indicators and rotation-activity-age relations (\S\ \ref{sec:216520}), we would expect to see almost two full rotations over the duration of the \TESS\ data. The lack of evident rotation signals therefore further supports the theory that the star is relatively inactive, as suggested by its \logrphk\ values. 

\subsection{Orbital parameters}\label{sec:216520orbit}

Having established that neither our observational cadence, stellar activity, nor stellar rotation are likely to be the cause of the \systwobPer\ day signal, we now test whether or not the combined RV data provides enough Bayesian evidence to support the existence of the suspect planet. We again apply a statistical model accounting for Keplerian signals and red noise, as well as for correlations between the radial velocities and the activity indicator data sets (see Section \ref{sec:methods}). We find that the \systwobPer\ day signal is well-supported by the data, with a Bayes factor of ln(BF$_{5}$) = 35.37, again well above the ln(BF$_{5}$) = 5 criteria for being identified as a statistically significant signal.

\begin{figure}
\includegraphics[width=.48\textwidth]{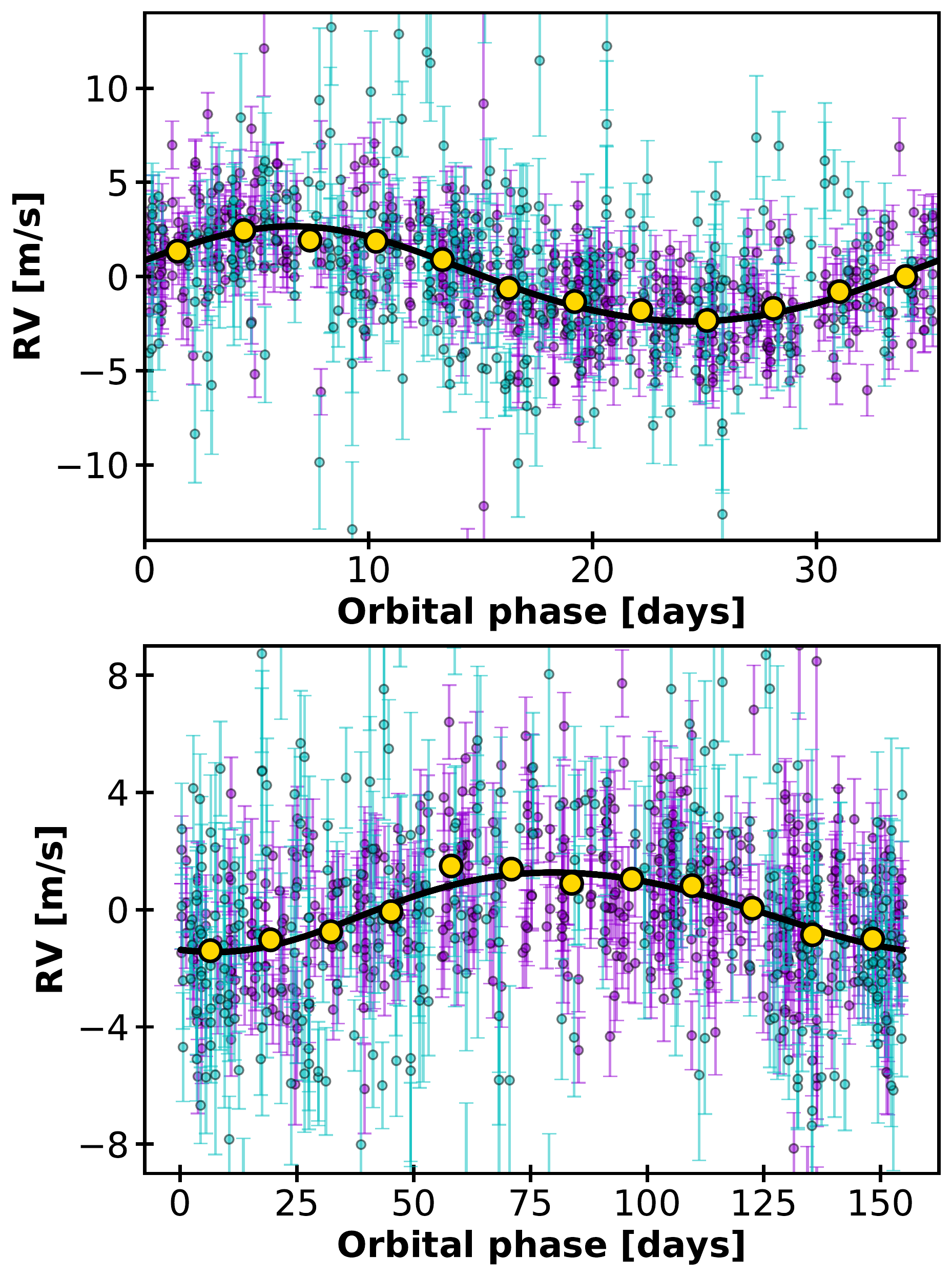}
\caption{\label{fig:HD216520PhasedRVs} Top panel: RV observations of HD~216520 phase folded to the best fit period of planet b, P$_b$ = \systwobPer\ days, with APF observations shown in cyan and Keck HIRES observations shown in purple. Bottom panel: Same as above, but folded to the best fit period of planet c, P$_c$ = \systwocPer\ days. In both cases the error bars include the excess white noise ``jitter'' from our analysis, and the black solid curve denotes the maximum \emph{a posteriori} Keplerian model. Yellow points depict the phase-binned RV data.}
\end{figure}

The resulting fit to the data reveals a two planet system, where the inner planet has a period of \systwobPerunc\ days, a semi-amplitude of K = \systwobKunc\ \ms, and an eccentricity of e = \systwobEccunc\ while the outer planet has a period of \systwocPerunc\ days, a semi-amplitude of K = \systwocKunc\ \ms, and an eccentricity of e = \systwocEccunc\ (Figure \ref{fig:HD216520PhasedRVs}, Table \ref{table:OrbitalParamsTable}). This corresponds to a \systwobMsiniunc\ M$_{\oplus}$ planet on an \systwobAunc\ AU orbit and an \systwocMsiniunc\ M$_{\oplus}$ planet on an \systwocAunc\ AU orbit, henceforth referred to as HD~216520~b and HD~216520~c, respectively.

\begin{figure}
\centering
\includegraphics[width=.45 \textwidth]{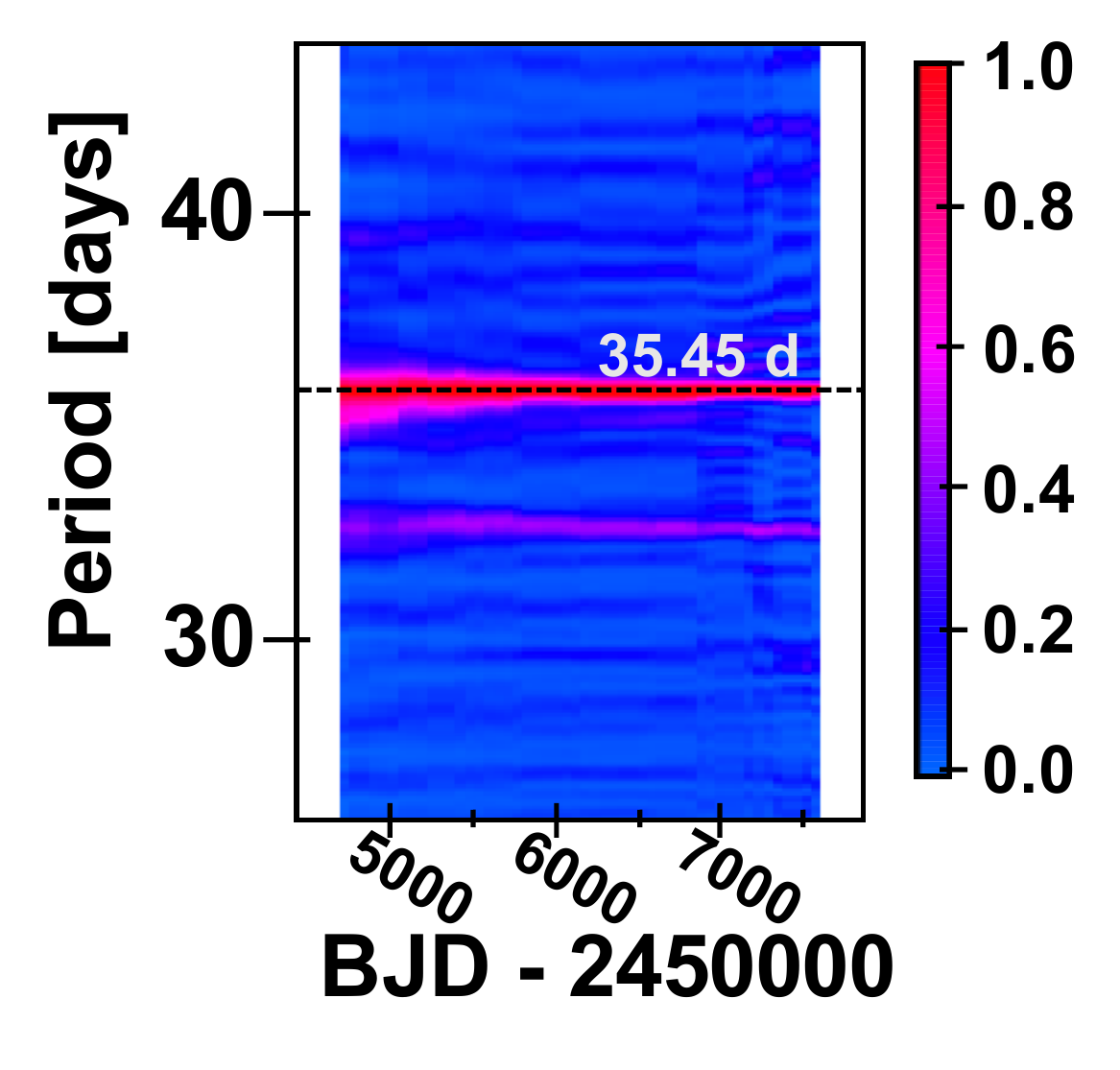}
\caption{\label{fig:HD216520_movingper} Moving MP-based periodogram  for the combined HD~216520 radial velocity data sets. The colors encode the scaled MP power, which is truncated to optimize the visualization of signals. The suspected \systwobPer\ day planet period is denoted with a horizontal dashed lines and shows robust detection through time bins encompassing more than 100 orbits.}
\end{figure}

To investigate the persistence of the proposed planetary signal over time, we create a Moving Periodogram (MP) using the MA(1) noise model of \citet{Feng2017b}. The Moving Periodogram is made by constructing Bayesian periodograms for the HD 216520 radial velocity data within a moving time window \citep[see][for a detailed description]{Feng2017b}. In doing so, we find that the \systwobPer\ day signal is detected robustly and repeatedly as more RV data points are added. Indeed, the signal in consistent through an observational baseline that covers more than 90 orbits of the suspected planet and more than 9 seasons of ground based RVs (Figure \ref{fig:HD216520_movingper}). If RV peaks are long-lived, and survive over numerous seasons of observation, then stellar activity is generally considered unlikely to be the source of the signal \citep[see, e.g.][]{Buchhave2016, Pinamonti2019}. As the moving MP does not show evidence of the \systwobPer\ day signal evolving over time nor show a broad distribution of period measurements over the decade long RV baseline, as we would expect were it caused by stellar variability, we take this as additional evidence that the signal is not due to surface variability on the host star. 

\subsection{Transit Search}\label{sec:216520Transit}

Using the transit probability equation from \citet{Winn2010} we calculate $P_{\rm{tra}}$ for each of the HD 216520 planet candidates. 
For planet b, $P_{\rm{tra}}$ = 1.8\% while for planet c the probability drops to $P_{\rm{tra}}$ = 0.6\%. 
Given the planets' best fit minimum masses (\systwobMsini\ \mearth\ and \systwocMsini\ \mearth\ for planets b and c, respectively) and the conditional distributions for planet radius given a measured planet mass presented in \citet{Ning2018}, we would expect these planets to have radii in the 2-4 \rearth\ range. A search of the three sectors of \TESS\ data does not reveal signs of transits for either planet, neither via a traditional box least squares (BLS) search nor via phase folding the photometry at the best fit RV periods.

\begin{figure}
\centering
\includegraphics[width=.48\textwidth]{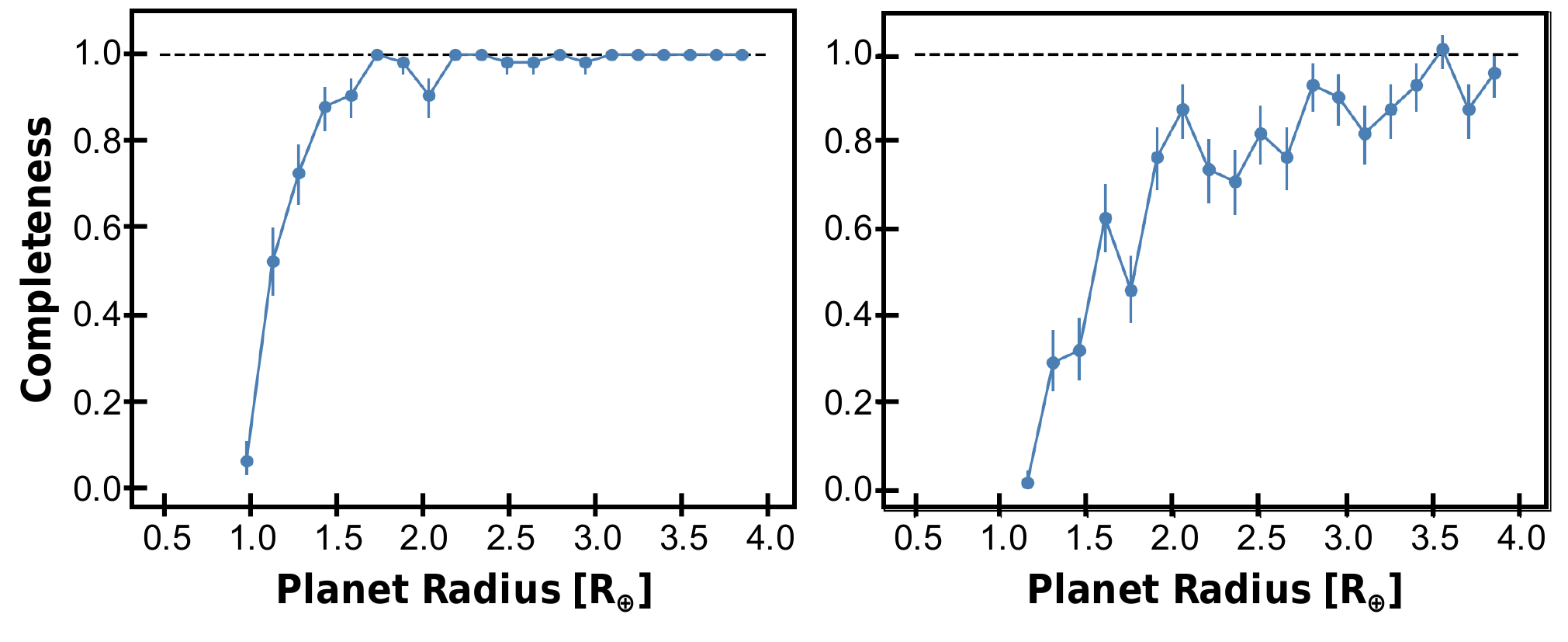}
\caption{\label{fig:HD216520_TESS_IR} Results of our injection recovery analysis using the HD 216520 \TESS\ data for planet b (left panel) and planet c (right panel). For planet b, we are able to detect the planet more than 80\% of the times when the planet radius is larger than 1.4 \rearth. For planet c, we are able to detect the single transit more than 80\% of the times when the planet radius is larger than 2.0 \rearth.}
\end{figure}

To test the likelihood that any potential transits may have been missed in the \TESS\ data, we performed an injection and recovery experiment to determine if we could detect these planets assuming they do transit (Figure \ref{fig:HD216520_TESS_IR}). We use the period posterior derived from the RV data, and injected the transits into the SPOC PDCSAP lightcurves assuming a uniform distribution of epochs within the TESS baseline, and a uniform distribution of impact parameters. We injected planets over a grid of radii between 1.0 and 4 \rearth, with a step size of 0.15 \rearth. We then detrended the light curves using the Kepler spline \citep{VanderburgJohnson2014}, and used the BLS method to search for the transits using the best fitted RV period. We define a detection of the injected planet if the signal is detected at the correct epoch and with a BLS pink noise to signal ratio larger than 10. For planet b, we are able to detect the planet more than 80\% of the time when the planet radius is larger than 1.4 \rearth. For planet c, we are able to detect the single transit more than 80\% of the time when the planet radius is larger than 2.0 \rearth.  Given the expected planet radii, we therefore conclude that planet b does not transit, and that planet c did not transit during the observational baseline covered by the \TESS\ photometry.

\subsection{Dynamic Stability}\label{sec:216520Stability}

A commonly used metric to assess the stability of two-planet systems is the separation of the planets in units of the mutual Hill radius
\begin{align}
    R_\mathrm{H,mut} = \left(\frac{a_1+a_2}{2}\right) \left( \frac{m_1 + m_2}{3 M_*} \right)^{1/3}
\end{align}
where $m_{1,2}$ and $a_{1,2}$ are the masses and semi-major axes of the planets respectively, and $M_*$ is the mass of the central star. Analytic calculations show that two low-eccentricity, low-inclination planets are ``Hill stable," meaning that they are protected from close encounters for all time, if their spacing in mutual Hill radii, $\Delta \equiv (a_2 - a_1)/R_\mathrm{H,mut}$, exceeds $2 \sqrt{3}$ \citep{Gladman1993, Chambers1996}.

While this criterion is only approximate, the spacing between the HD 216520 planets indicates that stability is likely not a concern. Using MAP parameters, the mutual Hill radius between the two planets is $R_\mathrm{H,mut} \approx 0.01 \, \mathrm{au}$, giving a spacing in mutual Hill radii of $\Delta \approx 31.6$.

As another check, we used the N-body integration package \texttt{rebound} \citep{ReinLiu2012} to integrate the system for $10^7$ orbital periods of the outer planet, i.e. roughly 4.2 Myr. We employed the \texttt{whfast} integrator \citep{ReinTamayo2015}, using a timestep of $0.01 \times \, (\text{inner planet's period})$. We see no indications of instability during the course of this integration. Over the course of the integration the variation in both planets' semi-major axes is $<10^{-5} \, \mathrm{au}$.


\section{GJ~686}\label{sec:GJ686}

GJ~686 is a V=9.62 \citep{ESA1997} M1.5V \citep{Lepine2013} star $d$ = 8.157\,$\pm$\,0.001 pc away in Hercules \citep[$\varpi$ = 122.5609\,$\pm$\,0.0346 mas][]{GaiaDR2} (Table \ref{table:stellarparams})\footnote{The proximity of GJ 686 was first 
reported and its parallax measured (108\,$\pm$\,11 mas) by \citet{Slocum1913}.}.
The star was recently reported to host an \msini\, = $7.1\,\pm\,0.9$ \mearth\, planet on a 15.53 day orbit by \citep[][]{Affer2019}, and the signal was subsequently confirmed by \citet[][]{Lalitha2019} (\msini\, = $6.24^{+0.58}_{-0.59}$\, \mearth). Here we present an updated orbital analysis of the system that includes additional data taken with the APF and PFS, and present an updated minimum mass with 7\%\, uncertainty. 

\subsection{Radial velocities}\label{sec:GJ686rvs}

\begin{figure}
\centering
\includegraphics[width=.48 \textwidth]{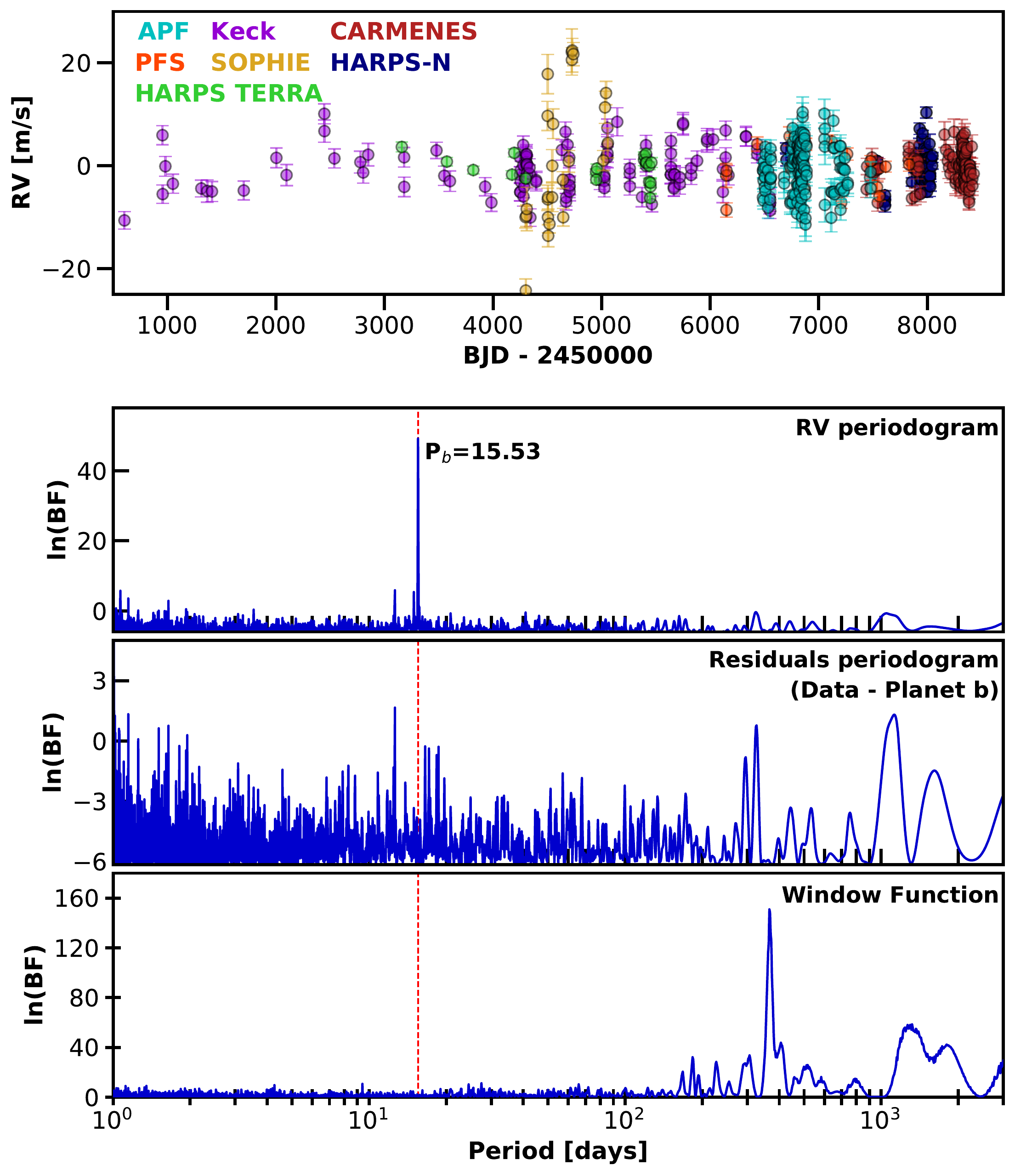}
\caption{\label{fig:GJ686vels} First panel: Unbinned radial velocity measurements of GJ~686 color coded by instrument. Second panel: Bayes factor periodogram of the RV data points showing the peak we detect at \systhreebPer\ days. Third panel: Residuals periodogram after the 15 day signal has been fitted and removed from the data. Broad peaks in the 1000-2000 day region remain. Fourth panel: Spectral window function of the combined RV data sets showing a lack of signals that could cause the 15.53 day signal.}
\end{figure}

The previously published data sets for GJ~686 include 114 unbinned Keck HIRES velocities (90 individual epochs) from June 1997 - September 2013, 20 HARPS velocities (19 individual epochs) from June 2004 - September 2010, 25 SOPHIE velocities obtained from July 2007 to August 2009, 64 HARPS-N velocities obtained from February 2014 - October 2017, and 100 velocities obtained with the visible arm of CARMENES from February 2016 - November 2018. These data are described in detail in \citet{Affer2019} \& \citet{Lalitha2019}. To these, we add 134 unbinned APF velocities (59 individual epochs) taken from July 2013 - March 2016, and 18 PFS velocities (18 individual epochs) taken from August 2012 - March 2017. Additionally, we reprocess the HARPS data included in the previous papers using the TERRA pipeline \citep{Anglada-Escude12a} before performing our own orbital fits. Searching the RV periodogram of the combined data set we find a strong, well defined peak at P = 15.53 days, matching the planet period found in the previously published works, and a lack of potential alias signals in the combined data's window function (Figure \ref{fig:GJ686vels}). After removing the 15 day signal, we detect an additional signal with a period of P $\sim$ 2000 days, which we (like \citet{Affer2019}) find to be evidence of a long-term activity cycle based on the Keck and APF activity indicators.

\subsection{Photometry and stellar rotation}\label{sec:GJ686photometry}

\begin{figure}
\centering
\includegraphics[width=.48 \textwidth]{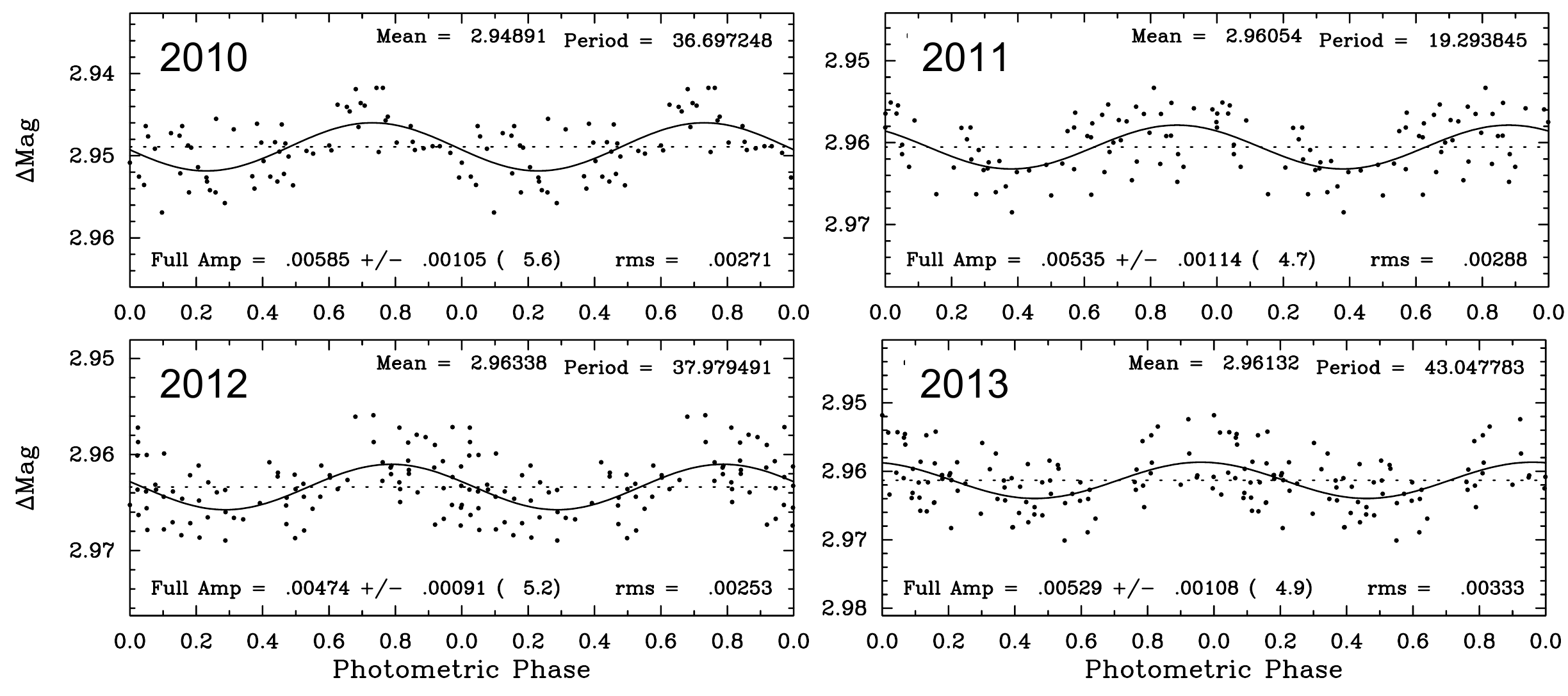}
\caption{\label{fig:GJ686Phot} The first four years of GJ~686 photometry, in which we detect coherent rotation signals. From these four data sets we derive a weighted mean rotation period of 38.732 $\pm$ 0.286 days.}
\end{figure}

A total of 508 photometric observations were obtained over 7 observing seasons, spanning 2010 through 2016, using the T12 0.8m APT at Fairborn Observatory. The comparison stars used in our data analysis were HD 158806 (an F6IV star with V = 6.92, B-V=0.46) and HD 159063 (a G0V star with V=6.98, B-V=0.53). The photometry shows that GJ~686 varies from year to year by roughly $\sim$1\%, consistent with low to moderate activity. When searching the photometric data for periodicity, we find significant signals only in the first four seasons.  The star appears to be "double spotted" in 2011, cutting the observed photometric period in half. A weighted mean of the four photometric periods (doubling that from the 2011 observing season) gives us a value of 38.732 $\pm$ 0.286 days (Table 1, Figure \ref{fig:GJ686Phot}), which we take to be our best estimate of the stellar rotation period.  We note that this is well separated from our proposed new planetary period.

\subsection{Activity indices}\label{sec:GJ686activity}

\begin{figure}
\centering
\includegraphics[width=.45 \textwidth]{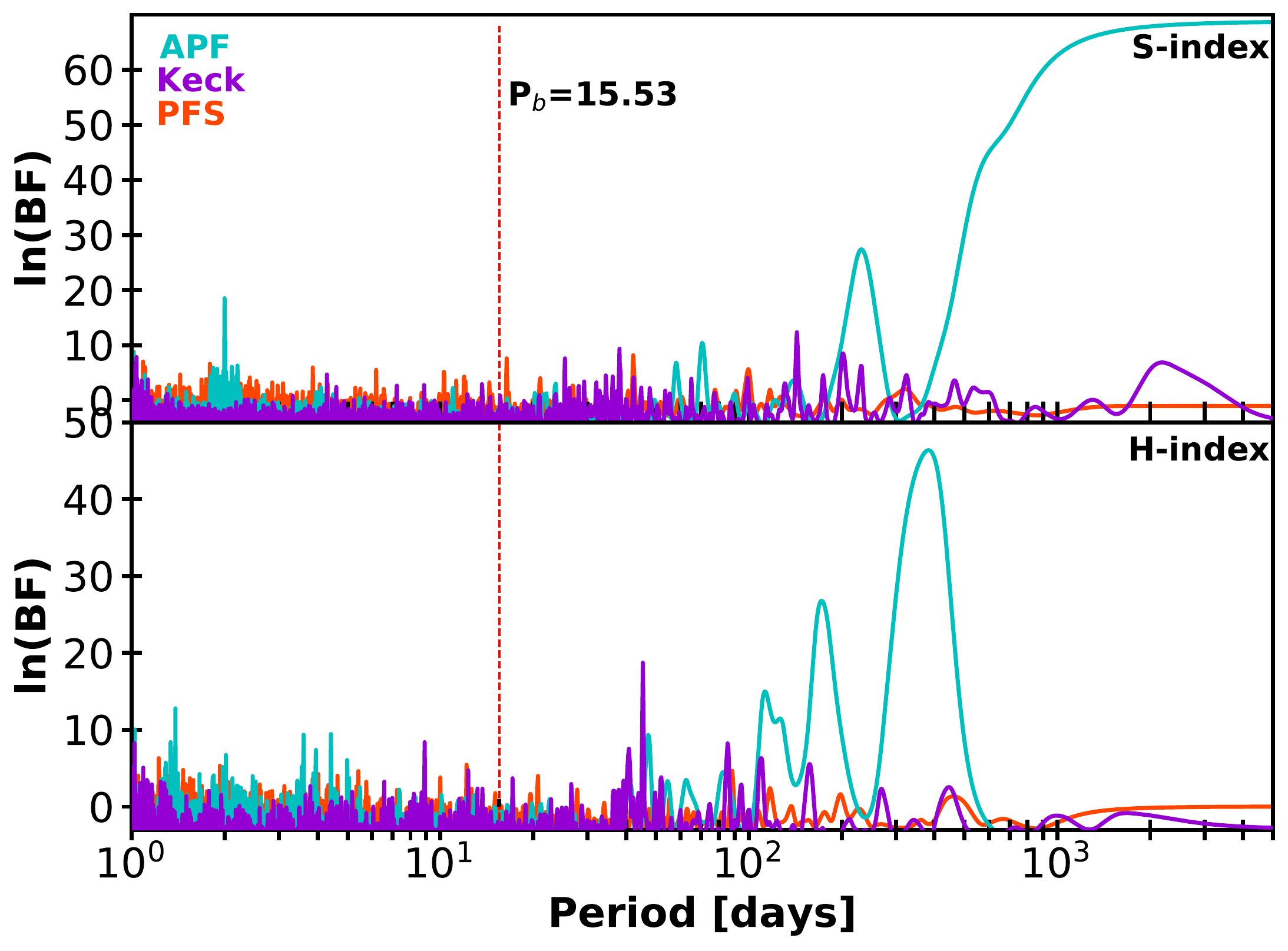}
\caption{\label{fig:GJ686activity_per} Top: Bayes factor periodogram of GJ~686 using the S-index values measured from the APF, Keck HIRES, and PFS data sets in cyan, purple, and orange, respectively. There are no prominent peaks in the vicinity of the 15.53 day planet period. Bottom: Same as above, but for the H-index measurements extracted from the Keck HIRES and PFS data in purple and orange, respectively.}
\end{figure}

When plotting the S- and H-index Bayes factor periodograms using measurements extracted from our APF, HIRES, and PFS spectra for GJ~686, we find that none of the periodograms of the various activity data sets show peaks at or near the 15.53 day planet period (Figure \ref{fig:GJ686activity_per}). We do, however, see a peak at P = 40.8 days, which is in rough agreement with the stellar rotation period we measure, and those presented by \citet{Affer2019} (P$_{rot}$ = 36.7 days) and \citet{Lalitha2019} (P$_{rot}$ = 38.4 days). When considering the 2000 day signal, which the HIRES S-index data shows clearly at the ln(BF$_{5}$)$\sim$10 level, we note that of the individual stellar activity data sets only the Keck HIRES indicators covers a long enough time baseline to be sensitive to such a long-period sinusoid. We identify this signal as a likely long-period magnetic cycle, and with a period of $\sim$5.5 years it aligns well with the typical long term variability timescales for early to mid M-dwarfs \citep{GomesdaSilva2012, Suarez-Mascareno2017}.

\subsection{Orbital parameters}\label{sec:GJ686orbit}

\begin{figure}
\includegraphics[width=.48\textwidth]{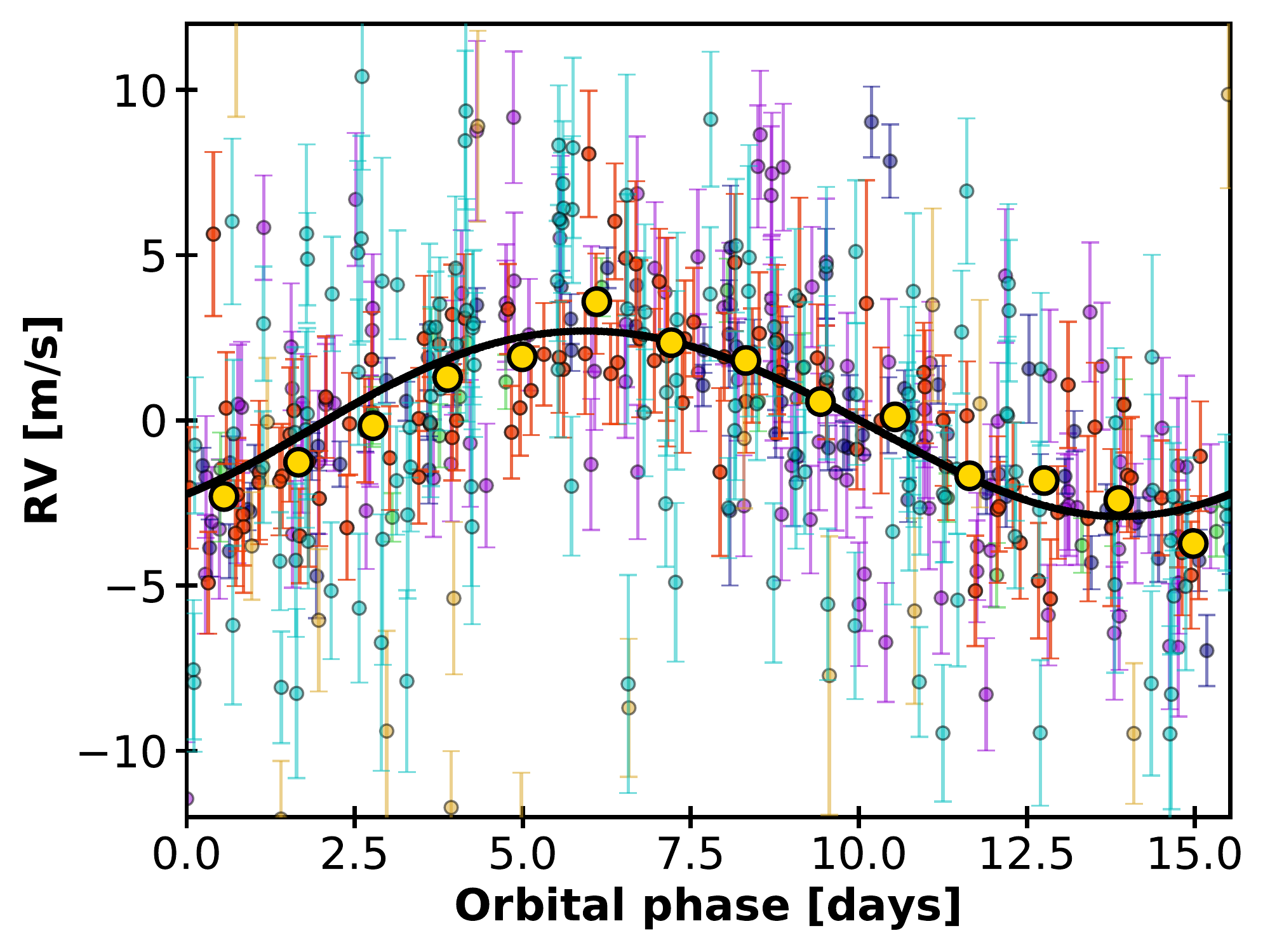}
\caption{\label{fig:GJ686PhasedRVs} Top panel: RV observations of GJ~686 (colors match those in Figure \ref{fig:GJ686vels}) phase folded to the best fit period of P = 15.53 days. The error bars include the excess white noise ``jitter'' from our analysis, and the black solid curve denotes the maximum a posteriori Keplerian model. Yellow points depict the phase-binned RV data.}
\end{figure}

Unsurprisingly, given its status as a previously published planet, we find that the 15.53 day signal is well supported by the combined RV data sets, with ln(BF$_{5}$) = 65.88. The resulting fit to the data reveals a planet with a period of \systhreebPer\ days, a semi-amplitude of K = \systhreebK\ \ms, and an eccentricity of e = \systhreebEcc\ (Figure \ref{fig:GJ686PhasedRVs}, Table \ref{table:OrbitalParamsTable}). This corresponds to a \systhreebMsini\ \mearth\ planet orbiting \systhreebA\ AU from its host star. These values are in good agreement with the previously published detections of GJ~686 b, and our RV semi-amplitude matches the smaller value measured in \citet{Lalitha2019} (K=3.02$^{+0.18}_{-0.20}$ \ms) more closely than the value measured by \citet{Affer2019} (K=3.29$^{+0.31}_{-0.32}$ \ms). 

\section{HD~180617 b}\label{sec:180617}

HD~180617 is an M3 dwarf star located just under six parsecs away from the Sun \citep{Gaia2016}. The star was recently found to host a 12.2 \msini\ planet on a 105.9 day orbit using data from Keck HIRES, HARPS, and CARMENES \citep{Kaminski2018}. Here we present an updated orbital fit that incorporates an additional 126 radial velocity measurements taken with the APF between July 2013 and July 2019. 

\subsection{Radial velocities}\label{sec:180617rvs}

\begin{figure}
\centering
\includegraphics[width=.48 \textwidth]{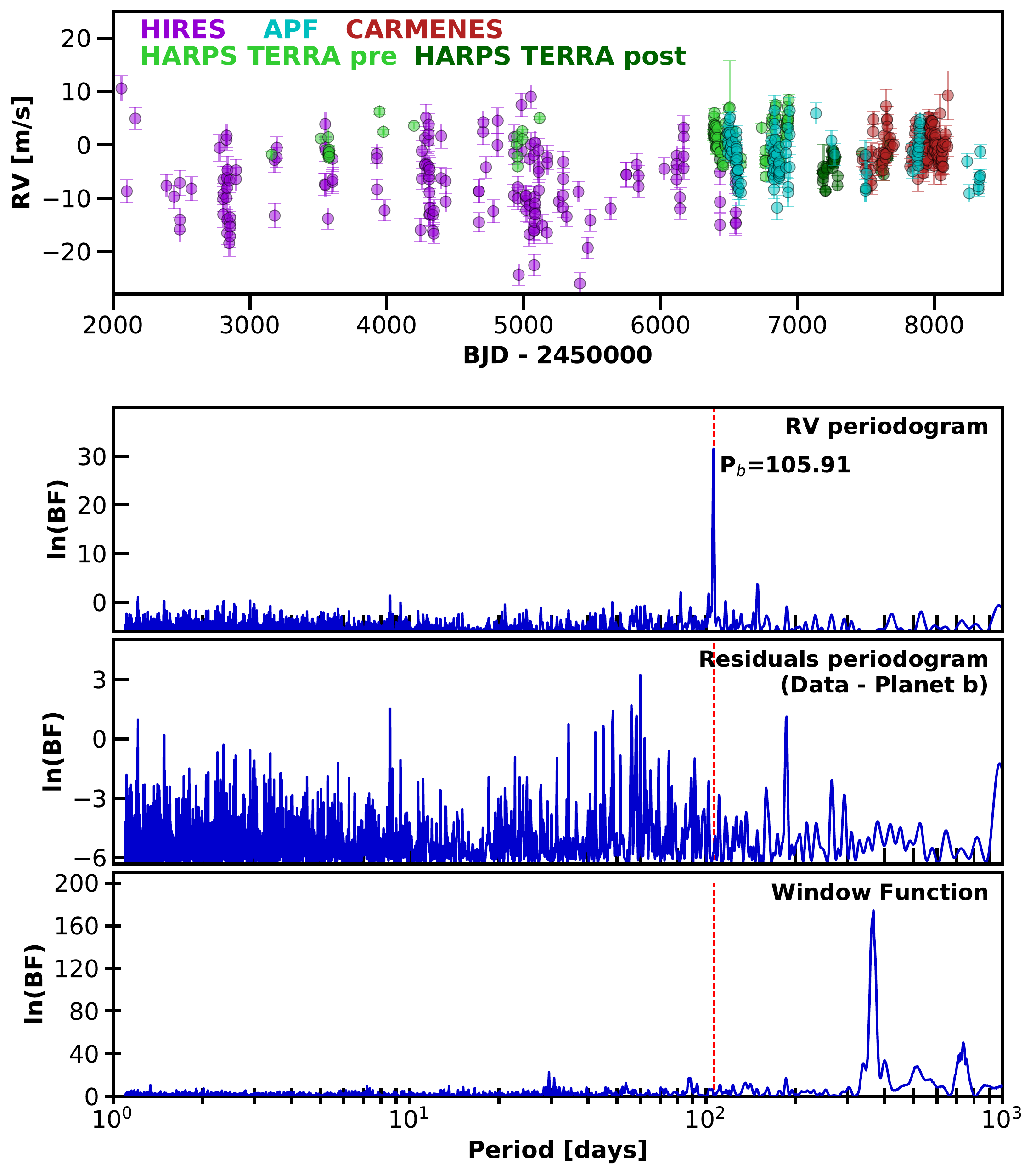}
\caption{\label{fig:HD180617vels} First Panel: Unbinned radial velocity measurements of HD~180617 taken with the APF (cyan), Keck HIRES (purple), and HARPS (green). Second Panel: Bayes factor periodogram of the RV data showing the planet's peak at \sysfourbPer\ days. Third panel: Residuals periodogram after the \sysfourbPer\ day signal has been fitted and removed. Fourth panel: Window function of the combined RV data set showing a clear lack of significant peaks at the proposed planet period.}
\end{figure}

The published RV data sets for this star include 158 unbinned HIRES velocities (134 individual epochs) taken from June 2001 to September 2014, 108 HARPS velocities taken before the fiber upgrade, 40 HARPS velocities taken after the fiber upgrade, and 124 CARMENES velocities. These RV data sets have mean uncertainties of 2.57, 0.85, 0.45, and 1.59 m s$^{-1}$, respectively, and are described in detail in \citet{Kaminski2018}. The APF data included here is comprised of 126 radial velocity measurements (58 individual epochs) taken between July 2013 and July 2019 that have a mean uncertainty of 1.65 m s$^{-1}$.  Searching  the  combined  data set, we find a strong, well-defined peak in the RV periodogram at P = \sysfourbPer\ days, which is consistent with the planet period detected in \citet{Kaminski2018}, and a corresponding lack of signals in the combined RV window function that could cause the RV peak (Figure \ref{fig:HD180617vels}).

\subsection{Activity indicators}\label{sec:180617activity}

\begin{figure}
\centering
\includegraphics[width=.45 \textwidth]{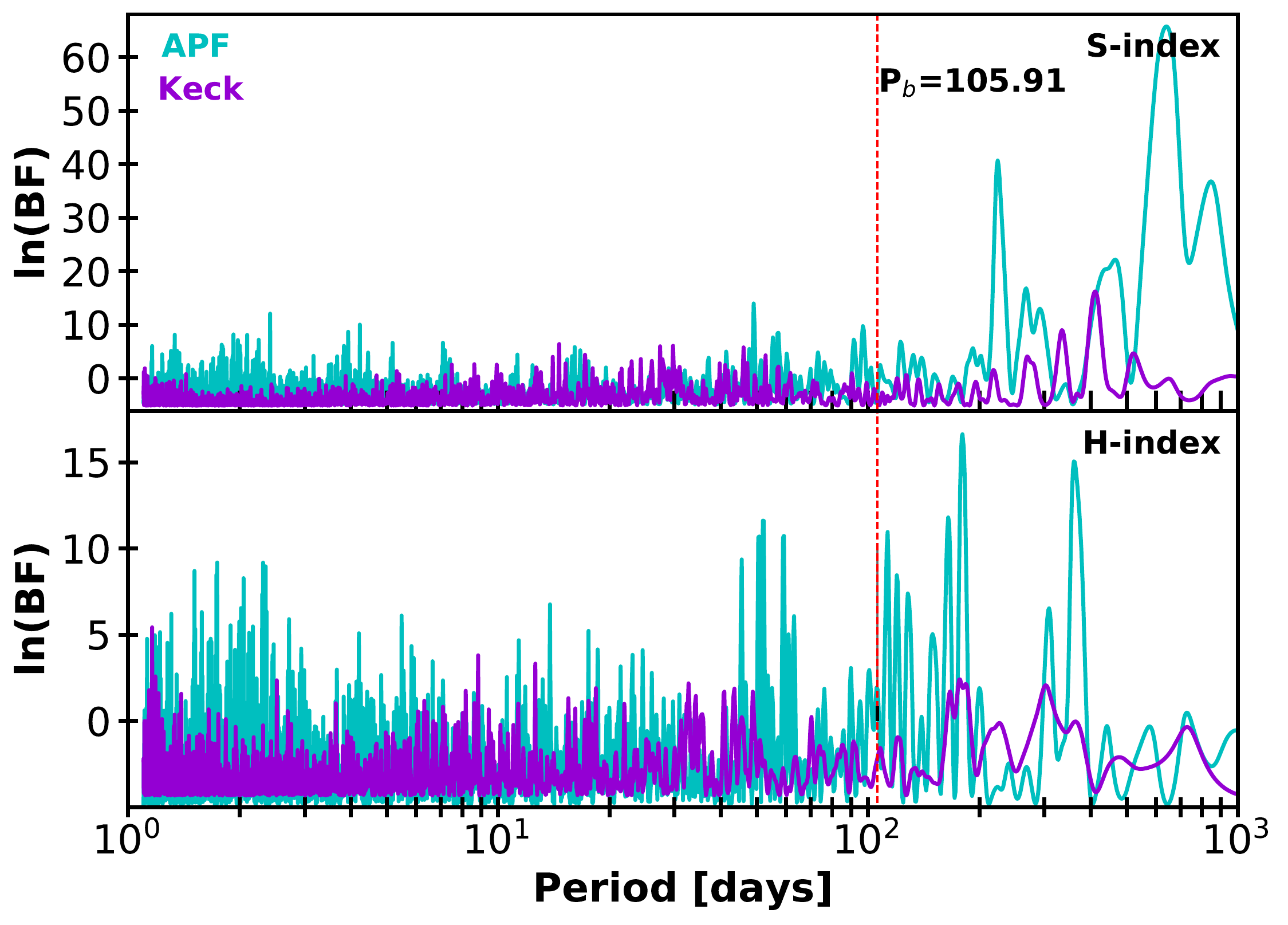}
\caption{\label{fig:HD180617activity_per} Top: Bayes factor periodogram of the S-index values of HD~180617 measured from the APF and Keck HIRES data sets in cyan and purple, respectively. There are no prominent peaks in the vicinity of the \sysfourbPer\ day planet period. Bottom: Same as above, but for the H-index measurements extracted from the APF and Keck HIRES spectra, which also show a lack of peaks at the planet's period.}
\end{figure}

S- and H-index activity indicators were extracted from each of the HIRES and APF spectra using the methods described in \S \ref{sec:methods} (Figure \ref{fig:HD180617activity_per}). We find no significant peaks at the period of the planet (P = \sysfourbPer\ days), however the data from the APF does include large peaks at $\sim$200 days, which match a peak seen in indicators sensitive to line-profile variations described in \citet{Kaminski2018}.

\subsection{Photometry and stellar rotation}\label{sec:180617photometry}
\begin{figure}
\centering
\includegraphics[width=.48 \textwidth]{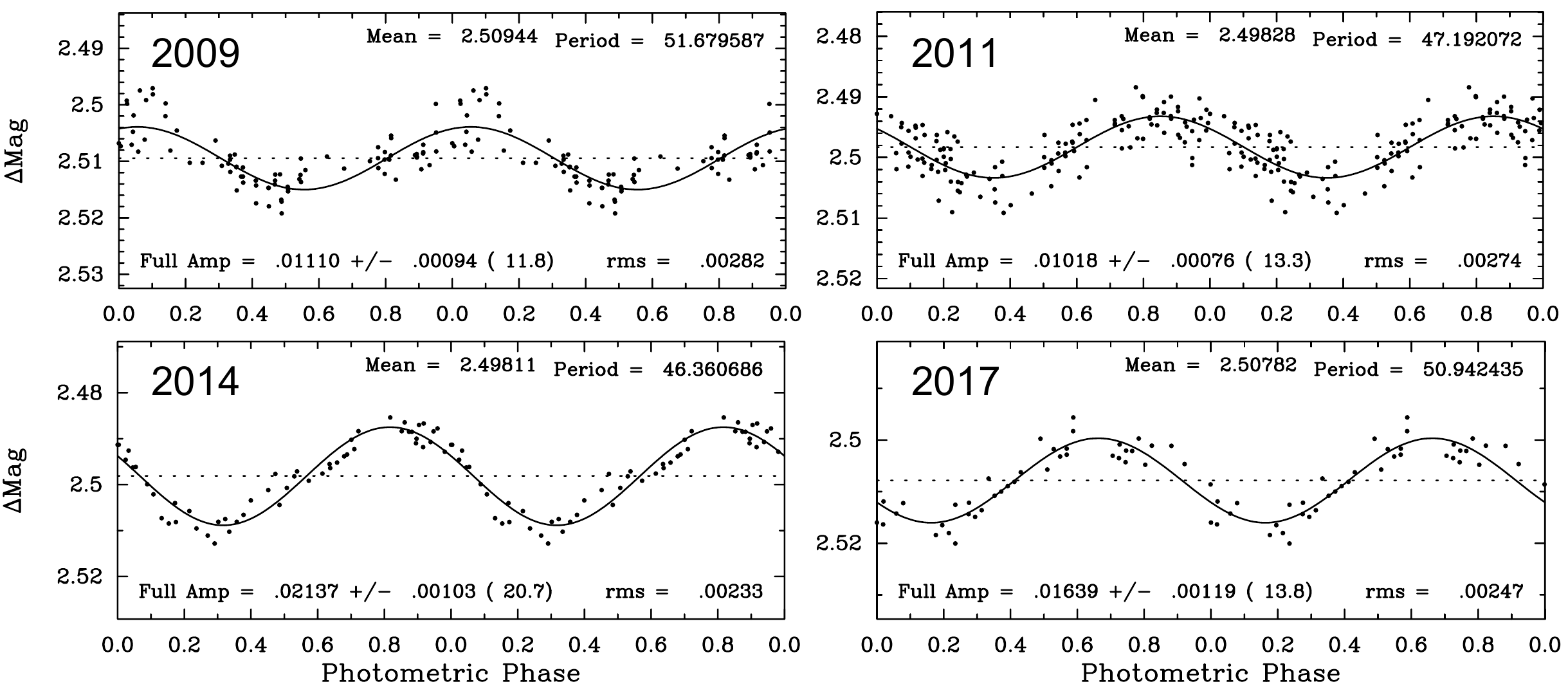}
\caption{\label{fig:HD180617Phot} APT photometry for four of the nine seasons during which HD 180617 was observed. Combining the results across all nine seasons produces a weighted mean rotation period of 50.60 $\pm$ 0.41 days.}
\end{figure}

A total of 707 observations were obtained between 2009 through 2017 using the T10 0.8m APT at Fairborn Observatory (Figure \ref{fig:HD180617Phot}). The comparison stars used in the photometric analysis were HD 183085 (an F2V star with V = 6.72, B-V=0.36) and HD 180945 (an F5V star with V=7.15 and B-V=0.47). We analyze each observing season for rotation and find periods for all nine seasons, with the star appearing to be ``double spotted" in 2010 and 2015. In those two years, the rotation period is taken to be twice the observed photometric period. Combining the results across all nine seasons produces a weighted mean rotation period of 50.60 $\pm$ 0.41 days. 

\subsection{Orbital parameters}\label{sec:180617orbit}

\begin{figure}
\includegraphics[width=.48\textwidth]{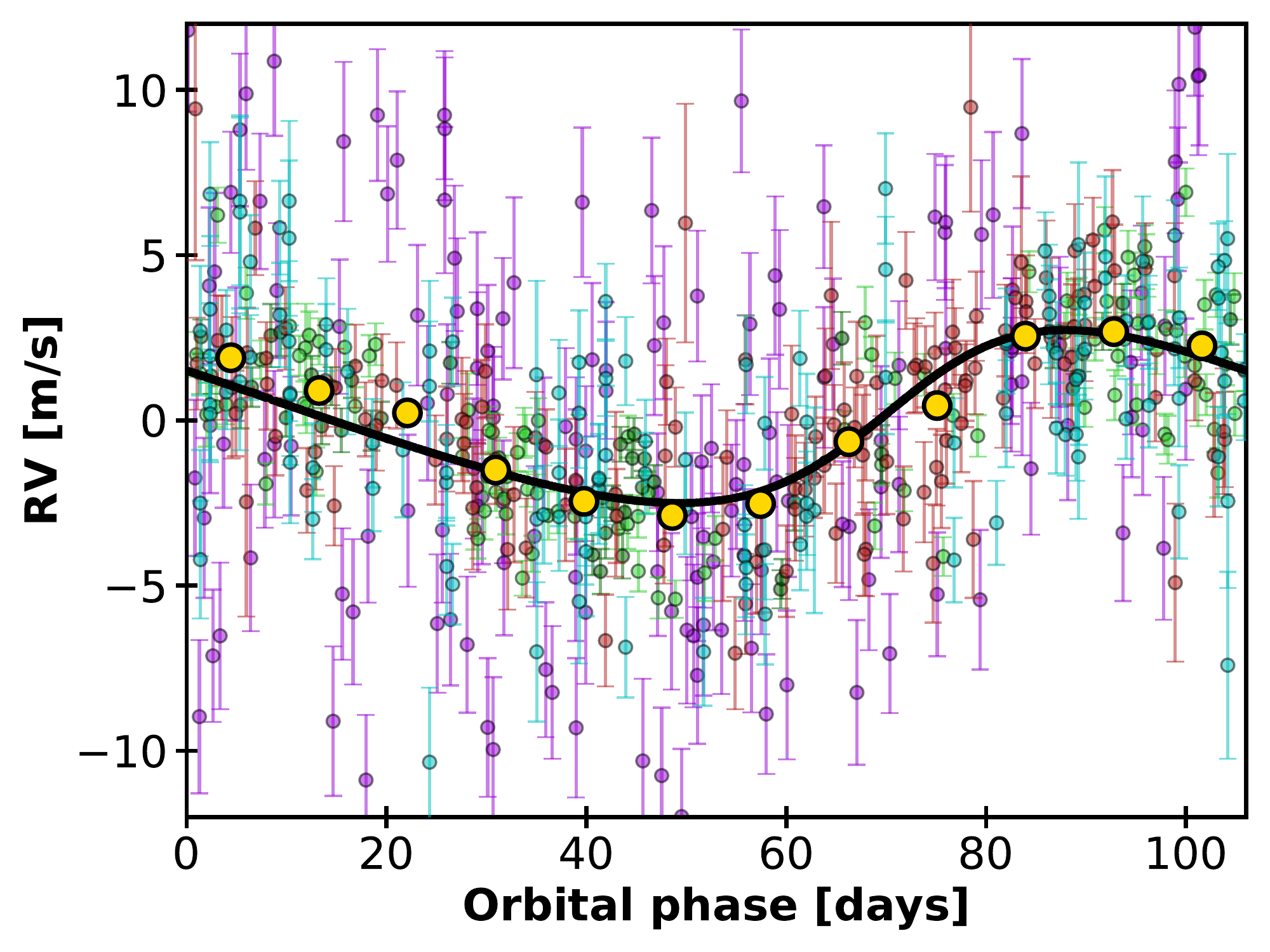}
\caption{\label{fig:HD180617PhasedRVs} RV observations of HD~180617 phase folded to the best fit period of P =  \sysfourbPer\ days (colors match those in Figure \ref{fig:HD180617vels}). The error bars include the excess white noise ``jitter'' from our analysis, and the black solid curve denotes the maximum a posteriori Keplerian model. Yellow points depict the phase-binned RV data.}
\end{figure}

We find that the best orbital fit to the combined RV data set for HD~180617 is a mildly eccentric (e = \sysfourbEcc), P = \sysfourbPer\ day orbit with a semi-amplitude K = \sysfourbK\ \ms\ (Figure \ref{fig:HD180617PhasedRVs}). This corresponds to a \sysfourbMsini\ \mearth\ mass planet in a \sysfourbA\ AU orbit around the host star, consistent with the findings of \citet{Kaminski2018}. A search of the RV residuals periodogram does not reveal any evidence for additional signals. 

\section{Discussion} \label{sec:Discussion}

In this study, we analyzed 34 Keck HIRES and 157 APF RV measurements of the K4V star HD~190007 along with 504 Keck HIRES and 300 APF RV measurements of the K0V star HD~216520. From these analyses we detect three new planets. Based on our derived orbital parameters, HD~190007~b and HD~216520~b are both close-in, sub-Neptune type planets while HD~216520~c is a longer period, temperate sub-Neptune planet. We also confirmed the orbital parameters of two previously published low-mass planets. First, we derived an updated orbital solution for the planet orbiting the M1.0V star GJ~686 reported by \citet{Affer2019} and \citet{Lalitha2019} using an additional 134 new RV APF measurements and 18 new PFS measurements. And second we updated the orbital solution of the planet orbiting the M3.0V star HD~180617 detailed in \citet{Kaminski2018}, adding 126 new APF measurements to the combined data set. 

\subsection{Benefits of long baseline, multi-facility data sets}
K- and M-dwarf stars make enticing radial velocity survey targets because 1) their lower stellar masses result in larger RV semi-amplitudes from an exoplanet with a set period and mass than would be induced on hotter, more massive stellar types and 2) their lower effective temperatures result in a higher number of stellar absorption lines that increase the RV information content in their spectra \citep{BeattyGaudi2015}. While, to our knowledge, HD~190007 and HD~216520 have only been included in the Keck HIRES and APF surveys described here, GJ~686 and HD~180617 have each been targeted by at least four independent surveys. 

This array of data sets spanning long temporal baselines allows for the derivation of very precise orbital models for each of the planets and also for the comparison of both planetary and stellar variability detections between different instruments and facilities. Our final orbital and planetary parameters for each of the planets are summarized in Table \ref{table:OrbitalParamsTable}. Each of the planetary \msini\ values is determined to the $\sigma_{K} \geq 10$ level, a benefit of having many precise RV observations taken over extended baselines. When comparing to the confirmed planets listed on the Exoplanet Archive, less than a quarter (roughly 22\%) of RV detected, sub-Neptune-mass planets have \msini\ measured to this level. Planets with such precise \msini\ values offer some of the best insights and constraints when modeling the formation, evolution, and dynamic interactions of planetary systems. Having such long baseline results that accurately constrain the stars' RV signal could also allow for the detection of N-body effects such as planet-planet scattering with more comprehensive modeling, similar to the use of transit timing variations in transit data.

\begin{table*}
\caption{Best fit orbital solution for each of the planets detailed above and the activity signals noted for HD 190007 and HD 216520. Reported values are the mean and standard deviation for each model parameter. Minimum mass and semi-major axis have been estimated by adopting the stellar masses listed for each star in Table \ref{table:stellarparams}}\label{table:OrbitalParamsTable}
\begin{center}
\begin{tabular}{llllllll}
\hline \hline
  & HD~190007 & HD~190007 & HD~216520 & HD~216520 & HD 216520 & GL~686 & HD~180617  \\
  & Planet b & Activity & Planet b & Planet c & Activity & Planet b & Planet b \\
\hline
$P$ (days) & \sysonebPerunc & \sysonecPerunc & \systwobPerunc & \systwocPerunc & \systwodPerunc & \systhreebPerunc & \sysfourbPerunc \\
$K$ (\ms) & \sysonebKunc & \sysonecKunc & \systwobKunc & \systwocKunc & \systwodKunc & \systhreebKunc & \sysfourbKunc \\
$e$ & \sysonebEccunc & \sysonecEccunc & \systwobEccunc & \systwocEccunc & \systwodEccunc & \systhreebEccunc & \sysfourbEccunc \\
$\omega$ (deg) & \sysonebOmegaunc & \sysonecOmegaunc & \systwobOmegaunc & \systwocOmegaunc & \systwodOmegaunc & \systhreebOmegaunc & \sysfourbOmegaunc \\
$M_{0}$ (deg) & \sysonebMounc & \sysonecMounc & \systwobMounc & \systwocMounc & \systwodMounc & \systhreebMounc & \sysfourbMounc \\
$ln(BF_{3})$ & 28.04 & 7.93 & 44.3 & 6.2 & 6.9 & 72.03 & 39.359\\
$ln(BF_{5})$ & 23.37 & 2.68 & 37.59 & -0.52 & 0.18 & 65.88 & 33.062\\
\hline
$m \sin i$ (M$_{\oplus}$) &  \sysonebMsiniunc & -- & \systwobMsiniunc & \systwocMsiniunc & -- & \systhreebMsiniunc & \sysfourbMsiniunc \\
$a$ (AU) & \sysonebAunc & -- & \systwobAunc & \systwocAunc & -- &\systhreebAunc & \sysfourbAunc \\
\hline \hline
\end{tabular}
\tablecomments{$M_{0}$ values are referenced to the first RV epoch for each star, which can be found in Tables 6-9.}
\end{center}
\end{table*}

We find that our final results for the previously published planets, GJ~686 b and HD~180617 b, are in good agreement with the earlier results found in \citet{Kaminski2018,Affer2019} and \citet{Lalitha2019}. For both planets our derived planet periods, eccentricities, semi-amplitudes, and \msini\ values are all within the 1-$\sigma$ uncertainties across all three existing publications. 

One particular strength of the APF telescope, with its ability to achieve nightly cadence on interesting RV target stars, is that the resulting data can help disentangle observational aliases. In particular, earlier versions of the HD~190007 data set showed strong signals at periods of both 11.72 and 1.09 days, which are daily aliases of one another. For example, earlier versions of the RV periodogram for HD~190007 showed an additional peak appeared at P = 1.09 days, corresponding to the 1 day alias ($f_{1.09d} = f_{11.72d} + f_{1day}$) of the 11.72 day signal. We had reason to suspect that the longer period signal was the true signature of the planet because the 1.09 day signal produced a notably lower ln(BF$_{5}$) value and required an orbital eccentricity of e=0.119$\pm$0.068. This non-zero eccentricity seemed unlikely for a planet on such a short period orbit, which we would expect to have circularized via tidal dissipation \citep{HaddenLithwick2017,VanEylenAlbrecht2015}. But it was challenging to make a robust determination of which signal was caused by the planet and which was the alias. By observing the star for an additional season and ensuring that it underwent high cadence observations (Figure \ref{fig:HD190007vels}) we were able to update the analysis and found that the 1.09 day signal had dramatically decreased in significance, further strengthening our argument that the 11.72 day signal presented here is the true Keplerian signature of the planet. This same affect of being able to discern between true signals and observational aliases can be achieved by combining data from different facilities that have some degree of longitudinal spread across the globe. This enables a broader range of short-period observational cadences based on telescope separations and will help remove power from the 1 day alias that plagues single-site observations.

\subsection{Potential for additional detection methods}
Having discovered these planets via the radial velocity method, we investigate whether they might make good candidates for additional detection/characterization methods such as transit, astrometric, and direct imaging observations. We calculate transit probabilities using \citet{Winn2010}, and astrometric semi-amplitudes and maximum projected separations using \citet{Perryman2011}. We find that due to the combination of low mass and relatively short orbital periods, none of the five planets detailed in this work are expected to produce an astrometric semi-amplitude larger than 5 micro-arcseconds placing them all beyond the reach of Gaia's detection threshold. Similarly, the projected on-sky separation of the HD~190007, HD~216520, and HD~180617 planets from their host stars are all below 0.03 arcseconds, making them inaccessible to even the upcoming generation of Extremely Large Telescopes. GJ~686~b is a slightly more promising case, with a maximum projected separation of 0.058 arcsecond, but even this is on the very edge of performance expected from ground-based, thirty meter class facilities.

For the transit probability calculations, we need to make an assumption about the planets' sizes. We compare the minimum masses listed in Table \ref{table:OrbitalParamsTable} with the conditional distributions for a planet's radius given its mass presented in Figure 7 of \citet{Ning2018}. We adopt a uniform radius assumption of R = 3 \rearth\ for all of our planets as they correspond most closely to the Mass = 10 \mearth\ distribution. Inserting this radius assumption along with the planets' measured orbital parameters into Equation 9 of \citet{Winn2010} produces transit probabilities well below 1\% for each of the planets.

While the radial velocity detections for all of these systems are robust we do not expect any of these planets to be particularly well-suited to additional methods of detection and characterization.

\subsection{Comparison to close-in \kepler-multi planet systems}
\begin{figure}
\includegraphics[width=.48\textwidth]{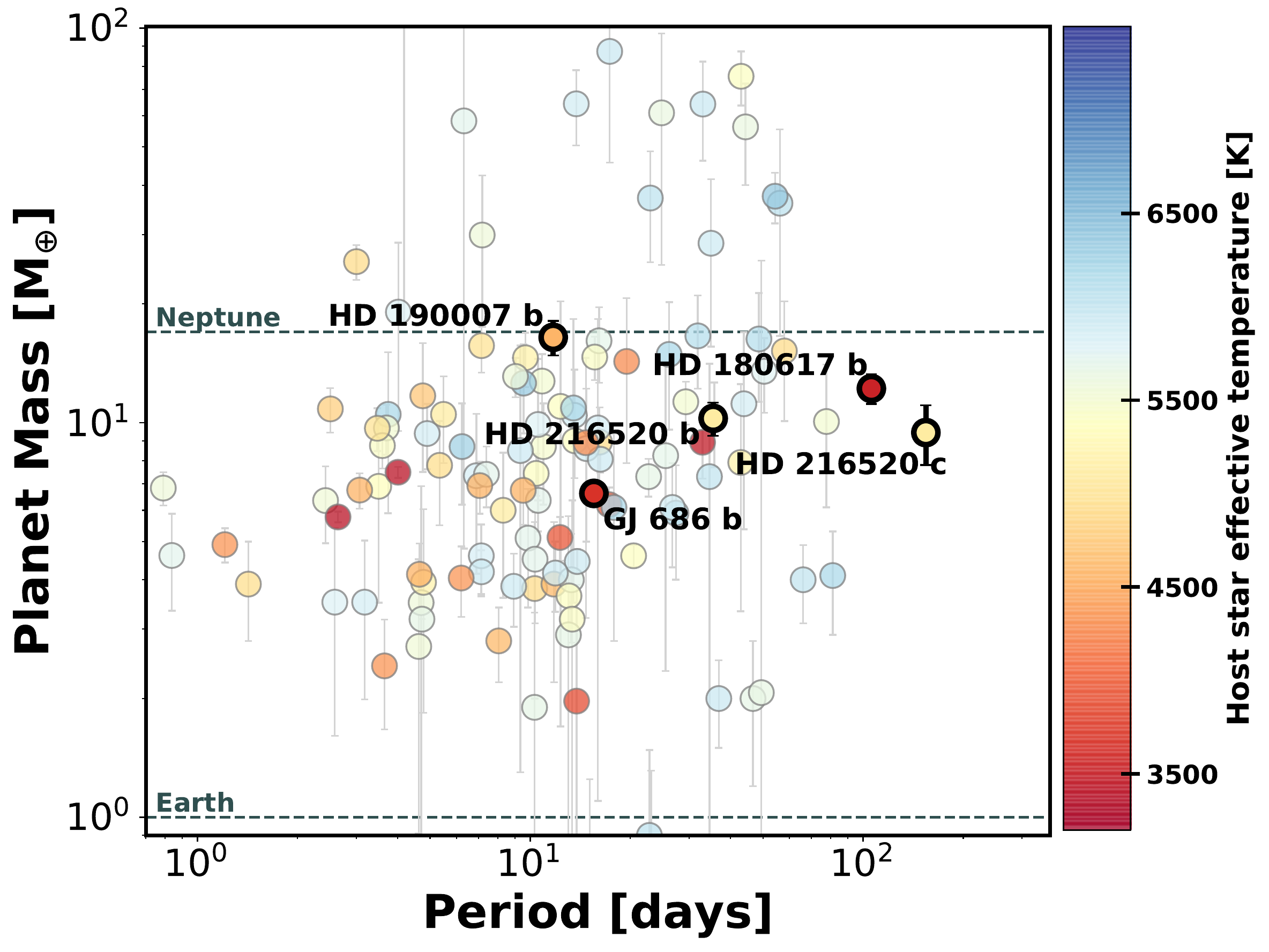}
\caption{\label{fig:KeplerK2multis} Out of the 136 \kepler\ and \textit{K2} sub-Neptune mass exoplanets that have orbital periods less than 100 days, 116 (85\%) are in multi-planet systems. Our two new and two updated planets, labeled here in black, land in a very similar region of period-mass parameter space as the \kepler\ multi-planet systems. If the planets in this paper were formed in a similar manner as the ones found by \kepler, this would suggest that additional, close-in and similar mass planets may be present in those systems. Note that our the new planets have only minimum mass measurements, as their inclination is unknown, and so their positions on this plot are lower limits.}
\end{figure}

With minimum masses just at or below that of Neptune and periods in the 10-100 day range two of the planets discovered here (HD~190007~b and HD~216520~b) are reminiscent of the numerous super-Earth and sub-Neptune planets detected in transit by the \kepler\ mission (Figure \ref{fig:KeplerK2multis}). In particular, \kepler\ revealed that about half of stars in our galaxy harbor the small (R$_{p} \leq 4\rearth$), close-in (P $\leq$ 100 days) planets, which are often found in tightly-spaced, multi-planet systems \citep{Lissauer2011, Latham2011, Lissauer2014, Rowe2014}. Most surprisingly, the multiple planets in the same systems tend to be similar in both mass and radius \citep{Millholland2017, Wang2017, Weiss2018}.

\begin{figure*}
\includegraphics[width=.98\textwidth]{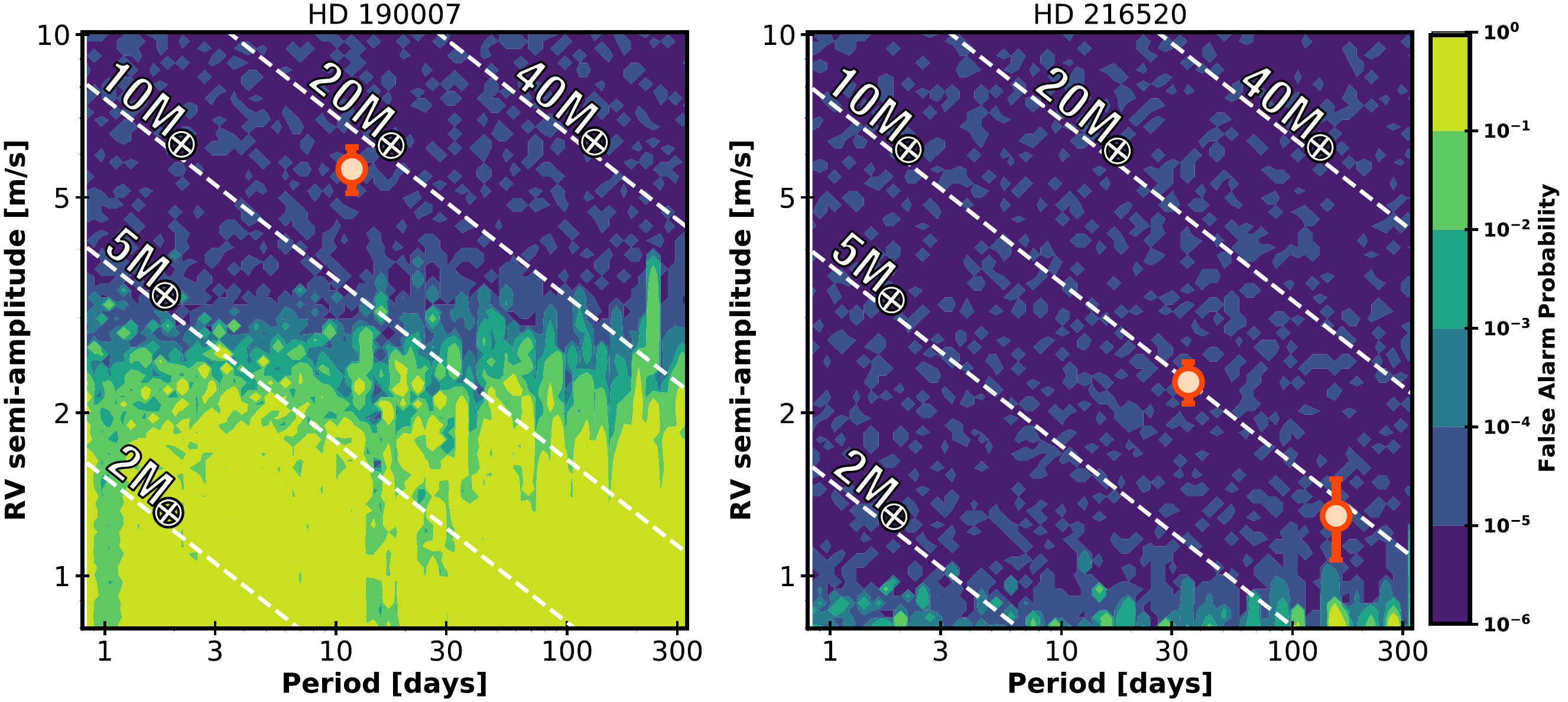}
\caption{\label{fig:IR_Contour} Injection and recovery contour maps for HD~190007 (left) and HD~216520 (right), with the newly discovered planets shown as orange circles. Coloration represents the FAP value of the LS periodogram at the injected period, with the darker colors (FAP \textless 0.01) representing planets that were successfully recovered. White lines show the RV semi-amplitude of a given planet mass as a function of period. For HD~190007 we are sensitive to planets with masses greater than 5-20\mearth\ depending on the period, while for HD~216520 we are sensitive to planets with masses greater than 1-10\mearth. The lack of a discovery of similar mass and period planets in these systems is in marked contrast with \kepler\ discoveries where $\sim$85\% of planets have neighbors that are close in mass, radius, and period.}
\end{figure*}

That is, statistically speaking, for a given short-period super-Earth or sub-Neptune planet such as HD~190007~b or HD~216520~b we would expect the existence of additional close-in planets and for those planets to have masses similar (within a factor of two) to that of the detected planet. 
While HD216520~c is very similar in mass to HD216520~b, much like the multi-planet systems found with Kepler, it does not have an orbital period within a factor of two of the inner planet as would be expected for a Kepler multi-planet system. We therefore examine the possibility that there may be additional planets with periods less than 100 days orbiting HD~190007 and HD~216520. Neither star's data set shows evidence of additional, significant signals in their respective RV residuals periodogram (Figures \ref{fig:HD190007vels} and \ref{fig:HD216520vels}). Yet this does not rule out the possibility that additional signals might be present at significance levels obscured by effects from stellar activity, our observing cadence, and our RV precision, among others. To test our data sets' ability to detect such signals we perform an injection and recovery test using the same residual RVs used to create those periodograms. First, we define period and semi-amplitude ranges similar to those of the small, close-in planets detected by \kepler: 0.5 \textless\ P \textless\ 300 days and 0.5 \textless\ K \textless\ 10 \ms. We then randomly draw a period and semi-amplitude value from a log distribution bounded by these end points and inject the corresponding Keplerian signal into the RV residuals. We generate a Lomb-Scargle periodogram of the resulting RVs and calculate the False Alarm Probability (FAP) of the periodogram at the highest peak within 10\% of the injected period. We define our recovery criteria such that cases when the resulting FAP value is below 0.01 are considered to be successful detections of the planet. We execute this process 100,000 times for each star's residuals data sets and visualize the results in Figure \ref{fig:IR_Contour}. The cool colored regions of the plot (FAP \textless\ 0.01) represent period and semi-amplitude combinations that we successfully recover. We note here that this process is an estimate of our ability to detect a periodic signal with a certain period. This is not the estimate of of our probability of recovering a Keplerian signal. Finding a periodic signal is a step in that process, so our detection probabilities are likely over-estimates of our ability to actually recover and characterize a planet's orbit.

In the case of HD~190007, for which we have 123 RV epochs taken over two decades, we are sensitive to planets with semi-amplitudes down to $\sim$3 \ms\ for periods out to one year. Given the star's mass, the 3 \ms\ sensitivity level corresponds to planets in the 4-28 \mearth\ range. As HD~190007 b has an \msini\ value of 16.46 \mearth\, we are most interested in our ability to detect ``similiar'' planets like those we'd expect to see in a \kepler\ -like system, which would fall in the 8-30 \mearth\ mass range. While planets with masses $\geq$20 \mearth\ should be detectable throughout the 1-100 day period range populated by the close-in \kepler\ multi-planet systems, those in the 8-20 \mearth\ range could be obscured at periods longer than 10 days.

For HD~216520, with its 369 RV epochs taken over nineteen years, the injection/recovery process suggests that we are sensitive to planets with semi-amplitudes down to $\sim$1 \ms\ for periods out to 300 days. A 1 \ms\ signal in the HD~216520 data set corresponds to a planet with mass ranging from 1.5-10 \mearth\ when considering periods less than 300 days. Given the 11.63\mearth\ \msini\ value of HD~216520 b, and again assuming that \kepler\ -like systems will have planets of similar masses, we would expect any additional planets in the system to have masses in the 5-20 \mearth\ range, making it likely that we would have noticed their presence in our existing data. 

Similar to our solar system, the \kepler\ multi-planet systems tend to be well organized - they have small mutual inclinations, circular orbits, and are generally well aligned with the host star's rotation. Since both HD~190007 b and HD~216520 b are on circular orbits, additional co-planar planets within these systems should be able to survive on similarly circular orbits beyond the Hill stability threshold. But as seen above, we rule out the presence of additional, close-in planets with masses similar to our newly detected planets in tightly-spaced stable orbits. 

One possibility is that because radial velocity detections measure a planet's minimum mass, and not its true mass, then one or both of these planets may actually be a more massive, gas giant on a highly inclined orbit. Gas giants are significantly less likely to be accompanied by either other close-in planets \citep{Steffen2012, Canas2019}, or planets with similar masses \citep{Wang2017}. HD~216520 in particular has an extremely small reported \vsini\ (0.2$\pm$0.05 k\ms) compared with stars that have similar \teff\ ($5082\,$K) lending some credence to the idea that the planet's orbital plane is far from the line of sight. As the orbits of most small planets align with their host star's equator \citep{Winn2015, Wang2018}, HD~216520's small \vsini\ value suggests that the planet is likely to be a highly inclined gas giant, otherwise the projected RV is too small to be detected. But while we selected the \citet{Brewer2016} \vsini\ value for consistency with the other stellar parameters reported in Table \ref{table:stellarparams} there are two additional \vsini\ measurements for this star in the literature. \citet{Luck2017} reports a \vsini\ value of 2.5 k\ms\ while \citet{Mishenina2008} finds \vsini\ = 1.4 k\ms. If one of these larger values is a more accurate measurement of the HD 216520's rotational velocity, which seems reasonable given the star's \teff, then it would be less likely that HD 216520 b is a gas giant masquerading as a Neptune.

An alternate explanation is that the planets detected in this paper using the RV technique and the planets detected by \kepler\ are not from the same population, and therefore do not share similar observational properties. There is an unsolved discrepancy of hot-Jupiters occurrence rate between the \kepler\ and RV samples, which is partially attributed to the fact that the \kepler\ sample has lower metallicity than the RV sample \citep{Guo2017} as the RV technique performs better on high-metallicity stars because of the increased information content in their stellar spectra. As pointed out by \citet{Brewer2018}, super-Earth/sub-Neptune's multiplicity is anti-correlated with host stellar metallicity. Metal-rich stars, HD~190007 for example ([Fe/H] = 0.16$\pm$0.05), are less likely to host multiple planet systems. This raises the interesting possibility that our two new planets detected here have followed a different path of planet formation than that taken by the majority of \kepler\ planets. Though our work has a small sample size, the addition of even a small number of planets helps to point the way towards a future verification or rebuttal of this picture.

\acknowledgments

\textcopyright 2020

The research was carried out in part at the Jet Propulsion Laboratory, California Institute of Technology, under a contract with the National Aeronautics and Space Administration (80NM0018D0004).

The work herein is based on observations obtained at the W. M. Keck Observatory, which is operated jointly by the University of California and the California Institute of Technology, and we thank the UC-Keck and NASA-Keck Time Assignment Committees for their support. We also wish to extend our special thanks to those of Hawaiian ancestry on whose sacred mountain of Mauna Kea we are privileged to be guests. Without their generous hospitality, the Keck observations presented herein would not have been possible. The work herein is also based on observations obtained with the Automated Planet Finder (APF) telescope and its Levy Spectrometer at Lick Observatory, along with data gathered with the 6.5 meter Magellan Telescopes located at Las Campanas Observatory, Chile.  G.W.H. acknowledges long-term support from NASA, NSF, Tennessee State University, and the State of Tennessee through its Centers of Excellence program.

This research has made use of the Keck Observatory Archive (KOA), which is operated by the W. M. Keck Observatory and the NASA Exoplanet Science Institute (NExScI), under contract with the National Aeronautics and Space Administration. We specifically acknowledge the 83 RVs used in the analysis of HD 216520 that are based on spectra taken by the California Planet Search team and archived in the Keck Observatory Archive. This research has made use of the SIMBAD database, operated at CDS, Strasbourg, France.

This publication makes use of VOSA, developed under the Spanish Virtual Observatory project supported by the Spanish MINECO through grant AyA2017-84089.
VOSA has been partially updated by using funding from the European Union's Horizon 2020 Research and Innovation Programme, under Grant Agreement nº 776403 (EXOPLANETS-A)

Simulations in this paper made use of the REBOUND code which is freely available at \url{http://github.com/hannorein/rebound}.

\vspace{5mm}
\facilities{UCO/Lick: The APF (Levy spectrograph), Magellan: Clay (Planet Finder Spectrograph), Keck I: (HIRES), TSU:AST (T4 and T12 APTs)}


\clearpage
\appendix
\section{Bayes Factor Periodograms}
Figures 23, 24, 25, and 26 depict the full set of Bayes Factor Periodograms for each of the four stars included in this study.

\begin{figure}
\centering
\includegraphics[width=.98 \textwidth]{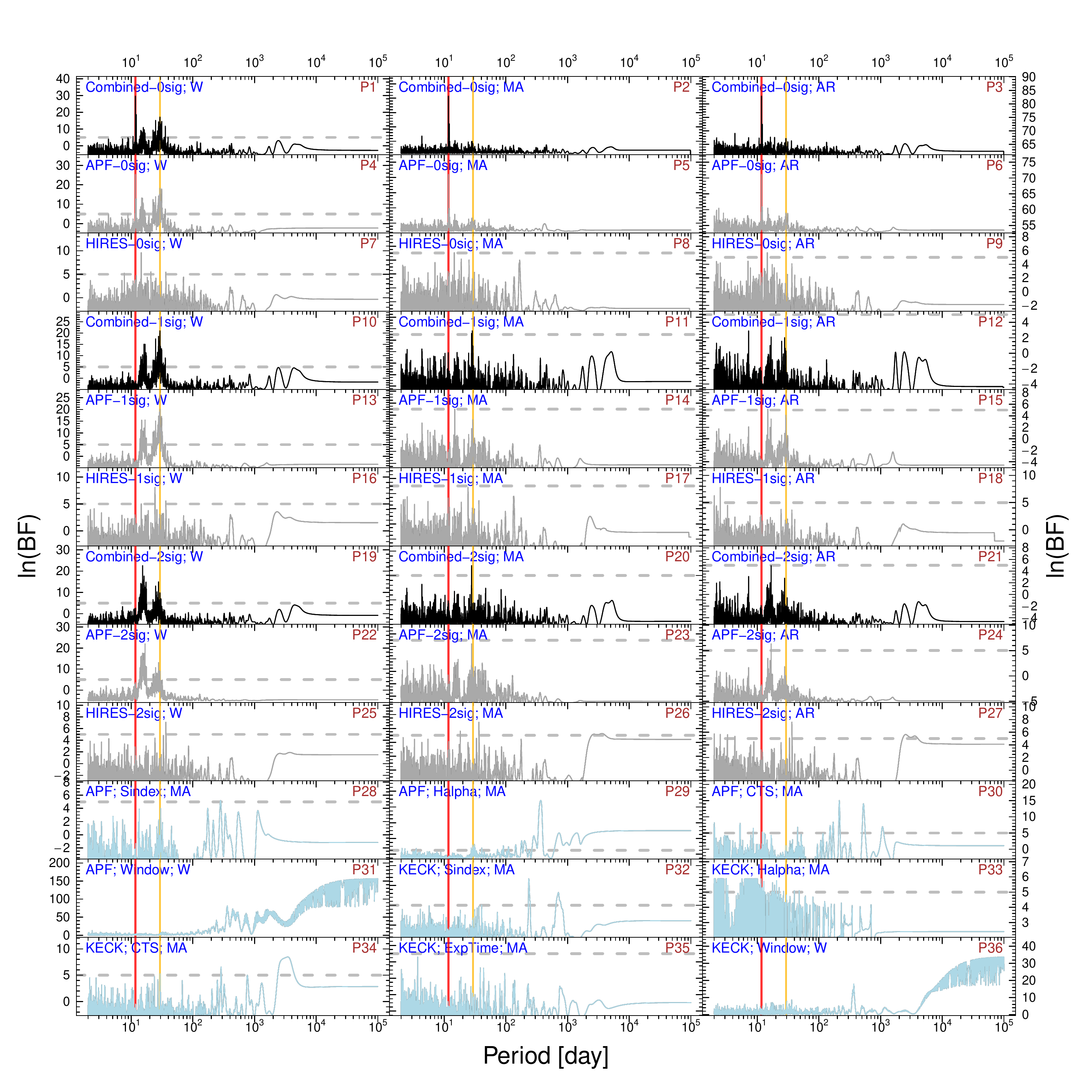}
\caption{\label{fig:BFP_HD190007} Full set of Bayes Factor Periodograms for HD~190007. The red line denotes the 11.72 day signal that we take to be the true planet signal, while the gold line shows the 29 day period that we find to be caused by stellar variability.}
\end{figure}

\begin{figure}
\centering
\includegraphics[width=.98 \textwidth]{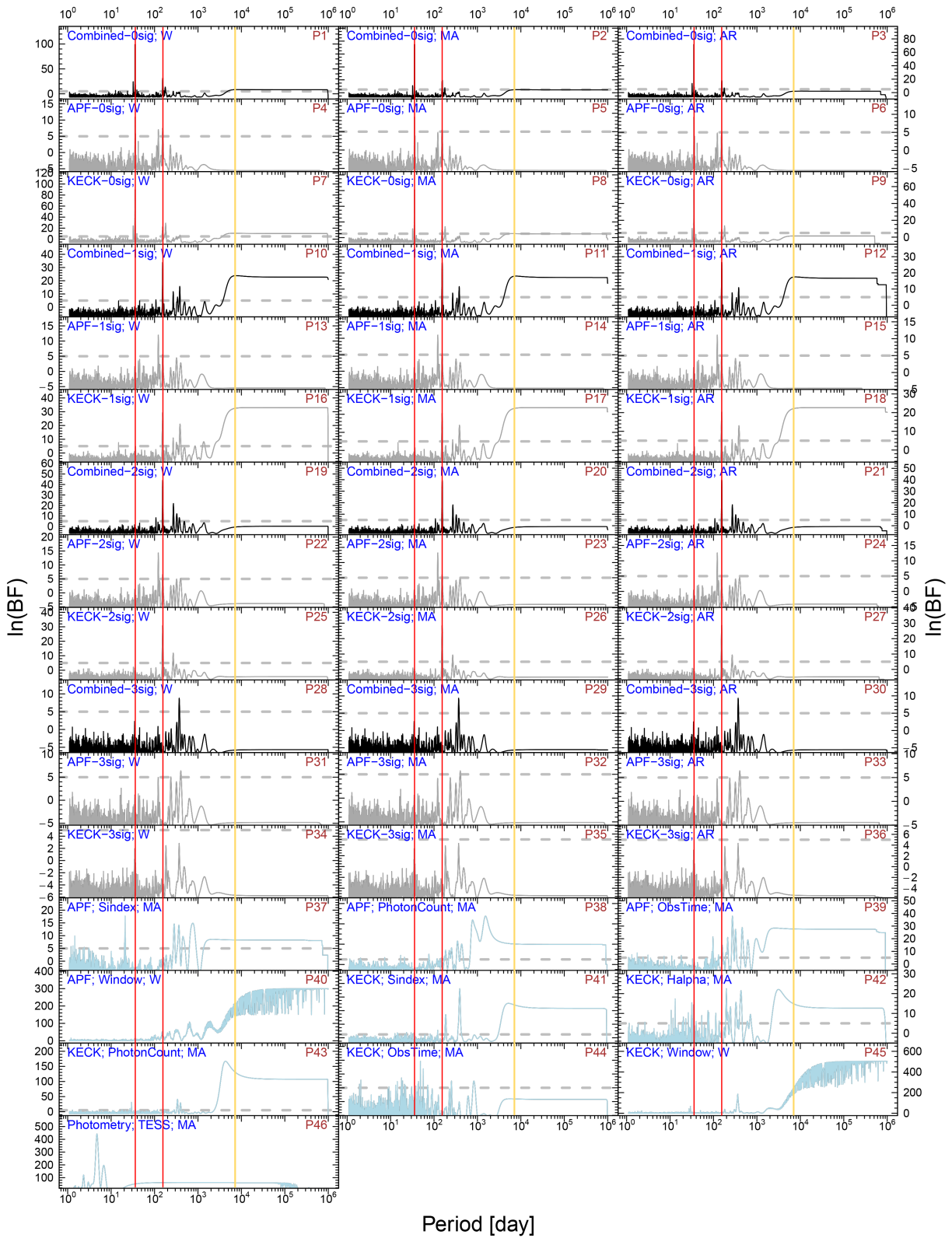}
\caption{\label{fig:BFP_HD216520} Full set of Bayes Factor Periodograms for HD~216520. The red lines denote the \systwobPer\ and \systwocPer\ day signals that we take to be planets. The gold line shows the additional \systwodPer\ day signal that we believe to be caused by the star's magnetic activity cycle.}
\end{figure}

\begin{figure}
\centering
\includegraphics[width=.75 \textwidth]{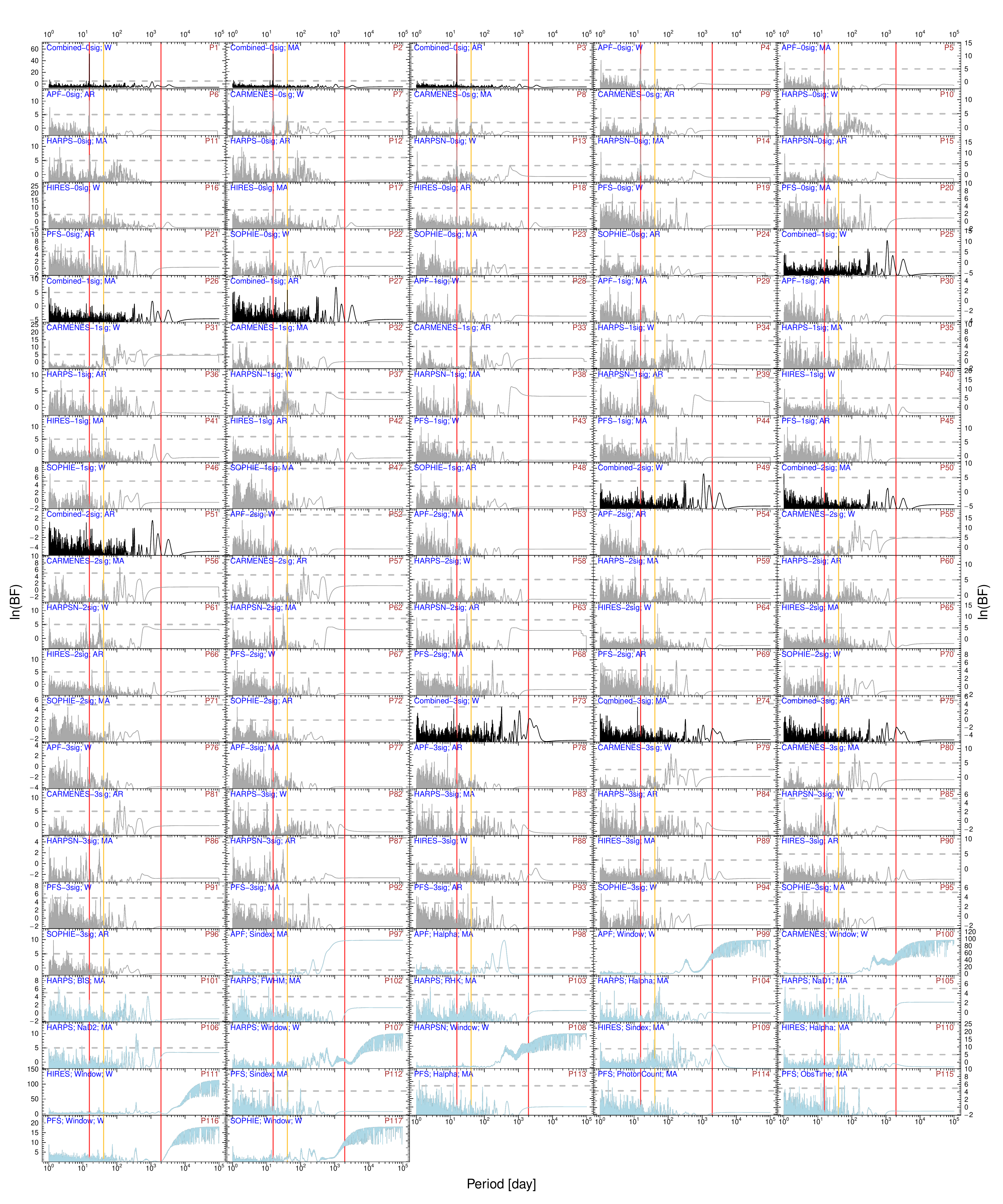}
\caption{\label{fig:BFP_GL686} Full set of Bayes Factor Periodograms for GL~686. The red line denotes the 15d planet signal originally identified in \citet{Affer2019} and \citet{Lalitha2019} and confirmed again in this work. The gold lines show the $\sim$40 day and $\sim$2,000 day signals that we identify as the star's rotational and magnetic activity cycles, respectively.}
\end{figure}

\begin{figure}
\centering
\includegraphics[width=.98 \textwidth]{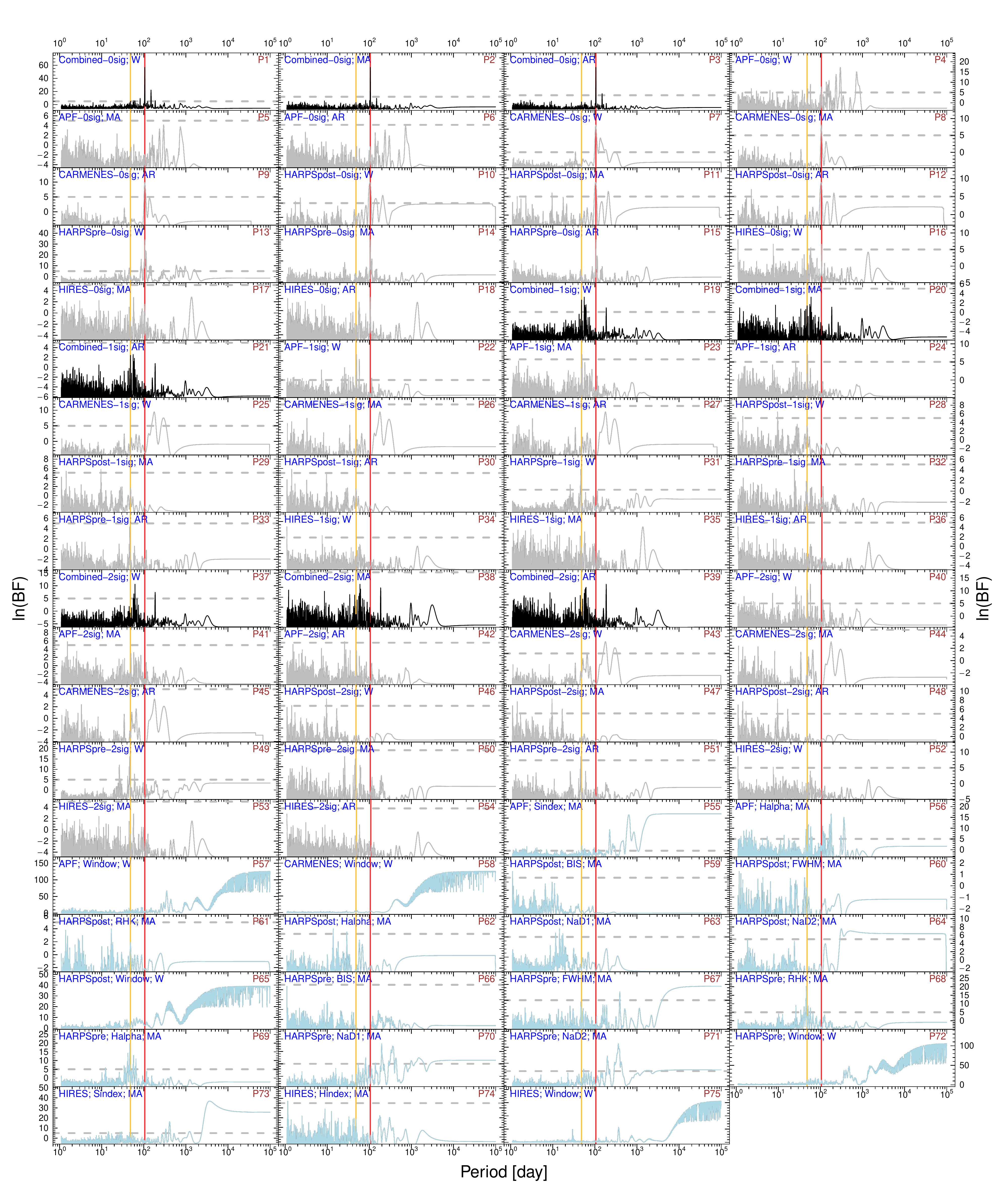}
\caption{\label{fig:BFP_HD180617} Full set of Bayes Factor Periodograms for HD~180617. The red line denotes the \sysfourbPer\ day planet signal originally identified in \citet{Kaminski2018} and confirmed in this work. The gold line shows the $\sim$50 day activity signal seen in the HARPS H-$\alpha$ measurements and the APT photometry that we take to be the star's rotation period.}
\end{figure}

\clearpage
\section{Radial Velocity Measurements}

Tables 6-9 contain the new RVs from HIRES, the APF, PFS, and HARPS TERRA presented in this paper. A portion of each table is shown here for guidance regarding their forms and content, while the full tables are published in the machine-readable format.

\begin{longtable}{l c c c c c}
\caption{New Radial Velocity data for HD 190007}\\
\hline
\hline
  \multicolumn{1}{l}{MJD } &
  \multicolumn{1}{c}{RV\,[\ms]} &
  \multicolumn{1}{c}{$\sigma_{\rm{RV}}$\,[\ms]} &
  \multicolumn{1}{c}{S-index} &
  \multicolumn{1}{c}{H-index} &
  \multicolumn{1}{c}{Instrument} \\
  \hline
  \endhead
   \hline
  \endfoot
\hline
2450984.05436	&	7.41	&	1.33	&	0.7281	&	-1	&	HIRES	\\
2451013.94476	&	-1.73	&	1.46	&	0.6926	&	-1	&	HIRES	\\
2451050.86802	&	-6.75	&	1.35	&	0.7218	&	-1	&	HIRES	\\
2451069.88122	&	-1.29	&	1.47	&	0.7261	&	-1	&	HIRES	\\
2451075.80608	&	7.09	&	1.21	&	0.6631	&	-1	&	HIRES	\\
2451341.05512	&	-7.35	&	1.52	&	0.6168	&	-1	&	HIRES	\\
2451368.86865	&	-5.04	&	1.57	&	0.5584	&	-1	&	HIRES	\\
2451409.92287	&	2.89	&	1.99	&	0.6589	&	-1	&	HIRES	\\
2451439.81141	&	3.88	&	1.54	&	0.6635	&	-1	&	HIRES	\\
2451439.81802	&	7.60	&	1.60	&	0.6523	&	-1	&	HIRES	\\
\hline\hline
\end{longtable}

\begin{longtable}{l c c c c c}
\caption{New Radial Velocity data for HD 216520}\\
\hline
\hline
  \multicolumn{1}{l}{MJD } &
  \multicolumn{1}{c}{RV\,[\ms]} &
  \multicolumn{1}{c}{$\sigma_{\rm{RV}}$\,[\ms]} &
  \multicolumn{1}{c}{S-index} &
  \multicolumn{1}{c}{H-index} &
  \multicolumn{1}{c}{Instrument} \\
  \hline
  \endhead
   \hline
  \endfoot
\hline
2452188.92575	&	-2.46	&	1.25	&	0.1696	&	-1	&	HIRES	\\
2452189.85472	&	-0.41	&	1.49	&	0.185	&	-1	&	HIRES	\\
2452235.68115	&	-1.11	&	1.38	&	0.1799	&	-1	&	HIRES	\\
2452446.08978	&	1.34	&	1.37	&	0.1814	&	-1	&	HIRES	\\
2452487.04685	&	-1.40	&	1.31	&	0.1802	&	-1	&	HIRES	\\
2452487.93822	&	-3.99	&	1.32	&	0.171	&	-1	&	HIRES	\\
2452488.96981	&	-2.27	&	1.25	&	0.1678	&	-1	&	HIRES	\\
2452537.89594	&	-2.53	&	1.38	&	0.1747	&	-1	&	HIRES	\\
2452574.81330	&	0.48	&	1.47	&	0.1363	&	-1	&	HIRES	\\
2452833.03093	&	3.99	&	1.57	&	0.1383	&	-1	&	HIRES	\\
\hline\hline
\end{longtable}

\begin{longtable}{l c c c c c}
\caption{New Radial Velocity data for GJ 686}\\
\hline
\hline
  \multicolumn{1}{l}{MJD } &
  \multicolumn{1}{c}{RV\,[m\,s$^{-1}$]} &
  \multicolumn{1}{c}{$\sigma_{\rm{RV}}$\,[\ms]} &
  \multicolumn{1}{c}{S-index} &
  \multicolumn{1}{c}{H-index} &
  \multicolumn{1}{c}{Instrument} \\
  \hline
  \endhead
   \hline
  \endfoot
\hline
2453159.74925	&	 3.65&	0.95	&	 0.54270	&	  0.28195	&	 TERRA 	\\
2453574.62254	&	 0.81&	0.82	&	 0.67370	&	  0.28953	&	 TERRA 	\\
2453817.86816	&	-0.83	&	0.75	&	 0.63931	&	  0.29228	&	 TERRA 	\\
2454174.87133	&	-1.72	&	0.77	&	 0.63735	&	  0.26124	&	 TERRA 	\\
2454194.90763	&	 2.51&	0.78	&	 0.73922	&	  0.28956	&	 TERRA 	\\
2454300.64483	&	-2.47	&	0.72	&	 0.65581	&	  0.28934	&	 TERRA 	\\
2454948.86154	&	-1.17	&	0.57	&	 0.66352	&	  0.23889	&	 TERRA 	\\
2454950.86206	&	-2.71	&	0.64	&	 0.65440	&	  0.23478	&	 TERRA 	\\
2454956.83071	&	-0.53	&	0.88	&	 0.56384	&	  0.23619	&	 TERRA 	\\
2455390.64094	&	1.13	&	1.10	&	 0.67622	&	  0.27807	&	 TERRA 	\\
\hline\hline
\end{longtable}

\begin{longtable}{l c c c c c}
\caption{New Radial Velocity data for HD 180617}\\
\hline
\hline
  \multicolumn{1}{l}{MJD } &
  \multicolumn{1}{c}{RV\,[m\,s$^{-1}$]} &
  \multicolumn{1}{c}{$\sigma_{\rm{RV}}$\,[\ms]} &
  \multicolumn{1}{c}{S-index} &
  \multicolumn{1}{c}{H-index} &
  \multicolumn{1}{c}{Instrument} \\
  \hline
  \endhead
   \hline
  \endfoot
\hline
2453159.81073	&	-1.89	&	0.68	&	0.93516	&	0.24810	&	TERRA	\\
2453517.83909	&	1.13	&	0.67	&	1.08339	&	0.21657	&	TERRA	\\
2453572.76530	&	1.38	&	1.52	&	1.15631	&	0.11502	&	TERRA	\\
2453573.72190	&	-1.81	&	0.64	&	1.07858	&	0.19700	&	TERRA	\\
2453574.68500	&	-1.82	&	0.41	&	1.11689	&	0.16847	&	TERRA	\\
2453575.65016	&	-1.55	&	0.29	&	1.01041	&	0.21838	&	TERRA	\\
2453576.65304	&	-2.48	&	0.75	&	1.00409	&	0.21559	&	TERRA	\\
2453577.67857	&	-0.95	&	0.47	&	1.06946	&	0.19882	&	TERRA	\\
2453578.67547	&	-1.44	&	0.37	&	0.99462	&	0.22405	&	TERRA	\\
2453579.65731	&	-2.24	&	0.45	&	0.98758	&	0.23099	&	TERRA	\\
\hline\hline
\end{longtable}


\begin{thebibliography}{}
\expandafter\ifx\csname natexlab\endcsname\relax\def\natexlab#1{#1}\fi
\providecommand{\url}[1]{\href{#1}{#1}}
\providecommand{\dodoi}[1]{doi:~\href{http://doi.org/#1}{\nolinkurl{#1}}}
\providecommand{\doeprint}[1]{\href{http://ascl.net/#1}{\nolinkurl{http://ascl.net/#1}}}
\providecommand{\doarXiv}[1]{\href{https://arxiv.org/abs/#1}{\nolinkurl{https://arxiv.org/abs/#1}}}

\bibitem[{ESA(1997)}]{ESA1997}
 1997, ESA Special Publication, Vol. 1200, {The HIPPARCOS and TYCHO catalogues.
  Astrometric and photometric star catalogues derived from the ESA HIPPARCOS
  Space Astrometry Mission}

\bibitem[{{Affer} {et~al.}(2019){Affer}, {Damasso}, {Micela}, {Poretti}, {Scand
  ariato}, {Maldonado}, {Lanza}, {Covino}, {Garrido Rubio}, {Gonz{\'a}lez
  Hern{\'a}ndez}, {Gratton}, {Leto}, {Maggio}, {Perger}, {Sozzetti},
  {Su{\'a}rez Mascare{\~n}o}, {Bonomo}, {Borsa}, {Claudi}, {Cosentino},
  {Desidera}, {Giacobbe}, {Molinari}, {Pedani}, {Pinamonti}, {Rebolo}, {Ribas},
  \& {Toledo-Padr{\'o}n}}]{Affer2019}
{Affer}, L., {Damasso}, M., {Micela}, G., {et~al.} 2019, \aap, 622, A193,
  \dodoi{10.1051/0004-6361/201834868}

\bibitem[{{Aguilera-G{\'o}mez} {et~al.}(2018){Aguilera-G{\'o}mez},
  {Ram{\'\i}rez}, \& {Chanam{\'e}}}]{Aguilera-Gomez2018}
{Aguilera-G{\'o}mez}, C., {Ram{\'\i}rez}, I., \& {Chanam{\'e}}, J. 2018, \aap,
  614, A55, \dodoi{10.1051/0004-6361/201732209}

\bibitem[{{Anders} {et~al.}(2019){Anders}, {Khalatyan}, {Chiappini}, {Queiroz},
  {Santiago}, {Jordi}, {Girardi}, {Brown}, {Matijevi{\v{c}}}, {Monari},
  {Cantat-Gaudin}, {Weiler}, {Khan}, {Miglio}, {Carrillo}, {Romero-G{\'o}mez},
  {Minchev}, {de Jong}, {Antoja}, {Ramos}, {Steinmetz}, \& {Enke}}]{Anders2019}
{Anders}, F., {Khalatyan}, A., {Chiappini}, C., {et~al.} 2019, \aap, 628, A94,
  \dodoi{10.1051/0004-6361/201935765}

\bibitem[{{Anglada-Escud{\'e}} \& {Butler}(2012)}]{Anglada-Escude12a}
{Anglada-Escud{\'e}}, G., \& {Butler}, R.~P. 2012, \apjs, 200, 15,
  \dodoi{10.1088/0067-0049/200/2/15}

\bibitem[{{Bai} {et~al.}(2019){Bai}, {Liu}, {Bai}, {Wang}, \& {Fan}}]{Bai2019}
{Bai}, Y., {Liu}, J., {Bai}, Z., {Wang}, S., \& {Fan}, D. 2019, \aj, 158, 93,
  \dodoi{10.3847/1538-3881/ab3048}

\bibitem[{{Baranne} {et~al.}(1996){Baranne}, {Queloz}, {Mayor}, {Adrianzyk},
  {Knispel}, {Kohler}, {Lacroix}, {Meunier}, {Rimbaud}, \& {Vin}}]{Baranne1996}
{Baranne}, A., {Queloz}, D., {Mayor}, M., {et~al.} 1996, \aaps, 119, 373

\bibitem[{{Bayo} {et~al.}(2008){Bayo}, {Rodrigo}, {Barrado Y Navascu{\'e}s},
  {Solano}, {Guti{\'e}rrez}, {Morales-Calder{\'o}n}, \& {Allard}}]{Bayo2008}
{Bayo}, A., {Rodrigo}, C., {Barrado Y Navascu{\'e}s}, D., {et~al.} 2008, \aap,
  492, 277, \dodoi{10.1051/0004-6361:200810395}

\bibitem[{{Beatty} \& {Gaudi}(2015)}]{BeattyGaudi2015}
{Beatty}, T.~G., \& {Gaudi}, B.~S. 2015, \pasp, 127, 1240,
  \dodoi{10.1086/684264}

\bibitem[{{Bensby} {et~al.}(2014){Bensby}, {Feltzing}, \& {Oey}}]{Bensby2014}
{Bensby}, T., {Feltzing}, S., \& {Oey}, M.~S. 2014, \aap, 562, A71,
  \dodoi{10.1051/0004-6361/201322631}

\bibitem[{{Bermejo} {et~al.}(2013){Bermejo}, {Asensio Ramos}, \& {Allende
  Prieto}}]{Bermejo2013}
{Bermejo}, J.~M., {Asensio Ramos}, A., \& {Allende Prieto}, C. 2013, VizieR
  Online Data Catalog, J/A+A/553/A95

\bibitem[{{Bianchi} {et~al.}(2017){Bianchi}, {Shiao}, \&
  {Thilker}}]{Bianchi2017}
{Bianchi}, L., {Shiao}, B., \& {Thilker}, D. 2017, \apjs, 230, 24,
  \dodoi{10.3847/1538-4365/aa7053}

\bibitem[{{Biazzo} {et~al.}(2007){Biazzo}, {Frasca}, {Catalano}, \&
  {Marilli}}]{Biazzo2007}
{Biazzo}, K., {Frasca}, A., {Catalano}, S., \& {Marilli}, E. 2007,
  Astronomische Nachrichten, 328, 938, \dodoi{10.1002/asna.200710781}

\bibitem[{{Bland-Hawthorn} \& {Gerhard}(2016)}]{Bland-Hawthorn2016}
{Bland-Hawthorn}, J., \& {Gerhard}, O. 2016, \araa, 54, 529,
  \dodoi{10.1146/annurev-astro-081915-023441}

\bibitem[{{Boeche} \& {Grebel}(2016)}]{Boeche2016}
{Boeche}, C., \& {Grebel}, E.~K. 2016, \aap, 587, A2,
  \dodoi{10.1051/0004-6361/201526758}

\bibitem[{{Bopp} \& {Espenak}(1977)}]{Bopp1977}
{Bopp}, B.~W., \& {Espenak}, F. 1977, \aj, 82, 916, \dodoi{10.1086/112146}

\bibitem[{{Bouchy} {et~al.}(2001){Bouchy}, {Pepe}, \& {Queloz}}]{Bouchy2001}
{Bouchy}, F., {Pepe}, F., \& {Queloz}, D. 2001, Astronomy \& Astrophysics, 374,
  733, \dodoi{10.1051/0004-6361:20010730}

\bibitem[{{Bouchy} {et~al.}(2011){Bouchy}, {H{\'e}brard}, {Delfosse}, {Udry},
  {Lagrange}, {Arnold}, {Boisse}, {Bonfils}, {Debondt}, {Diaz}, {Eggenberger},
  {Ehrenreich}, {Forveille}, {Lovis}, {Moutou}, {Pepe}, {Perrier}, {Queloz},
  {Santerne}, {Santos}, \& {S{\'e}gransan}}]{Bouchy2011}
{Bouchy}, F., {H{\'e}brard}, G., {Delfosse}, X., {et~al.} 2011, in EPSC-DPS
  Joint Meeting 2011, Vol. 2011, 240

\bibitem[{{Bourges} {et~al.}(2017){Bourges}, {Mella}, {Lafrasse}, {Duvert},
  {Chelli}, {Le Bouquin}, {Delfosse}, \& {Chesneau}}]{Bourges2017}
{Bourges}, L., {Mella}, G., {Lafrasse}, S., {et~al.} 2017, VizieR Online Data
  Catalog, II/346

\bibitem[{{Bovy}(2016)}]{Bovy2016}
{Bovy}, J. 2016, \apj, 817, 49, \dodoi{10.3847/0004-637X/817/1/49}

\bibitem[{{Brandt} \& {Huang}(2015)}]{Brandt2015}
{Brandt}, T.~D., \& {Huang}, C.~X. 2015, \apj, 807, 24,
  \dodoi{10.1088/0004-637X/807/1/24}

\bibitem[{{Brewer} {et~al.}(2016){Brewer}, {Fischer}, {Valenti}, \&
  {Piskunov}}]{Brewer2016}
{Brewer}, J.~M., {Fischer}, D.~A., {Valenti}, J.~A., \& {Piskunov}, N. 2016,
  \apjs, 225, 32, \dodoi{10.3847/0067-0049/225/2/32}

\bibitem[{{Brewer} {et~al.}(2018){Brewer}, {Wang}, {Fischer}, \&
  {Foreman-Mackey}}]{Brewer2018}
{Brewer}, J.~M., {Wang}, S., {Fischer}, D.~A., \& {Foreman-Mackey}, D. 2018,
  \apjl, 867, L3, \dodoi{10.3847/2041-8213/aae710}

\bibitem[{{Bryan} {et~al.}(2019){Bryan}, {Knutson}, {Lee}, {Fulton}, {Batygin},
  {Ngo}, \& {Meshkat}}]{Bryan2019}
{Bryan}, M.~L., {Knutson}, H.~A., {Lee}, E.~J., {et~al.} 2019, \aj, 157, 52,
  \dodoi{10.3847/1538-3881/aaf57f}

\bibitem[{{Buchhave} {et~al.}(2016){Buchhave}, {Dressing}, {Dumusque}, {Rice},
  {Vanderburg}, {Mortier}, {Lopez-Morales}, {Lopez}, {Lundkvist}, {Kjeldsen},
  {Affer}, {Bonomo}, {Charbonneau}, {Collier Cameron}, {Cosentino}, {Figueira},
  {Fiorenzano}, {Harutyunyan}, {Haywood}, {Johnson}, {Latham}, {Lovis},
  {Malavolta}, {Mayor}, {Micela}, {Molinari}, {Motalebi}, {Nascimbeni}, {Pepe},
  {Phillips}, {Piotto}, {Pollacco}, {Queloz}, {Sasselov}, {S{\'e}gransan},
  {Sozzetti}, {Udry}, \& {Watson}}]{Buchhave2016}
{Buchhave}, L.~A., {Dressing}, C.~D., {Dumusque}, X., {et~al.} 2016, \aj, 152,
  160, \dodoi{10.3847/0004-6256/152/6/160}

\bibitem[{{Burt} {et~al.}(2015){Burt}, {Holden}, {Hanson}, {Laughlin}, {Vogt},
  {Butler}, {Keiser}, \& {Deich}}]{Burt2015}
{Burt}, J., {Holden}, B., {Hanson}, R., {et~al.} 2015, Journal of Astronomical
  Telescopes, Instruments, and Systems

\bibitem[{{Burt} {et~al.}(2020){Burt}, {Nielsen}, {Quinn}, {Mamajek},
  {Matthews}, {Zhou}, {Seidel}, {Huang}, {Lopez}, {Soto}, {Otegi}, {Stassun},
  {Kreidberg}, {Collins}, {Eastman}, {Rodriguez}, {Vanderburg}, {Halverson},
  {Teske}, {Wang}, {Butler}, {Bouchy}, {Dumusque}, {Segransen}, {Shectman},
  {Crane}, {Feng}, {Montet}, {Feinstein}, {Beletski}, {Flowers}, {G{\"u}nther},
  {Daylan}, {Collins}, {Conti}, {Gan}, {Jensen}, {Kielkopf}, {Tan}, {Helled},
  {Dorn}, {Haldemann}, {Lissauer}, {Ricker}, {Vanderspek}, {Latham}, {Seager},
  {Winn}, {Jenkins}, {Twicken}, {Smith}, {Tenenbaum}, {Cartwright}, {Barclay},
  {Pepper}, {Esquerdo}, \& {Fong}}]{Burt2020}
{Burt}, J.~A., {Nielsen}, L.~D., {Quinn}, S.~N., {et~al.} 2020, \aj, 160, 153,
  \dodoi{10.3847/1538-3881/abac0c}

\bibitem[{{Butler} {et~al.}(1996){Butler}, {Marcy}, {Williams}, {McCarthy},
  {Dosanjh}, \& {Vogt}}]{Butler1996}
{Butler}, R.~P., {Marcy}, G.~W., {Williams}, E., {et~al.} 1996, \pasp, 108,
  500, \dodoi{10.1086/133755}

\bibitem[{{Butler} {et~al.}(2017){Butler}, {Vogt}, {Laughlin}, {Burt},
  {Rivera}, {Tuomi}, {Teske}, {Arriagada}, {Diaz}, {Holden}, \&
  {Keiser}}]{Butler2017}
{Butler}, R.~P., {Vogt}, S.~S., {Laughlin}, G., {et~al.} 2017, \aj, 153, 208,
  \dodoi{10.3847/1538-3881/aa66ca}

\bibitem[{{Ca{\~n}as} {et~al.}(2019){Ca{\~n}as}, {Wang}, {Mahadevan}, {Bender},
  {De Lee}, {Fleming}, {Garc{\'\i}a-Hern{\'a}ndez}, {Hearty}, {Majewski},
  {Roman-Lopes}, {Schneider}, \& {Stassun}}]{Canas2019}
{Ca{\~n}as}, C.~I., {Wang}, S., {Mahadevan}, S., {et~al.} 2019, \apjl, 870,
  L17, \dodoi{10.3847/2041-8213/aafa1e}

\bibitem[{{Casagrande} {et~al.}(2006){Casagrande}, {Portinari}, \&
  {Flynn}}]{Casagrande2006}
{Casagrande}, L., {Portinari}, L., \& {Flynn}, C. 2006, \mnras, 373, 13,
  \dodoi{10.1111/j.1365-2966.2006.10999.x}

\bibitem[{{Casagrande} {et~al.}(2010){Casagrande}, {Ram{\'\i}rez},
  {Mel{\'e}ndez}, {Bessell}, \& {Asplund}}]{Casagrande2010}
{Casagrande}, L., {Ram{\'\i}rez}, I., {Mel{\'e}ndez}, J., {Bessell}, M., \&
  {Asplund}, M. 2010, \aap, 512, A54, \dodoi{10.1051/0004-6361/200913204}

\bibitem[{{Chambers} {et~al.}(1996){Chambers}, {Wetherill}, \&
  {Boss}}]{Chambers1996}
{Chambers}, J.~E., {Wetherill}, G.~W., \& {Boss}, A.~P. 1996, \icarus, 119,
  261, \dodoi{10.1006/icar.1996.0019}

\bibitem[{{Chelli} {et~al.}(2016){Chelli}, {Duvert}, {Bourg{\`e}s}, {Mella},
  {Lafrasse}, {Bonneau}, \& {Chesneau}}]{Chelli2016}
{Chelli}, A., {Duvert}, G., {Bourg{\`e}s}, L., {et~al.} 2016, \aap, 589, A112,
  \dodoi{10.1051/0004-6361/201527484}

\bibitem[{{Chen} {et~al.}(2014){Chen}, {Girardi}, {Bressan}, {Marigo},
  {Barbieri}, \& {Kong}}]{Chen2014}
{Chen}, Y., {Girardi}, L., {Bressan}, A., {et~al.} 2014, \mnras, 444, 2525,
  \dodoi{10.1093/mnras/stu1605}

\bibitem[{{Christiansen} {et~al.}(2017){Christiansen}, {Vanderburg}, {Burt},
  {Fulton}, {Batygin}, {Benneke}, {Brewer}, {Charbonneau}, {Ciardi}, {Collier
  Cameron}, {Coughlin}, {Crossfield}, {Dressing}, {Greene}, {Howard}, {Latham},
  {Molinari}, {Mortier}, {Mullally}, {Pepe}, {Rice}, {Sinukoff}, {Sozzetti},
  {Thompson}, {Udry}, {Vogt}, {Barman}, {Batalha}, {Bouchy}, {Buchhave},
  {Butler}, {Cosentino}, {Dupuy}, {Ehrenreich}, {Fiorenzano}, {Hansen},
  {Henning}, {Hirsch}, {Holden}, {Isaacson}, {Johnson}, {Knutson}, {Kosiarek},
  {L{\'o}pez-Morales}, {Lovis}, {Malavolta}, {Mayor}, {Micela}, {Motalebi},
  {Petigura}, {Phillips}, {Piotto}, {Rogers}, {Sasselov}, {Schlieder},
  {S{\'e}gransan}, {Watson}, \& {Weiss}}]{Christiansen2017}
{Christiansen}, J.~L., {Vanderburg}, A., {Burt}, J., {et~al.} 2017, \aj, 154,
  122, \dodoi{10.3847/1538-3881/aa832d}

\bibitem[{{Cosentino} {et~al.}(2012){Cosentino}, {Lovis}, {Pepe}, {Collier
  Cameron}, {Latham}, {Molinari}, {Udry}, {Bezawada}, {Black}, {Born},
  {Buchschacher}, {Charbonneau}, {Figueira}, {Fleury}, {Galli}, {Gallie},
  {Gao}, {Ghedina}, {Gonzalez}, {Gonzalez}, {Guerra}, {Henry}, {Horne},
  {Hughes}, {Kelly}, {Lodi}, {Lunney}, {Maire}, {Mayor}, {Micela}, {Ordway},
  {Peacock}, {Phillips}, {Piotto}, {Pollacco}, {Queloz}, {Rice}, {Riverol},
  {Riverol}, {San Juan}, {Sasselov}, {Segransan}, {Sozzetti}, {Sosnowska},
  {Stobie}, {Szentgyorgyi}, {Vick}, \& {Weber}}]{Cosentino2012}
{Cosentino}, R., {Lovis}, C., {Pepe}, F., {et~al.} 2012, in Society of
  Photo-Optical Instrumentation Engineers (SPIE) Conference Series, Vol. 8446,
  Society of Photo-Optical Instrumentation Engineers (SPIE) Conference Series,
  1

\bibitem[{{Crane}(2010)}]{Crane2010}
{Crane}, J. 2010, in Astronomy of Exoplanets with Precise Radial Velocities, 19

\bibitem[{{Crane} {et~al.}(2006){Crane}, {Shectman}, \& {Butler}}]{Crane2006}
{Crane}, J.~D., {Shectman}, S.~A., \& {Butler}, R.~P. 2006, Society of
  Photo-Optical Instrumentation Engineers (SPIE) Conference Series, Vol. 6269,
  {The Carnegie Planet Finder Spectrograph}, 626931

\bibitem[{{Crane} {et~al.}(2008){Crane}, {Shectman}, {Butler}, {Thompson}, \&
  {Burley}}]{Crane2008}
{Crane}, J.~D., {Shectman}, S.~A., {Butler}, R.~P., {Thompson}, I.~B., \&
  {Burley}, G.~S. 2008, Society of Photo-Optical Instrumentation Engineers
  (SPIE) Conference Series, Vol. 7014, {The Carnegie Planet Finder
  Spectrograph: a status report}, 701479

\bibitem[{{Cummings} {et~al.}(2017){Cummings}, {Deliyannis}, {Maderak}, \&
  {Steinhauer}}]{Cummings2017}
{Cummings}, J.~D., {Deliyannis}, C.~P., {Maderak}, R.~M., \& {Steinhauer}, A.
  2017, \aj, 153, 128, \dodoi{10.3847/1538-3881/aa5b86}

\bibitem[{{Curtis} {et~al.}(2019){Curtis}, {Ag{\"u}eros}, {Douglas}, \&
  {Meibom}}]{Curtis2019}
{Curtis}, J.~L., {Ag{\"u}eros}, M.~A., {Douglas}, S.~T., \& {Meibom}, S. 2019,
  \apj, 879, 49, \dodoi{10.3847/1538-4357/ab2393}

\bibitem[{{Cutri} \& {et al.}(2012)}]{Cutri2012}
{Cutri}, R.~M., \& {et al.} 2012, VizieR Online Data Catalog, II/311

\bibitem[{{Cutri} {et~al.}(2003){Cutri}, {Skrutskie}, {van Dyk}, {Beichman},
  {Carpenter}, {Chester}, {Cambresy}, {Evans}, {Fowler}, {Gizis}, {Howard},
  {Huchra}, {Jarrett}, {Kopan}, {Kirkpatrick}, {Light}, {Marsh}, {McCallon},
  {Schneider}, {Stiening}, {Sykes}, {Weinberg}, {Wheaton}, {Wheelock}, \&
  {Zacarias}}]{Cutri2003}
{Cutri}, R.~M., {Skrutskie}, M.~F., {van Dyk}, S., {et~al.} 2003, VizieR Online
  Data Catalog, 2246

\bibitem[{{Dawson} \& {Fabrycky}(2010)}]{Dawson2010}
{Dawson}, R.~I., \& {Fabrycky}, D.~C. 2010, \apj, 722, 937,
  \dodoi{10.1088/0004-637X/722/1/937}

\bibitem[{{D{\'{\i}}az} {et~al.}(2018){D{\'{\i}}az}, {Jenkins}, {Tuomi},
  {Butler}, {Soto}, {Teske}, {Feng}, {Shectman}, {Arriagada}, {Crane},
  {Thompson}, \& {Vogt}}]{diaz2018}
{D{\'{\i}}az}, M.~R., {Jenkins}, J.~S., {Tuomi}, M., {et~al.} 2018, \aj, 155,
  126, \dodoi{10.3847/1538-3881/aaa896}

\bibitem[{{Dumusque} {et~al.}(2017){Dumusque}, {Borsa}, {Damasso},
  {D{\'{\i}}az}, {Gregory}, {Hara}, {Hatzes}, {Rajpaul}, {Tuomi}, {Aigrain},
  {Anglada-Escud{\'e}}, {Bonomo}, {Bou{\'e}}, {Dauvergne}, {Frustagli},
  {Giacobbe}, {Haywood}, {Jones}, {Laskar}, {Pinamonti}, {Poretti}, {Rainer},
  {S{\'e}gransan}, {Sozzetti}, \& {Udry}}]{Dumusque2017}
{Dumusque}, X., {Borsa}, F., {Damasso}, M., {et~al.} 2017, \aap, 598, A133,
  \dodoi{10.1051/0004-6361/201628671}

\bibitem[{{Duncan} {et~al.}(1991){Duncan}, {Vaughan}, {Wilson}, {Preston},
  {Frazer}, {Lanning}, {Misch}, {Mueller}, {Soyumer}, {Woodard}, {Baliunas},
  {Noyes}, {Hartmann}, {Porter}, {Zwaan}, {Middelkoop}, {Rutten}, \&
  {Mihalas}}]{Duncan1991}
{Duncan}, D.~K., {Vaughan}, A.~H., {Wilson}, O.~C., {et~al.} 1991, \apjs, 76,
  383, \dodoi{10.1086/191572}

\bibitem[{{Dutra-Ferreira} {et~al.}(2016){Dutra-Ferreira}, {Pasquini},
  {Smiljanic}, {Porto de Mello}, \& {Steffen}}]{Dutra-Ferreira2016}
{Dutra-Ferreira}, L., {Pasquini}, L., {Smiljanic}, R., {Porto de Mello}, G.~F.,
  \& {Steffen}, M. 2016, \aap, 585, A75, \dodoi{10.1051/0004-6361/201526783}

\bibitem[{{Egeland}(2018)}]{Egeland2018}
{Egeland}, R. 2018, \apjs, 236, 19, \dodoi{10.3847/1538-4365/aab771}

\bibitem[{{Eker} {et~al.}(2018){Eker}, {Bak{\i}{\textcommabelow s}}, {Bilir},
  {Soydugan}, {Steer}, {Soydugan}, {Bak{\i}{\textcommabelow s}},
  {Ali{\c{c}}avu{\textcommabelow s}}, {Aslan}, \& {Alpsoy}}]{Eker2018}
{Eker}, Z., {Bak{\i}{\textcommabelow s}}, V., {Bilir}, S., {et~al.} 2018,
  \mnras, 479, 5491, \dodoi{10.1093/mnras/sty1834}

\bibitem[{{Feng} {et~al.}(2017{\natexlab{a}}){Feng}, {Tuomi}, \&
  {Jones}}]{feng2017}
{Feng}, F., {Tuomi}, M., \& {Jones}, H.~R.~A. 2017{\natexlab{a}}, \aap, 605,
  A103, \dodoi{10.1051/0004-6361/201730406}

\bibitem[{{Feng} {et~al.}(2017{\natexlab{b}}){Feng}, {Tuomi}, \&
  {Jones}}]{Feng2017b}
---. 2017{\natexlab{b}}, \mnras, 470, 4794, \dodoi{10.1093/mnras/stx1126}

\bibitem[{{Feng} {et~al.}(2016){Feng}, {Tuomi}, {Jones}, {Butler}, \&
  {Vogt}}]{feng2016}
{Feng}, F., {Tuomi}, M., {Jones}, H.~R.~A., {Butler}, R.~P., \& {Vogt}, S.
  2016, \mnras, 461, 2440, \dodoi{10.1093/mnras/stw1478}

\bibitem[{{Feng} {et~al.}(2019){Feng}, {Crane}, {Xuesong Wang}, {Teske},
  {Shectman}, {D{\'\i}az}, {Thompson}, {Jones}, \& {Butler}}]{Feng2019}
{Feng}, F., {Crane}, J.~D., {Xuesong Wang}, S., {et~al.} 2019, \apjs, 242, 25,
  \dodoi{10.3847/1538-4365/ab1b16}

\bibitem[{{Franchini} {et~al.}(2014){Franchini}, {Morossi}, {di Marcantonio},
  {Malagnini}, \& {Chavez}}]{Franchini2014}
{Franchini}, M., {Morossi}, C., {di Marcantonio}, P., {Malagnini}, M.~L., \&
  {Chavez}, M. 2014, \mnras, 442, 220, \dodoi{10.1093/mnras/stu873}

\bibitem[{{Frasca} {et~al.}(2015){Frasca}, {Biazzo}, {Lanzafame}, {Alcal{\'a}},
  {Brugaletta}, {Klutsch}, {Stelzer}, {Sacco}, {Spina}, {Jeffries}, {Montes},
  {Alfaro}, {Barentsen}, {Bonito}, {Gameiro}, {L{\'o}pez-Santiago}, {Pace},
  {Pasquini}, {Prisinzano}, {Sousa}, {Gilmore}, {Randich}, {Micela},
  {Bragaglia}, {Flaccomio}, {Bayo}, {Costado}, {Franciosini}, {Hill},
  {Hourihane}, {Jofr{\'e}}, {Lardo}, {Maiorca}, {Masseron}, {Morbidelli}, \&
  {Worley}}]{Frasca2015}
{Frasca}, A., {Biazzo}, K., {Lanzafame}, A.~C., {et~al.} 2015, \aap, 575, A4,
  \dodoi{10.1051/0004-6361/201424409}

\bibitem[{{Gagn{\'e}} {et~al.}(2018){Gagn{\'e}}, {Mamajek}, {Malo}, {Riedel},
  {Rodriguez}, {Lafreni{\`e}re}, {Faherty}, {Roy-Loubier}, {Pueyo}, {Robin}, \&
  {Doyon}}]{Gagne2018}
{Gagn{\'e}}, J., {Mamajek}, E.~E., {Malo}, L., {et~al.} 2018, \apj, 856, 23,
  \dodoi{10.3847/1538-4357/aaae09}

\bibitem[{{Gaia Collaboration} {et~al.}(2016){Gaia Collaboration}, {Brown},
  {Vallenari}, {Prusti}, {de Bruijne}, {Mignard}, {Drimmel}, {Babusiaux},
  {Bailer-Jones}, {Bastian}, \& et~al.}]{Gaia2016}
{Gaia Collaboration}, {Brown}, A.~G.~A., {Vallenari}, A., {et~al.} 2016, \aap,
  595, A2, \dodoi{10.1051/0004-6361/201629512}

\bibitem[{{Gaia Collaboration} {et~al.}(2018){Gaia Collaboration}, {Brown},
  {Vallenari}, {Prusti}, {de Bruijne}, {Babusiaux}, {Bailer-Jones}, {Biermann},
  {Evans}, {Eyer}, {Jansen}, {Jordi}, {Klioner}, {Lammers}, {Lindegren},
  {Luri}, {Mignard}, {Panem}, {Pourbaix}, {Randich}, {Sartoretti}, {Siddiqui},
  {Soubiran}, {van Leeuwen}, {Walton}, {Arenou}, {Bastian}, {Cropper},
  {Drimmel}, {Katz}, {Lattanzi}, {Bakker}, {Cacciari}, {Casta{\~n}eda},
  {Chaoul}, {Cheek}, {De Angeli}, {Fabricius}, {Guerra}, {Holl}, {Masana},
  {Messineo}, {Mowlavi}, {Nienartowicz}, {Panuzzo}, {Portell}, {Riello},
  {Seabroke}, {Tanga}, {Th{\'e}venin}, {Gracia-Abril}, {Comoretto},
  {Garcia-Reinaldos}, {Teyssier}, {Altmann}, {Andrae}, {Audard},
  {Bellas-Velidis}, {Benson}, {Berthier}, {Blomme}, {Burgess}, {Busso},
  {Carry}, {Cellino}, {Clementini}, {Clotet}, {Creevey}, {Davidson}, {De
  Ridder}, {Delchambre}, {Dell'Oro}, {Ducourant},
  {Fern{\'a}ndez-Hern{\'a}ndez}, {Fouesneau}, {Fr{\'e}mat}, {Galluccio},
  {Garc{\'\i}a-Torres}, {Gonz{\'a}lez-N{\'u}{\~n}ez}, {Gonz{\'a}lez-Vidal},
  {Gosset}, {Guy}, {Halbwachs}, {Hambly}, {Harrison}, {Hern{\'a}ndez},
  {Hestroffer}, {Hodgkin}, {Hutton}, {Jasniewicz}, {Jean-Antoine-Piccolo},
  {Jordan}, {Korn}, {Krone-Martins}, {Lanzafame}, {Lebzelter}, {L{\"o}ffler},
  {Manteiga}, {Marrese}, {Mart{\'\i}n-Fleitas}, {Moitinho}, {Mora}, {Muinonen},
  {Osinde}, {Pancino}, {Pauwels}, {Petit}, {Recio-Blanco}, {Richards},
  {Rimoldini}, {Robin}, {Sarro}, {Siopis}, {Smith}, {Sozzetti}, {S{\"u}veges},
  {Torra}, {van Reeven}, {Abbas}, {Abreu Aramburu}, {Accart}, {Aerts},
  {Altavilla}, {{\'A}lvarez}, {Alvarez}, {Alves}, {Anderson}, {Andrei},
  {Anglada Varela}, {Antiche}, {Antoja}, {Arcay}, {Astraatmadja}, {Bach},
  {Baker}, {Balaguer-N{\'u}{\~n}ez}, {Balm}, {Barache}, {Barata}, {Barbato},
  {Barblan}, {Barklem}, {Barrado}, {Barros}, {Barstow}, {Bartholom{\'e}
  Mu{\~n}oz}, {Bassilana}, {Becciani}, {Bellazzini}, {Berihuete}, {Bertone},
  {Bianchi}, {Bienaym{\'e}}, {Blanco-Cuaresma}, {Boch}, {Boeche}, {Bombrun},
  {Borrachero}, {Bossini}, {Bouquillon}, {Bourda}, {Bragaglia}, {Bramante},
  {Breddels}, {Bressan}, {Brouillet}, {Br{\"u}semeister}, {Brugaletta},
  {Bucciarelli}, {Burlacu}, {Busonero}, {Butkevich}, {Buzzi}, {Caffau},
  {Cancelliere}, {Cannizzaro}, {Cantat-Gaudin}, {Carballo}, {Carlucci},
  {Carrasco}, {Casamiquela}, {Castellani}, {Castro-Ginard}, {Charlot},
  {Chemin}, {Chiavassa}, {Cocozza}, {Costigan}, {Cowell}, {Crifo}, {Crosta},
  {Crowley}, {Cuypers}, {Dafonte}, {Damerdji}, {Dapergolas}, {David}, {David},
  {de Laverny}, {De Luise}, {De March}, {de Martino}, {de Souza}, {de Torres},
  {Debosscher}, {del Pozo}, {Delbo}, {Delgado}, {Delgado}, {Di Matteo},
  {Diakite}, {Diener}, {Distefano}, {Dolding}, {Drazinos}, {Dur{\'a}n},
  {Edvardsson}, {Enke}, {Eriksson}, {Esquej}, {Eynard Bontemps}, {Fabre},
  {Fabrizio}, {Faigler}, {Falc{\~a}o}, {Farr{\`a}s Casas}, {Federici},
  {Fedorets}, {Fernique}, {Figueras}, {Filippi}, {Findeisen}, {Fonti},
  {Fraile}, {Fraser}, {Fr{\'e}zouls}, {Gai}, {Galleti}, {Garabato},
  {Garc{\'\i}a-Sedano}, {Garofalo}, {Garralda}, {Gavel}, {Gavras}, {Gerssen},
  {Geyer}, {Giacobbe}, {Gilmore}, {Girona}, {Giuffrida}, {Glass}, {Gomes},
  {Granvik}, {Gueguen}, {Guerrier}, {Guiraud}, {Guti{\'e}rrez-S{\'a}nchez},
  {Haigron}, {Hatzidimitriou}, {Hauser}, {Haywood}, {Heiter}, {Helmi}, {Heu},
  {Hilger}, {Hobbs}, {Hofmann}, {Holland}, {Huckle}, {Hypki}, {Icardi},
  {Jan{\ss}en}, {Jevardat de Fombelle}, {Jonker}, {Juh{\'a}sz}, {Julbe},
  {Karampelas}, {Kewley}, {Klar}, {Kochoska}, {Kohley}, {Kolenberg},
  {Kontizas}, {Kontizas}, {Koposov}, {Kordopatis}, {Kostrzewa-Rutkowska},
  {Koubsky}, {Lambert}, {Lanza}, {Lasne}, {Lavigne}, {Le Fustec}, {Le
  Poncin-Lafitte}, {Lebreton}, {Leccia}, {Leclerc}, {Lecoeur-Taibi},
  {Lenhardt}, {Leroux}, {Liao}, {Licata}, {Lindstr{\o}m}, {Lister}, {Livanou},
  {Lobel}, {L{\'o}pez}, {Managau}, {Mann}, {Mantelet}, {Marchal}, {Marchant},
  {Marconi}, {Marinoni}, {Marschalk{\'o}}, {Marshall}, {Martino}, {Marton},
  {Mary}, {Massari}, {Matijevi{\v{c}}}, {Mazeh}, {McMillan}, {Messina},
  {Michalik}, {Millar}, {Molina}, {Molinaro}, {Moln{\'a}r}, {Montegriffo},
  {Mor}, {Morbidelli}, {Morel}, {Morris}, {Mulone}, {Muraveva}, {Musella},
  {Nelemans}, {Nicastro}, {Noval}, {O'Mullane}, {Ord{\'e}novic},
  {Ord{\'o}{\~n}ez-Blanco}, {Osborne}, {Pagani}, {Pagano}, {Pailler},
  {Palacin}, {Palaversa}, {Panahi}, {Pawlak}, {Piersimoni}, {Pineau}, {Plachy},
  {Plum}, {Poggio}, {Poujoulet}, {Pr{\v{s}}a}, {Pulone}, {Racero}, {Ragaini},
  {Rambaux}, {Ramos-Lerate}, {Regibo}, {Reyl{\'e}}, {Riclet}, {Ripepi}, {Riva},
  {Rivard}, {Rixon}, {Roegiers}, {Roelens}, {Romero-G{\'o}mez}, {Rowell},
  {Royer}, {Ruiz-Dern}, {Sadowski}, {Sagrist{\`a} Sell{\'e}s}, {Sahlmann},
  {Salgado}, {Salguero}, {Sanna}, {Santana-Ros}, {Sarasso}, {Savietto},
  {Schultheis}, {Sciacca}, {Segol}, {Segovia}, {S{\'e}gransan}, {Shih},
  {Siltala}, {Silva}, {Smart}, {Smith}, {Solano}, {Solitro}, {Sordo}, {Soria
  Nieto}, {Souchay}, {Spagna}, {Spoto}, {Stampa}, {Steele},
  {Steidelm{\"u}ller}, {Stephenson}, {Stoev}, {Suess}, {Surdej}, {Szabados},
  {Szegedi-Elek}, {Tapiador}, {Taris}, {Tauran}, {Taylor}, {Teixeira},
  {Terrett}, {Teyssand ier}, {Thuillot}, {Titarenko}, {Torra Clotet}, {Turon},
  {Ulla}, {Utrilla}, {Uzzi}, {Vaillant}, {Valentini}, {Valette}, {van Elteren},
  {Van Hemelryck}, {van Leeuwen}, {Vaschetto}, {Vecchiato}, {Veljanoski},
  {Viala}, {Vicente}, {Vogt}, {von Essen}, {Voss}, {Votruba}, {Voutsinas},
  {Walmsley}, {Weiler}, {Wertz}, {Wevers}, {Wyrzykowski}, {Yoldas},
  {{\v{Z}}erjal}, {Ziaeepour}, {Zorec}, {Zschocke}, {Zucker}, {Zurbach}, \&
  {Zwitter}}]{GaiaDR2}
---. 2018, \aap, 616, A1, \dodoi{10.1051/0004-6361/201833051}

\bibitem[{{Gladman}(1993)}]{Gladman1993}
{Gladman}, B. 1993, \icarus, 106, 247, \dodoi{10.1006/icar.1993.1169}

\bibitem[{{Gomes da Silva} {et~al.}(2011){Gomes da Silva}, {Santos}, {Bonfils},
  {Delfosse}, {Forveille}, \& {Udry}}]{Gomes2011}
{Gomes da Silva}, J., {Santos}, N.~C., {Bonfils}, X., {et~al.} 2011, \aap, 534,
  A30, \dodoi{10.1051/0004-6361/201116971}

\bibitem[{{Gomes da Silva} {et~al.}(2012){Gomes da Silva}, {Santos}, {Bonfils},
  {Delfosse}, {Forveille}, {Udry}, {Dumusque}, \& {Lovis}}]{GomesdaSilva2012}
---. 2012, \aap, 541, A9, \dodoi{10.1051/0004-6361/201118598}

\bibitem[{{Gonz{\'a}lez Hern{\'a}ndez} \&
  {Bonifacio}(2009)}]{GonzalezHernandez2009}
{Gonz{\'a}lez Hern{\'a}ndez}, J.~I., \& {Bonifacio}, P. 2009, \aap, 497, 497,
  \dodoi{10.1051/0004-6361/200810904}

\bibitem[{{Gossage} {et~al.}(2018){Gossage}, {Conroy}, {Dotter}, {Choi},
  {Rosenfield}, {Cargile}, \& {Dolphin}}]{Gossage2018}
{Gossage}, S., {Conroy}, C., {Dotter}, A., {et~al.} 2018, \apj, 863, 67,
  \dodoi{10.3847/1538-4357/aad0a0}

\bibitem[{{Gray} {et~al.}(2006){Gray}, {Corbally}, {Garrison}, {McFadden},
  {Bubar}, {McGahee}, {O'Donoghue}, \& {Knox}}]{Gray2006}
{Gray}, R.~O., {Corbally}, C.~J., {Garrison}, R.~F., {et~al.} 2006, \aj, 132,
  161, \dodoi{10.1086/504637}

\bibitem[{{Gray} {et~al.}(2003){Gray}, {Corbally}, {Garrison}, {McFadden}, \&
  {Robinson}}]{Gray2003}
{Gray}, R.~O., {Corbally}, C.~J., {Garrison}, R.~F., {McFadden}, M.~T., \&
  {Robinson}, P.~E. 2003, \aj, 126, 2048, \dodoi{10.1086/378365}

\bibitem[{{Guo} {et~al.}(2017){Guo}, {Johnson}, {Mann}, {Kraus}, {Curtis}, \&
  {Latham}}]{Guo2017}
{Guo}, X., {Johnson}, J.~A., {Mann}, A.~W., {et~al.} 2017, \apj, 838, 25,
  \dodoi{10.3847/1538-4357/aa6004}

\bibitem[{{Haario} {et~al.}(2006){Haario}, {Laine}, {Mira}, \&
  {Saksman}}]{Haario2006}
{Haario}, H., {Laine}, M., {Mira}, A., \& {Saksman}, E. 2006, Stat Comp., 16

\bibitem[{{Hadden} \& {Lithwick}(2017)}]{HaddenLithwick2017}
{Hadden}, S., \& {Lithwick}, Y. 2017, \aj, 154, 5,
  \dodoi{10.3847/1538-3881/aa71ef}

\bibitem[{{Henry}(1999)}]{Henry1999a}
{Henry}, G.~W. 1999, \pasp, 111, 845, \dodoi{10.1086/316388}

\bibitem[{{Henry} {et~al.}(1999){Henry}, {Marcy}, {Butler}, \&
  {Vogt}}]{Henry1999b}
{Henry}, G.~W., {Marcy}, G., {Butler}, R.~P., \& {Vogt}, S.~S. 1999, \iaucirc,
  7307, 1

\bibitem[{{Henry} {et~al.}(2000){Henry}, {Marcy}, {Butler}, \&
  {Vogt}}]{Henry2000}
{Henry}, G.~W., {Marcy}, G.~W., {Butler}, R.~P., \& {Vogt}, S.~S. 2000, \apjl,
  529, L41, \dodoi{10.1086/312458}

\bibitem[{{Hinkel} {et~al.}(2017){Hinkel}, {Mamajek}, {Turnbull}, {Osby},
  {Shkolnik}, {Smith}, {Klimasewski}, {Somers}, \& {Desch}}]{Hinkel2017}
{Hinkel}, N.~R., {Mamajek}, E.~E., {Turnbull}, M.~C., {et~al.} 2017, \apj, 848,
  34, \dodoi{10.3847/1538-4357/aa8b0f}

\bibitem[{{Houdebine}(2012)}]{Houdebine2012}
{Houdebine}, E.~R. 2012, \mnras, 421, 3189,
  \dodoi{10.1111/j.1365-2966.2012.20649.x}

\bibitem[{{Houdebine} {et~al.}(2019){Houdebine}, {Mullan}, {Doyle}, {de La
  Vieuville}, {Butler}, \& {Paletou}}]{Houdebine2019}
{Houdebine}, {\'E}.~R., {Mullan}, D.~J., {Doyle}, J.~G., {et~al.} 2019, \aj,
  158, 56, \dodoi{10.3847/1538-3881/ab23fe}

\bibitem[{{Houdebine} {et~al.}(2016){Houdebine}, {Mullan}, {Paletou}, \&
  {Gebran}}]{Houdebine2016}
{Houdebine}, E.~R., {Mullan}, D.~J., {Paletou}, F., \& {Gebran}, M. 2016, \apj,
  822, 97, \dodoi{10.3847/0004-637X/822/2/97}

\bibitem[{{Howard} {et~al.}(2010){Howard}, {Johnson}, {Marcy}, {Fischer},
  {Wright}, {Bernat}, {Henry}, {Peek}, {Isaacson}, {Apps}, {Endl}, {Cochran},
  {Valenti}, {Anderson}, \& {Piskunov}}]{Howard2010}
{Howard}, A.~W., {Johnson}, J.~A., {Marcy}, G.~W., {et~al.} 2010, \apj, 721,
  1467, \dodoi{10.1088/0004-637X/721/2/1467}

\bibitem[{{Isaacson} \& {Fischer}(2010)}]{Isaacson2010}
{Isaacson}, H., \& {Fischer}, D. 2010, \apj, 725, 875,
  \dodoi{10.1088/0004-637X/725/1/875}

\bibitem[{{Jenkins} {et~al.}(2016){Jenkins}, {Twicken}, {McCauliff},
  {Campbell}, {Sanderfer}, {Lung}, {Mansouri-Samani}, {Girouard}, {Tenenbaum},
  {Klaus}, {Smith}, {Caldwell}, {Chacon}, {Henze}, {Heiges}, {Latham},
  {Morgan}, {Swade}, {Rinehart}, \& {Vanderspek}}]{Jenkins2016}
{Jenkins}, J.~M., {Twicken}, J.~D., {McCauliff}, S., {et~al.} 2016, Society of
  Photo-Optical Instrumentation Engineers (SPIE) Conference Series, Vol. 9913,
  {The TESS science processing operations center}, 99133E

\bibitem[{{Kaminski} {et~al.}(2018){Kaminski}, {Trifonov}, {Caballero},
  {Quirrenbach}, {Ribas}, {Reiners}, {Amado}, {Zechmeister}, {Dreizler},
  {Perger}, {Tal-Or}, {Bonfils}, {Mayor}, {Astudillo-Defru}, {Bauer},
  {B{\'e}jar}, {Cifuentes}, {Colom{\'e}}, {Cort{\'e}s-Contreras}, {Delfosse},
  {D{\'\i}ez-Alonso}, {Forveille}, {Guenther}, {Hatzes}, {Henning}, {Jeffers},
  {K{\"u}rster}, {Lafarga}, {Luque}, {Mandel}, {Montes}, {Morales},
  {Passegger}, {Pedraz}, {Reffert}, {Sadegi}, {Schweitzer}, {Seifert}, {Stahl},
  \& {Udry}}]{Kaminski2018}
{Kaminski}, A., {Trifonov}, T., {Caballero}, J.~A., {et~al.} 2018, \aap, 618,
  A115, \dodoi{10.1051/0004-6361/201833354}

\bibitem[{{Kass} \& {Raftery}(1995)}]{KassRaftery1995}
{Kass}, \& {Raftery}. 1995, "Journal of the American statistical association",
  90

\bibitem[{{Katz} {et~al.}(2011){Katz}, {Soubiran}, {Cayrel}, {Barbuy}, {Friel},
  {Bienaym{\'e}}, \& {Perrin}}]{Katz2011}
{Katz}, D., {Soubiran}, C., {Cayrel}, R., {et~al.} 2011, \aap, 525, A90,
  \dodoi{10.1051/0004-6361/201014840}

\bibitem[{{Kazarovets} {et~al.}(2003){Kazarovets}, {Kireeva}, {Samus}, \&
  {Durlevich}}]{Kazarovets2003}
{Kazarovets}, E.~V., {Kireeva}, N.~N., {Samus}, N.~N., \& {Durlevich}, O.~V.
  2003, Information Bulletin on Variable Stars, 5422, 1

\bibitem[{{Kervella} {et~al.}(2019){Kervella}, {Arenou}, {Mignard}, \&
  {Th{\'e}venin}}]{Kervella2019}
{Kervella}, P., {Arenou}, F., {Mignard}, F., \& {Th{\'e}venin}, F. 2019, \aap,
  623, A72, \dodoi{10.1051/0004-6361/201834371}

\bibitem[{{K{\'o}sp{\'a}l} {et~al.}(2009){K{\'o}sp{\'a}l}, {Ardila},
  {Mo{\'o}r}, \& {{\'A}brah{\'a}m}}]{Kospal2009}
{K{\'o}sp{\'a}l}, {\'A}., {Ardila}, D.~R., {Mo{\'o}r}, A., \&
  {{\'A}brah{\'a}m}, P. 2009, \apjl, 700, L73,
  \dodoi{10.1088/0004-637X/700/2/L73}

\bibitem[{{Lalitha} {et~al.}(2019){Lalitha}, {Baroch}, {Morales}, {Passegger},
  {Bauer}, {Cardona Guill{\'e}n}, {Dreizler}, {Oshagh}, {Reiners}, {Ribas},
  {Caballero}, {Quirrenbach}, {Amado}, {B{\'e}jar}, {Colom{\'e}},
  {Cort{\'e}s-Contreras}, {Galad{\'\i}-Enr{\'\i}quez}, {Gonz{\'a}lez-Cuesta},
  {Guenther}, {Hagen}, {Henning}, {Herrero}, {Husser}, {Jeffers}, {Kaminski},
  {K{\"u}rster}, {Lafarga}, {Lodieu}, {L{\'o}pez-Gonz{\'a}lez}, {Montes},
  {Perger}, {Rosich}, {Rodr{\'\i}guez}, {Rodr{\'\i}guez-L{\'o}pez}, {Schmitt},
  {Tal-Or}, \& {Zechmeister}}]{Lalitha2019}
{Lalitha}, S., {Baroch}, D., {Morales}, J.~C., {et~al.} 2019, \aap, 627, A116,
  \dodoi{10.1051/0004-6361/201935534}

\bibitem[{{Latham} {et~al.}(2011){Latham}, {Rowe}, {Quinn}, {Batalha},
  {Borucki}, {Brown}, {Bryson}, {Buchhave}, {Caldwell}, {Carter},
  {Christiansen}, {Ciardi}, {Cochran}, {Dunham}, {Fabrycky}, {Ford}, {Gautier},
  {Gilliland}, {Holman}, {Howell}, {Ibrahim}, {Isaacson}, {Jenkins}, {Koch},
  {Lissauer}, {Marcy}, {Quintana}, {Ragozzine}, {Sasselov}, {Shporer},
  {Steffen}, {Welsh}, \& {Wohler}}]{Latham2011}
{Latham}, D.~W., {Rowe}, J.~F., {Quinn}, S.~N., {et~al.} 2011, \apjl, 732, L24,
  \dodoi{10.1088/2041-8205/732/2/L24}

\bibitem[{{L{\'e}pine} \& {Gaidos}(2011)}]{Lepine2011}
{L{\'e}pine}, S., \& {Gaidos}, E. 2011, \aj, 142, 138,
  \dodoi{10.1088/0004-6256/142/4/138}

\bibitem[{{L{\'e}pine} {et~al.}(2013){L{\'e}pine}, {Hilton}, {Mann}, {Wilde},
  {Rojas-Ayala}, {Cruz}, \& {Gaidos}}]{Lepine2013}
{L{\'e}pine}, S., {Hilton}, E.~J., {Mann}, A.~W., {et~al.} 2013, \aj, 145, 102,
  \dodoi{10.1088/0004-6256/145/4/102}

\bibitem[{{Liddle}(2007)}]{liddle2007}
{Liddle}, A.~R. 2007, \mnras, 377, L74,
  \dodoi{10.1111/j.1745-3933.2007.00306.x}

\bibitem[{{Lindegren} {et~al.}(2018){Lindegren}, {Hern{\'a}ndez}, {Bombrun},
  {Klioner}, {Bastian}, {Ramos-Lerate}, {de Torres}, {Steidelm{\"u}ller},
  {Stephenson}, {Hobbs}, {Lammers}, {Biermann}, {Geyer}, {Hilger}, {Michalik},
  {Stampa}, {McMillan}, {Casta{\~n}eda}, {Clotet}, {Comoretto}, {Davidson},
  {Fabricius}, {Gracia}, {Hambly}, {Hutton}, {Mora}, {Portell}, {van Leeuwen},
  {Abbas}, {Abreu}, {Altmann}, {Andrei}, {Anglada}, {Balaguer-N{\'u}{\~n}ez},
  {Barache}, {Becciani}, {Bertone}, {Bianchi}, {Bouquillon}, {Bourda},
  {Br{\"u}semeister}, {Bucciarelli}, {Busonero}, {Buzzi}, {Cancelliere},
  {Carlucci}, {Charlot}, {Cheek}, {Crosta}, {Crowley}, {de Bruijne}, {de
  Felice}, {Drimmel}, {Esquej}, {Fienga}, {Fraile}, {Gai}, {Garralda},
  {Gonz{\'a}lez-Vidal}, {Guerra}, {Hauser}, {Hofmann}, {Holl}, {Jordan},
  {Lattanzi}, {Lenhardt}, {Liao}, {Licata}, {Lister}, {L{\"o}ffler},
  {Marchant}, {Martin-Fleitas}, {Messineo}, {Mignard}, {Morbidelli}, {Poggio},
  {Riva}, {Rowell}, {Salguero}, {Sarasso}, {Sciacca}, {Siddiqui}, {Smart},
  {Spagna}, {Steele}, {Taris}, {Torra}, {van Elteren}, {van Reeven}, \&
  {Vecchiato}}]{Lindegren2018}
{Lindegren}, L., {Hern{\'a}ndez}, J., {Bombrun}, A., {et~al.} 2018, \aap, 616,
  A2, \dodoi{10.1051/0004-6361/201832727}

\bibitem[{{Lissauer} {et~al.}(2014){Lissauer}, {Dawson}, \&
  {Tremaine}}]{Lissauer2014}
{Lissauer}, J.~J., {Dawson}, R.~I., \& {Tremaine}, S. 2014, \nat, 513, 336,
  \dodoi{10.1038/nature13781}

\bibitem[{{Lissauer} {et~al.}(2011){Lissauer}, {Ragozzine}, {Fabrycky},
  {Steffen}, {Ford}, {Jenkins}, {Shporer}, {Holman}, {Rowe}, {Quintana},
  {Batalha}, {Borucki}, {Bryson}, {Caldwell}, {Carter}, {Ciardi}, {Dunham},
  {Fortney}, {Gautier}, {Howell}, {Koch}, {Latham}, {Marcy}, {Morehead}, \&
  {Sasselov}}]{Lissauer2011}
{Lissauer}, J.~J., {Ragozzine}, D., {Fabrycky}, D.~C., {et~al.} 2011, \apjs,
  197, 8, \dodoi{10.1088/0067-0049/197/1/8}

\bibitem[{{Liu} {et~al.}(2016){Liu}, {Yong}, {Asplund}, {Ram{\'\i}rez}, \&
  {Mel{\'e}ndez}}]{Liu2016}
{Liu}, F., {Yong}, D., {Asplund}, M., {Ram{\'\i}rez}, I., \& {Mel{\'e}ndez}, J.
  2016, \mnras, 457, 3934, \dodoi{10.1093/mnras/stw247}

\bibitem[{{Lockwood} {et~al.}(1997){Lockwood}, {Skiff}, \&
  {Radick}}]{Lockwood1997}
{Lockwood}, G.~W., {Skiff}, B.~A., \& {Radick}, R.~R. 1997, \apj, 485, 789,
  \dodoi{10.1086/304453}

\bibitem[{{Lodieu} {et~al.}(2019){Lodieu}, {P{\'e}rez-Garrido}, {Smart}, \&
  {Silvotti}}]{Lodieu2019}
{Lodieu}, N., {P{\'e}rez-Garrido}, A., {Smart}, R.~L., \& {Silvotti}, R. 2019,
  \aap, 628, A66, \dodoi{10.1051/0004-6361/201935533}

\bibitem[{{L{\'o}pez-Morales} {et~al.}(2006){L{\'o}pez-Morales}, {Morrell},
  {Butler}, \& {Seager}}]{LopezMorales2006}
{L{\'o}pez-Morales}, M., {Morrell}, N.~I., {Butler}, R.~P., \& {Seager}, S.
  2006, \pasp, 118, 1506, \dodoi{10.1086/508904}

\bibitem[{{Luck}(2017)}]{Luck2017}
{Luck}, R.~E. 2017, \aj, 153, 21, \dodoi{10.3847/1538-3881/153/1/21}

\bibitem[{{Luck} \& {Heiter}(2005)}]{Luck2005}
{Luck}, R.~E., \& {Heiter}, U. 2005, \aj, 129, 1063, \dodoi{10.1086/427250}

\bibitem[{{Mackereth} \& {Bovy}(2018)}]{Mackereth2018}
{Mackereth}, J.~T., \& {Bovy}, J. 2018, \pasp, 130, 114501,
  \dodoi{10.1088/1538-3873/aadcdd}

\bibitem[{{Mamajek} \& {Hillenbrand}(2008)}]{MamajekHillenbrand2008}
{Mamajek}, E.~E., \& {Hillenbrand}, L.~A. 2008, \apj, 687, 1264,
  \dodoi{10.1086/591785}

\bibitem[{{Mann} {et~al.}(2015){Mann}, {Feiden}, {Gaidos}, {Boyajian}, \& {von
  Braun}}]{Mann2015}
{Mann}, A.~W., {Feiden}, G.~A., {Gaidos}, E., {Boyajian}, T., \& {von Braun},
  K. 2015, \apj, 804, 64, \dodoi{10.1088/0004-637X/804/1/64}

\bibitem[{{Mart{\'\i}n} {et~al.}(2018){Mart{\'\i}n}, {Lodieu}, {Pavlenko}, \&
  {B{\'e}jar}}]{Martin2018}
{Mart{\'\i}n}, E.~L., {Lodieu}, N., {Pavlenko}, Y., \& {B{\'e}jar}, V. J.~S.
  2018, \apj, 856, 40, \dodoi{10.3847/1538-4357/aaaeb8}

\bibitem[{{Mayor} {et~al.}(2003){Mayor}, {Pepe}, {Queloz}, {Bouchy},
  {Rupprecht}, {Lo Curto}, {Avila}, {Benz}, {Bertaux}, {Bonfils}, {Dall},
  {Dekker}, {Delabre}, {Eckert}, {Fleury}, {Gilliotte}, {Gojak}, {Guzman},
  {Kohler}, {Lizon}, {Longinotti}, {Lovis}, {Megevand}, {Pasquini}, {Reyes},
  {Sivan}, {Sosnowska}, {Soto}, {Udry}, {van Kesteren}, {Weber}, \&
  {Weilenmann}}]{Mayor2003}
{Mayor}, M., {Pepe}, F., {Queloz}, D., {et~al.} 2003, The Messenger, 114, 20

\bibitem[{{Meschiari} {et~al.}(2009){Meschiari}, {Wolf}, {Rivera}, {Laughlin},
  {Vogt}, \& {Butler}}]{Meschiari2009}
{Meschiari}, S., {Wolf}, A.~S., {Rivera}, E., {et~al.} 2009, Publications of
  the ASP, 121, 1016, \dodoi{10.1086/605730}

\bibitem[{{Mikolaitis} {et~al.}(2019){Mikolaitis}, {Drazdauskas},
  {Minkevi{\v{c}}i{\={u}}t{\.{e}}}, {Stonkut{\.{e}}},
  {Tautvai{\v{s}}ien{\.{e}}}, {Klebonas}, {Bagdonas}, {Pak{\v{s}}tien{\.{e}}},
  \& {Janulis}}]{Mikolaitis2019}
{Mikolaitis}, {\v{S}}., {Drazdauskas}, A., {Minkevi{\v{c}}i{\={u}}t{\.{e}}},
  R., {et~al.} 2019, \aap, 628, A49, \dodoi{10.1051/0004-6361/201835004}

\bibitem[{{Millholland} {et~al.}(2017){Millholland}, {Wang}, \&
  {Laughlin}}]{Millholland2017}
{Millholland}, S., {Wang}, S., \& {Laughlin}, G. 2017, \apjl, 849, L33,
  \dodoi{10.3847/2041-8213/aa9714}

\bibitem[{{Mishenina} {et~al.}(2013){Mishenina}, {Pignatari}, {Korotin},
  {Soubiran}, {Charbonnel}, {Thielemann}, {Gorbaneva}, \&
  {Basak}}]{Mishenina2013}
{Mishenina}, T.~V., {Pignatari}, M., {Korotin}, S.~A., {et~al.} 2013, \aap,
  552, A128, \dodoi{10.1051/0004-6361/201220687}

\bibitem[{{Mishenina} {et~al.}(2008){Mishenina}, {Soubiran}, {Bienaym{\'e}},
  {Korotin}, {Belik}, {Usenko}, \& {Kovtyukh}}]{Mishenina2008}
{Mishenina}, T.~V., {Soubiran}, C., {Bienaym{\'e}}, O., {et~al.} 2008, \aap,
  489, 923, \dodoi{10.1051/0004-6361:200810360}

\bibitem[{{Mishenina} {et~al.}(2012){Mishenina}, {Soubiran}, {Kovtyukh},
  {Katsova}, \& {Livshits}}]{Mishenina2012}
{Mishenina}, T.~V., {Soubiran}, C., {Kovtyukh}, V.~V., {Katsova}, M.~M., \&
  {Livshits}, M.~A. 2012, \aap, 547, A106, \dodoi{10.1051/0004-6361/201118412}

\bibitem[{{Nelson} {et~al.}(2020){Nelson}, {Ford}, {Buchner}, {Cloutier},
  {D{\'\i}az}, {Faria}, {Hara}, {Rajpaul}, \& {Rukdee}}]{Nelson2020}
{Nelson}, B.~E., {Ford}, E.~B., {Buchner}, J., {et~al.} 2020, \aj, 159, 73,
  \dodoi{10.3847/1538-3881/ab5190}

\bibitem[{{Ning} {et~al.}(2018){Ning}, {Wolfgang}, \& {Ghosh}}]{Ning2018}
{Ning}, B., {Wolfgang}, A., \& {Ghosh}, S. 2018, \apj, 869, 5,
  \dodoi{10.3847/1538-4357/aaeb31}

\bibitem[{{Olspert} {et~al.}(2018){Olspert}, {Lehtinen}, {K{\"a}pyl{\"a}},
  {Pelt}, \& {Grigorievskiy}}]{Olspert2018}
{Olspert}, N., {Lehtinen}, J.~J., {K{\"a}pyl{\"a}}, M.~J., {Pelt}, J., \&
  {Grigorievskiy}, A. 2018, \aap, 619, A6, \dodoi{10.1051/0004-6361/201732525}

\bibitem[{{Pepe} {et~al.}(2002){Pepe}, {Mayor}, {Rupprecht}, {Avila},
  {Ballester}, {Beckers}, {Benz}, {Bertaux}, {Bouchy}, {Buzzoni}, {Cavadore},
  {Deiries}, {Dekker}, {Delabre}, {D'Odorico}, {Eckert}, {Fischer}, {Fleury},
  {George}, {Gilliotte}, {Gojak}, {Guzman}, {Koch}, {Kohler}, {Kotzlowski},
  {Lacroix}, {Le Merrer}, {Lizon}, {Lo Curto}, {Longinotti}, {Megevand},
  {Pasquini}, {Petitpas}, {Pichard}, {Queloz}, {Reyes}, {Richaud}, {Sivan},
  {Sosnowska}, {Soto}, {Udry}, {Ureta}, {van Kesteren}, {Weber}, {Weilenmann},
  {Wicenec}, {Wieland}, {Christensen-Dalsgaard}, {Dravins}, {Hatzes},
  {K{\"u}rster}, {Paresce}, \& {Penny}}]{Pepe2002}
{Pepe}, F., {Mayor}, M., {Rupprecht}, G., {et~al.} 2002, The Messenger, 110, 9

\bibitem[{{Pepe} {et~al.}(2004){Pepe}, {Mayor}, {Queloz}, {Benz}, {Bonfils},
  {Bouchy}, {Lo Curto}, {Lovis}, {M{\'e}gevand}, {Moutou}, {Naef}, {Rupprecht},
  {Santos}, {Sivan}, {Sosnowska}, \& {Udry}}]{Pepe2004}
{Pepe}, F., {Mayor}, M., {Queloz}, D., {et~al.} 2004, \aap, 423, 385,
  \dodoi{10.1051/0004-6361:20040389}

\bibitem[{{Perryman}(2011)}]{Perryman2011}
{Perryman}, M. 2011, {The Exoplanet Handbook}

\bibitem[{{Pinamonti} {et~al.}(2019){Pinamonti}, {Sozzetti}, {Giacobbe},
  {Damasso}, {Scandariato}, {Perger}, {Gonz{\'a}lez Hern{\'a}ndez}, {Lanza},
  {Maldonado}, {Micela}, {Su{\'a}rez Mascare{\~n}o}, {Toledo-Padr{\'o}n},
  {Affer}, {Benatti}, {Bignamini}, {Bonomo}, {Claudi}, {Cosentino}, {Desidera},
  {Maggio}, {Martinez Fiorenzano}, {Pagano}, {Piotto}, {Rainer}, {Rebolo}, \&
  {Ribas}}]{Pinamonti2019}
{Pinamonti}, M., {Sozzetti}, A., {Giacobbe}, P., {et~al.} 2019, \aap, 625,
  A126, \dodoi{10.1051/0004-6361/201834969}

\bibitem[{{Quirrenbach} {et~al.}(2014){Quirrenbach}, {Amado}, {Caballero},
  {Mundt}, {Reiners}, {Ribas}, {Seifert}, {Abril}, {Aceituno},
  {Alonso-Floriano}, {Ammler-von Eiff}, {Antona Jim{\'e}nez}, {Anwand
  -Heerwart}, {Azzaro}, {Bauer}, {Barrado}, {Becerril}, {B{\'e}jar},
  {Ben{\'\i}tez}, {Berdi{\~n}as}, {C{\'a}rdenas}, {Casal}, {Claret},
  {Colom{\'e}}, {Cort{\'e}s-Contreras}, {Czesla}, {Doellinger}, {Dreizler},
  {Feiz}, {Fern{\'a}ndez}, {Galad{\'\i}}, {G{\'a}lvez-Ortiz},
  {Garc{\'\i}a-Piquer}, {Garc{\'\i}a-Vargas}, {Garrido}, {Gesa}, {G{\'o}mez
  Galera}, {Gonz{\'a}lez {\'A}lvarez}, {Gonz{\'a}lez Hern{\'a}ndez},
  {Gr{\"o}zinger}, {Gu{\`a}rdia}, {Guenther}, {de Guindos},
  {Guti{\'e}rrez-Soto}, {Hagen}, {Hatzes}, {Hauschildt}, {Helmling}, {Henning},
  {Hermann}, {Hern{\'a}ndez Casta{\~n}o}, {Herrero}, {Hidalgo}, {Holgado},
  {Huber}, {Huber}, {Jeffers}, {Joergens}, {de Juan}, {Kehr}, {Klein},
  {K{\"u}rster}, {Lamert}, {Lalitha}, {Laun}, {Lemke}, {Lenzen}, {L{\'o}pez del
  Fresno}, {L{\'o}pez Mart{\'\i}}, {L{\'o}pez-Santiago}, {Mall}, {Mandel},
  {Mart{\'\i}n}, {Mart{\'\i}n-Ruiz}, {Mart{\'\i}nez-Rodr{\'\i}guez}, {Marvin},
  {Mathar}, {Mirabet}, {Montes}, {Morales Mu{\~n}oz}, {Moya}, {Naranjo},
  {Ofir}, {Oreiro}, {Pall{\'e}}, {Panduro}, {Passegger}, {P{\'e}rez-Calpena},
  {P{\'e}rez Medialdea}, {Perger}, {Pluto}, {Ram{\'o}n}, {Rebolo}, {Redondo},
  {Reffert}, {Reinhardt}, {Rhode}, {Rix}, {Rodler}, {Rodr{\'\i}guez},
  {Rodr{\'\i}guez-L{\'o}pez}, {Rodr{\'\i}guez-P{\'e}rez}, {Rohloff}, {Rosich},
  {S{\'a}nchez-Blanco}, {S{\'a}nchez Carrasco}, {Sanz-Forcada}, {Sarmiento},
  {Sch{\"a}fer}, {Schiller}, {Schmidt}, {Schmitt}, {Solano}, {Stahl}, {Storz},
  {St{\"u}rmer}, {Su{\'a}rez}, {Ulbrich}, {Veredas}, {Wagner}, {Winkler},
  {Zapatero Osorio}, {Zechmeister}, {Abell{\'a}n de Paco},
  {Anglada-Escud{\'e}}, {del Burgo}, {Klutsch}, {Lizon}, {L{\'o}pez-Morales},
  {Morales}, {Perryman}, {Tulloch}, \& {Xu}}]{Quirrenbach2014}
{Quirrenbach}, A., {Amado}, P.~J., {Caballero}, J.~A., {et~al.} 2014, in
  Society of Photo-Optical Instrumentation Engineers (SPIE) Conference Series,
  Vol. 9147, \procspie, 91471F

\bibitem[{{Quirrenbach} {et~al.}(2018){Quirrenbach}, {Amado}, {Ribas},
  {Reiners}, {Caballero}, {Seifert}, {Aceituno}, {Azzaro}, {Baroch}, {Barrado},
  {Bauer}, {Becerril}, {B{\`e}jar}, {Ben{\'\i}tez}, {Brinkm{\"o}ller}, {Cardona
  Guill{\'e}n}, {Cifuentes}, {Colom{\'e}}, {Cort{\'e}s-Contreras}, {Czesla},
  {Dreizler}, {Fr{\"o}lich}, {Fuhrmeister}, {Galad{\'\i}-Enr{\'\i}quez},
  {Gonz{\'a}lez Hern{\'a}ndez}, {Gonz{\'a}lez Peinado}, {Guenther}, {de
  Guindos}, {Hagen}, {Hatzes}, {Hauschildt}, {Helmling}, {Henning}, {Herbort},
  {Hern{\'a}ndez Casta{\~n}o}, {Herrero}, {Hintz}, {Jeffers}, {Johnson}, {de
  Juan}, {Kaminski}, {Klahr}, {K{\"u}rster}, {Lafarga}, {Sairam}, {Lamp{\'o}n},
  {Lara}, {Launhardt}, {L{\'o}pez del Fresno}, {L{\'o}pez-Puertas}, {Luque},
  {Mandel}, {Marfil}, {Mart{\'\i}n}, {Mart{\'\i}n-Ruiz}, {Mathar}, {Montes},
  {Morales}, {Nagel}, {Nortmann}, {Nowak}, {Pall{\'e}}, {Passegger}, {Pavlov},
  {Pedraz}, {P{\'e}rez-Medialdea}, {Perger}, {Rebolo}, {Reffert},
  {Rodr{\'\i}guez}, {Rodr{\'\i}guez L{\'o}pez}, {Rosich}, {Sabotta}, {Sadegi},
  {Salz}, {S{\'a}nchez-L{\'o}pez}, {Sanz-Forcada}, {Sarkis}, {Sch{\"a}fer},
  {Schiller}, {Schmitt}, {Sch{\"o}fer}, {Schweitzer}, {Shulyak}, {Solano},
  {Stahl}, {Tala Pinto}, {Trifonov}, {Zapatero Osorio}, {Yan}, {Zechmeister},
  {Abell{\'a}n}, {Abril}, {Alonso-Floriano}, {Ammler-von Eiff},
  {Anglada-Escud{\'e}}, {Anwand-Heerwart}, {Arroyo-Torres}, {Berdi{\~n}as},
  {Bergondy}, {Bl{\"u}mcke}, {del Burgo}, {Cano}, {Carro}, {C{\'a}rdenas},
  {Casal}, {Claret}, {D{\'\i}ez-Alonso}, {Doellinger}, {Dorda}, {Feiz},
  {Fern{\'a}ndez}, {Ferro}, {Gaisn{\'e}}, {Gallardo}, {G{\'a}lvez-Ortiz},
  {Garc{\'\i}a-Piquer}, {Garc{\'\i}a-Vargas}, {Garrido}, {Gesa}, {G{\'o}mez
  Galera}, {Gonz{\'a}lez-{\'A}lvarez}, {Gonz{\'a}lez-Cuesta}, {Grohnert},
  {Gr{\"o}zinger}, {Gu{\`a}rdia}, {Guijarro}, {Hedrosa}, {Hermann}, {Hermelo},
  {Hern{\'a}ndez Arab{\'\i}}, {Hern{\'a}ndez Hernando}, {Hidalgo}, {Holgado},
  {Huber}, {Huber}, {Huke}, {Kehr}, {Kim}, {Klein}, {Kl{\"u}ter}, {Klutsch},
  {Labarga}, {Labiche}, {Lamert}, {Laun}, {L{\'a}zaro}, {Lemke}, {Lenzen},
  {Llamas}, {Lizon}, {Lodieu}, {L{\'o}pez Gonz{\'a}lez}, {L{\'o}pez-Morales},
  {L{\'o}pez Salas}, {L{\'o}pez-Santiago}, {Mag{\'a}n Madinabeitia}, {Mall},
  {Mancini}, {Mar{\'\i}n Molina}, {Mart{\'\i}nez-Rodr{\'\i}guez}, {Maroto
  Fern{\'a}ndez}, {Marvin}, {Mirabet}, {Moreno-Raya}, {Moya}, {Mundt},
  {Naranjo}, {Panduro}, {Pascual}, {P{\'e}rez-Calpena}, {Perryman}, {Pluto},
  {Ram{\'o}n}, {Redondo}, {Reinhart}, {Rhode}, {Rix}, {Rodler}, {Rohloff},
  {S{\'a}nchez-Blanco}, {S{\'a}nchez Carrasco}, {Sarmiento}, {Schmidt},
  {Storz}, {Strachan}, {St{\"u}rmer}, {Su{\'a}rez}, {Tabernero}, {Tal-Or},
  {Tulloch}, {Ulbrich}, {Veredas}, {Vico Linares}, {Vidal-Dasilva},
  {Vilardell}, {Wagner}, {Winkler}, {Wolthoff}, {Xu}, \&
  {Zhao}}]{Quirrenbach2018}
{Quirrenbach}, A., {Amado}, P.~J., {Ribas}, I., {et~al.} 2018, in Society of
  Photo-Optical Instrumentation Engineers (SPIE) Conference Series, Vol. 10702,
  \procspie, 107020W

\bibitem[{{Radick} {et~al.}(1998){Radick}, {Lockwood}, {Skiff}, \&
  {Baliunas}}]{Radick1998}
{Radick}, R.~R., {Lockwood}, G.~W., {Skiff}, B.~A., \& {Baliunas}, S.~L. 1998,
  \apjs, 118, 239, \dodoi{10.1086/313135}

\bibitem[{{Rajpaul} {et~al.}(2016){Rajpaul}, {Aigrain}, \&
  {Roberts}}]{Rajpaul2016}
{Rajpaul}, V., {Aigrain}, S., \& {Roberts}, S. 2016, \mnras, 456, L6,
  \dodoi{10.1093/mnrasl/slv164}

\bibitem[{{Ram{\'\i}rez} {et~al.}(2012){Ram{\'\i}rez}, {Fish}, {Lambert}, \&
  {Allende Prieto}}]{Ramirez2012}
{Ram{\'\i}rez}, I., {Fish}, J.~R., {Lambert}, D.~L., \& {Allende Prieto}, C.
  2012, \apj, 756, 46, \dodoi{10.1088/0004-637X/756/1/46}

\bibitem[{{Rein} \& {Liu}(2012)}]{ReinLiu2012}
{Rein}, H., \& {Liu}, S.~F. 2012, \aap, 537, A128,
  \dodoi{10.1051/0004-6361/201118085}

\bibitem[{{Rein} \& {Tamayo}(2015)}]{ReinTamayo2015}
{Rein}, H., \& {Tamayo}, D. 2015, \mnras, 452, 376,
  \dodoi{10.1093/mnras/stv1257}

\bibitem[{{Reiners} {et~al.}(2018){Reiners}, {Zechmeister}, {Caballero},
  {Ribas}, {Morales}, {Jeffers}, {Sch{\"o}fer}, {Tal-Or}, {Quirrenbach},
  {Amado}, {Kaminski}, {Seifert}, {Abril}, {Aceituno}, {Alonso-Floriano},
  {Ammler-von Eiff}, {Antona}, {Anglada-Escud{\'e}}, {Anwand-Heerwart},
  {Arroyo-Torres}, {Azzaro}, {Baroch}, {Barrado}, {Bauer}, {Becerril},
  {B{\'e}jar}, {Ben{\'{\i}}tez}, {Berdi{\~n}as}, {Bergond}, {Bl{\"u}mcke},
  {Brinkm{\"o}ller}, {del Burgo}, {Cano}, {C{\'a}rdenas V{\'a}zquez}, {Casal},
  {Cifuentes}, {Claret}, {Colom{\'e}}, {Cort{\'e}s-Contreras}, {Czesla},
  {D{\'{\i}}ez-Alonso}, {Dreizler}, {Feiz}, {Fern{\'a}ndez}, {Ferro},
  {Fuhrmeister}, {Galad{\'{\i}}-Enr{\'{\i}}quez}, {Garcia-Piquer},
  {Garc{\'{\i}}a Vargas}, {Gesa}, {G{\'o}mez Galera}, {Gonz{\'a}lez
  Hern{\'a}ndez}, {Gonz{\'a}lez-Peinado}, {Gr{\"o}zinger}, {Grohnert},
  {Gu{\`a}rdia}, {Guenther}, {Guijarro}, {de Guindos}, {Guti{\'e}rrez-Soto},
  {Hagen}, {Hatzes}, {Hauschildt}, {Hedrosa}, {Helmling}, {Henning}, {Hermelo},
  {Hern{\'a}ndez Arab{\'{\i}}}, {Hern{\'a}ndez Casta{\~n}o}, {Hern{\'a}ndez
  Hernando}, {Herrero}, {Huber}, {Huke}, {Johnson}, {de Juan}, {Kim}, {Klein},
  {Kl{\"u}ter}, {Klutsch}, {K{\"u}rster}, {Lafarga}, {Lamert}, {Lamp{\'o}n},
  {Lara}, {Laun}, {Lemke}, {Lenzen}, {Launhardt}, {L{\'o}pez del Fresno},
  {L{\'o}pez-Gonz{\'a}lez}, {L{\'o}pez-Puertas}, {L{\'o}pez Salas},
  {L{\'o}pez-Santiago}, {Luque}, {Mag{\'a}n Madinabeitia}, {Mall}, {Mancini},
  {Mandel}, {Marfil}, {Mar{\'{\i}}n Molina}, {Maroto Fern{\'a}ndez},
  {Mart{\'{\i}}n}, {Mart{\'{\i}}n-Ruiz}, {Marvin}, {Mathar}, {Mirabet},
  {Montes}, {Moreno-Raya}, {Moya}, {Mundt}, {Nagel}, {Naranjo}, {Nortmann},
  {Nowak}, {Ofir}, {Oreiro}, {Pall{\'e}}, {Panduro}, {Pascual}, {Passegger},
  {Pavlov}, {Pedraz}, {P{\'e}rez-Calpena}, {P{\'e}rez Medialdea}, {Perger},
  {Perryman}, {Pluto}, {Rabaza}, {Ram{\'o}n}, {Rebolo}, {Redondo}, {Reffert},
  {Reinhart}, {Rhode}, {Rix}, {Rodler}, {Rodr{\'{\i}}guez},
  {Rodr{\'{\i}}guez-L{\'o}pez}, {Rodr{\'{\i}}guez Trinidad}, {Rohloff},
  {Rosich}, {Sadegi}, {S{\'a}nchez-Blanco}, {S{\'a}nchez Carrasco},
  {S{\'a}nchez-L{\'o}pez}, {Sanz-Forcada}, {Sarkis}, {Sarmiento},
  {Sch{\"a}fer}, {Schmitt}, {Schiller}, {Schweitzer}, {Solano}, {Stahl},
  {Strachan}, {St{\"u}rmer}, {Su{\'a}rez}, {Tabernero}, {Tala}, {Trifonov},
  {Tulloch}, {Ulbrich}, {Veredas}, {Vico Linares}, {Vilardell}, {Wagner},
  {Winkler}, {Wolthoff}, {Xu}, {Yan}, \& {Zapatero Osorio}}]{Reiners2018}
{Reiners}, A., {Zechmeister}, M., {Caballero}, J.~A., {et~al.} 2018, \aap, 612,
  A49, \dodoi{10.1051/0004-6361/201732054}

\bibitem[{{Ricker} {et~al.}(2014){Ricker}, {Winn}, {Vanderspek}, {Latham},
  {Bakos}, {Bean}, {Berta-Thompson}, {Brown}, {Buchhave}, {Butler}, {Butler},
  {Chaplin}, {Charbonneau}, {Christensen-Dalsgaard}, {Clampin}, {Deming},
  {Doty}, {De Lee}, {Dressing}, {Dunham}, {Endl}, {Fressin}, {Ge}, {Henning},
  {Holman}, {Howard}, {Ida}, {Jenkins}, {Jernigan}, {Johnson}, {Kaltenegger},
  {Kawai}, {Kjeldsen}, {Laughlin}, {Levine}, {Lin}, {Lissauer}, {MacQueen},
  {Marcy}, {McCullough}, {Morton}, {Narita}, {Paegert}, {Palle}, {Pepe},
  {Pepper}, {Quirrenbach}, {Rinehart}, {Sasselov}, {Sato}, {Seager},
  {Sozzetti}, {Stassun}, {Sullivan}, {Szentgyorgyi}, {Torres}, {Udry}, \&
  {Villasenor}}]{Ricker2014}
{Ricker}, G.~R., {Winn}, J.~N., {Vanderspek}, R., {et~al.} 2014, in Society of
  Photo-Optical Instrumentation Engineers (SPIE) Conference Series, Vol. 9143,
  Society of Photo-Optical Instrumentation Engineers (SPIE) Conference Series,
  20

\bibitem[{{Rowan} {et~al.}(2016){Rowan}, {Meschiari}, {Laughlin}, {Vogt},
  {Butler}, {Burt}, {Wang}, {Holden}, {Hanson}, {Arriagada}, {Keiser}, {Teske},
  \& {Diaz}}]{Rowan2016}
{Rowan}, D., {Meschiari}, S., {Laughlin}, G., {et~al.} 2016, \apj, 817, 104,
  \dodoi{10.3847/0004-637X/817/2/104}

\bibitem[{{Rowe} {et~al.}(2014){Rowe}, {Bryson}, {Marcy}, {Lissauer},
  {Jontof-Hutter}, {Mullally}, {Gilliland}, {Issacson}, {Ford}, {Howell},
  {Borucki}, {Haas}, {Huber}, {Steffen}, {Thompson}, {Quintana}, {Barclay},
  {Still}, {Fortney}, {Gautier}, {Hunter}, {Caldwell}, {Ciardi}, {Devore},
  {Cochran}, {Jenkins}, {Agol}, {Carter}, \& {Geary}}]{Rowe2014}
{Rowe}, J.~F., {Bryson}, S.~T., {Marcy}, G.~W., {et~al.} 2014, \apj, 784, 45,
  \dodoi{10.1088/0004-637X/784/1/45}

\bibitem[{{Schweitzer} {et~al.}(2019){Schweitzer}, {Passegger}, {Cifuentes},
  {B{\'e}jar}, {Cort{\'e}s-Contreras}, {Caballero}, {del Burgo}, {Czesla},
  {K{\"u}rster}, {Montes}, {Zapatero Osorio}, {Ribas}, {Reiners},
  {Quirrenbach}, {Amado}, {Aceituno}, {Anglada-Escud{\'e}}, {Bauer},
  {Dreizler}, {Jeffers}, {Guenther}, {Henning}, {Kaminski}, {Lafarga},
  {Marfil}, {Morales}, {Schmitt}, {Seifert}, {Solano}, {Tabernero}, \&
  {Zechmeister}}]{Schweitzer2019}
{Schweitzer}, A., {Passegger}, V.~M., {Cifuentes}, C., {et~al.} 2019, \aap,
  625, A68, \dodoi{10.1051/0004-6361/201834965}

\bibitem[{{Slocum} \& {Mitchell}(1913)}]{Slocum1913}
{Slocum}, F., \& {Mitchell}, S.~A. 1913, \apj, 38, 1, \dodoi{10.1086/142012}

\bibitem[{{Smith} {et~al.}(2012){Smith}, {Stumpe}, {Van Cleve}, {Jenkins},
  {Barclay}, {Fanelli}, {Girouard}, {Kolodziejczak}, {McCauliff}, {Morris}, \&
  {Twicken}}]{Smith2012}
{Smith}, J.~C., {Stumpe}, M.~C., {Van Cleve}, J.~E., {et~al.} 2012, \pasp, 124,
  1000, \dodoi{10.1086/667697}

\bibitem[{{Soubiran} {et~al.}(2013){Soubiran}, {Jasniewicz}, {Chemin}, {Crifo},
  {Udry}, {Hestroffer}, \& {Katz}}]{Soubiran2013}
{Soubiran}, C., {Jasniewicz}, G., {Chemin}, L., {et~al.} 2013, \aap, 552, A64,
  \dodoi{10.1051/0004-6361/201220927}

\bibitem[{{Stassun} {et~al.}(2018){Stassun}, {Oelkers}, {Pepper}, {Paegert},
  {De Lee}, {Torres}, {Latham}, {Charpinet}, {Dressing}, {Huber}, {Kane},
  {L{\'e}pine}, {Mann}, {Muirhead}, {Rojas-Ayala}, {Silvotti}, {Fleming},
  {Levine}, \& {Plavchan}}]{Stassun2018}
{Stassun}, K.~G., {Oelkers}, R.~J., {Pepper}, J., {et~al.} 2018, \aj, 156, 102,
  \dodoi{10.3847/1538-3881/aad050}

\bibitem[{{Stassun} {et~al.}(2019){Stassun}, {Oelkers}, {Paegert}, {Torres},
  {Pepper}, {De Lee}, {Collins}, {Latham}, {Muirhead}, {Chittidi},
  {Rojas-Ayala}, {Fleming}, {Rose}, {Tenenbaum}, {Ting}, {Kane}, {Barclay},
  {Bean}, {Brassuer}, {Charbonneau}, {Ge}, {Lissauer}, {Mann}, {McLean},
  {Mullally}, {Narita}, {Plavchan}, {Ricker}, {Sasselov}, {Seager}, {Sharma},
  {Shiao}, {Sozzetti}, {Stello}, {Vanderspek}, {Wallace}, \&
  {Winn}}]{Stassun2019}
{Stassun}, K.~G., {Oelkers}, R.~J., {Paegert}, M., {et~al.} 2019, \aj, 158,
  138, \dodoi{10.3847/1538-3881/ab3467}

\bibitem[{{Steffen} {et~al.}(2012){Steffen}, {Ragozzine}, {Fabrycky}, {Carter},
  {Ford}, {Holman}, {Rowe}, {Welsh}, {Borucki}, {Boss}, {Ciardi}, \&
  {Quinn}}]{Steffen2012}
{Steffen}, J.~H., {Ragozzine}, D., {Fabrycky}, D.~C., {et~al.} 2012,
  Proceedings of the National Academy of Science, 109, 7982,
  \dodoi{10.1073/pnas.1120970109}

\bibitem[{{Stevens} {et~al.}(2017){Stevens}, {Stassun}, \&
  {Gaudi}}]{Stevens2017}
{Stevens}, D.~J., {Stassun}, K.~G., \& {Gaudi}, B.~S. 2017, \aj, 154, 259,
  \dodoi{10.3847/1538-3881/aa957b}

\bibitem[{{Stumpe} {et~al.}(2014){Stumpe}, {Smith}, {Catanzarite}, {Van Cleve},
  {Jenkins}, {Twicken}, \& {Girouard}}]{Stumpe2014}
{Stumpe}, M.~C., {Smith}, J.~C., {Catanzarite}, J.~H., {et~al.} 2014, \pasp,
  126, 100, \dodoi{10.1086/674989}

\bibitem[{{Stumpe} {et~al.}(2012){Stumpe}, {Smith}, {Van Cleve}, {Twicken},
  {Barclay}, {Fanelli}, {Girouard}, {Jenkins}, {Kolodziejczak}, {McCauliff}, \&
  {Morris}}]{Stumpe2012}
{Stumpe}, M.~C., {Smith}, J.~C., {Van Cleve}, J.~E., {et~al.} 2012, \pasp, 124,
  985, \dodoi{10.1086/667698}

\bibitem[{{Su{\'a}rez Mascare{\~n}o} {et~al.}(2017){Su{\'a}rez Mascare{\~n}o},
  {Rebolo}, {Gonz{\'a}lez Hern{\'a}ndez}, \& {Esposito}}]{Suarez-Mascareno2017}
{Su{\'a}rez Mascare{\~n}o}, A., {Rebolo}, R., {Gonz{\'a}lez Hern{\'a}ndez},
  J.~I., \& {Esposito}, M. 2017, \mnras, 468, 4772,
  \dodoi{10.1093/mnras/stx771}

\bibitem[{{Takeda} {et~al.}(2007){Takeda}, {Ford}, {Sills}, {Rasio}, {Fischer},
  \& {Valenti}}]{Takeda2007}
{Takeda}, G., {Ford}, E.~B., {Sills}, A., {et~al.} 2007, \apjs, 168, 297,
  \dodoi{10.1086/509763}

\bibitem[{{Teske} {et~al.}(2018){Teske}, {Wang}, {Wolfgang}, {Dai}, {Shectman},
  {Butler}, {Crane}, \& {Thompson}}]{Teske2018}
{Teske}, J.~K., {Wang}, S., {Wolfgang}, A., {et~al.} 2018, \aj, 155, 148,
  \dodoi{10.3847/1538-3881/aaab56}

\bibitem[{{Torres}(2010)}]{Torres2010}
{Torres}, G. 2010, \aj, 140, 1158, \dodoi{10.1088/0004-6256/140/5/1158}

\bibitem[{{Torres}(2019)}]{Torres2019}
---. 2019, \apj, 883, 105, \dodoi{10.3847/1538-4357/ab3a30}

\bibitem[{{Tuomi} {et~al.}(2018){Tuomi}, {Jones}, {Barnes},
  {Anglada-Escud{\'e}}, {Butler}, {Kiraga}, \& {Vogt}}]{Tuomi2018}
{Tuomi}, M., {Jones}, H.~R.~A., {Barnes}, J.~R., {et~al.} 2018, \aj, 155, 192,
  \dodoi{10.3847/1538-3881/aab09c}

\bibitem[{{Tuomi} {et~al.}(2013){Tuomi}, {Jones}, {Jenkins}, {Tinney},
  {Butler}, {Vogt}, {Barnes}, {Wittenmyer}, {O'Toole}, {Horner}, {Bailey},
  {Carter}, {Wright}, {Salter}, \& {Pinfield}}]{Tuomi2013}
{Tuomi}, M., {Jones}, H.~R.~A., {Jenkins}, J.~S., {et~al.} 2013, \aap, 551,
  A79, \dodoi{10.1051/0004-6361/201220509}

\bibitem[{{Van Eylen} \& {Albrecht}(2015)}]{VanEylenAlbrecht2015}
{Van Eylen}, V., \& {Albrecht}, S. 2015, \apj, 808, 126,
  \dodoi{10.1088/0004-637X/808/2/126}

\bibitem[{{Vanderburg} \& {Johnson}(2014)}]{VanderburgJohnson2014}
{Vanderburg}, A., \& {Johnson}, J.~A. 2014, \pasp, 126, 948,
  \dodoi{10.1086/678764}

\bibitem[{{Vican}(2012)}]{Vican2012}
{Vican}, L. 2012, \aj, 143, 135, \dodoi{10.1088/0004-6256/143/6/135}

\bibitem[{{Vogt} {et~al.}(1994){Vogt}, {Allen}, {Bigelow}, {Bresee}, {Brown},
  {Cantrall}, {Conrad}, {Couture}, {Delaney}, {Epps}, {Hilyard}, {Hilyard},
  {Horn}, {Jern}, {Kanto}, {Keane}, {Kibrick}, {Lewis}, {Osborne},
  {Pardeilhan}, {Pfister}, {Ricketts}, {Robinson}, {Stover}, {Tucker}, {Ward},
  \& {Wei}}]{Vogt1994}
{Vogt}, S.~S., {Allen}, S.~L., {Bigelow}, B.~C., {et~al.} 1994, in Society of
  Photo-Optical Instrumentation Engineers (SPIE) Conference Series, Vol. 2198,
  Instrumentation in Astronomy VIII, ed. D.~L. {Crawford} \& E.~R. {Craine},
  362

\bibitem[{{Vogt} {et~al.}(2014){Vogt}, {Radovan}, {Kibrick}, {Butler},
  {Alcott}, {Allen}, {Arriagada}, {Bolte}, {Burt}, {Cabak}, {Chloros},
  {Cowley}, {Deich}, {Dupraw}, {Earthman}, {Epps}, {Faber}, {Fischer}, {Gates},
  {Hilyard}, {Holden}, {Johnston}, {Keiser}, {Kanto}, {Katsuki}, {Laiterman},
  {Lanclos}, {Laughlin}, {Lewis}, {Lockwood}, {Lynam}, {Marcy}, {McLean},
  {Miller}, {Misch}, {Peck}, {Pfister}, {Phillips}, {Rivera}, {Sandford},
  {Saylor}, {Stover}, {Thompson}, {Walp}, {Ward}, {Wareham}, {Wei}, \&
  {Wright}}]{Vogt2014}
{Vogt}, S.~S., {Radovan}, M., {Kibrick}, R., {et~al.} 2014, Publications of the
  ASP, 126, 359, \dodoi{10.1086/676120}

\bibitem[{{Vogt} {et~al.}(2017){Vogt}, {Butler}, {Burt}, {Tuomi}, {Laughlin},
  {Holden}, {Teske}, {Shectman}, {Crane}, {D{\'{\i}}az}, {Thompson},
  {Arriagada}, \& {Keiser}}]{Vogt2017}
{Vogt}, S.~S., {Butler}, R.~P., {Burt}, J., {et~al.} 2017, \aj, 154, 181,
  \dodoi{10.3847/1538-3881/aa8b61}

\bibitem[{{Wang}(2017)}]{Wang2017}
{Wang}, S. 2017, Research Notes of the American Astronomical Society, 1, 26,
  \dodoi{10.3847/2515-5172/aa9be5}

\bibitem[{{Wang} {et~al.}(2018){Wang}, {Addison}, {Fischer}, {Brewer},
  {Isaacson}, {Howard}, \& {Laughlin}}]{Wang2018}
{Wang}, S., {Addison}, B., {Fischer}, D.~A., {et~al.} 2018, \aj, 155, 70,
  \dodoi{10.3847/1538-3881/aaa2fb}

\bibitem[{{Weiss} {et~al.}(2018){Weiss}, {Marcy}, {Petigura}, {Fulton},
  {Howard}, {Winn}, {Isaacson}, {Morton}, {Hirsch}, {Sinukoff}, {Cumming},
  {Hebb}, \& {Cargile}}]{Weiss2018}
{Weiss}, L.~M., {Marcy}, G.~W., {Petigura}, E.~A., {et~al.} 2018, \aj, 155, 48,
  \dodoi{10.3847/1538-3881/aa9ff6}

\bibitem[{{Winn}(2010)}]{Winn2010}
{Winn}, J.~N. 2010, {Exoplanet Transits and Occultations}, 55--77

\bibitem[{{Winn} \& {Fabrycky}(2015)}]{Winn2015}
{Winn}, J.~N., \& {Fabrycky}, D.~C. 2015, \araa, 53, 409,
  \dodoi{10.1146/annurev-astro-082214-122246}

\bibitem[{{Wright} {et~al.}(2004){Wright}, {Marcy}, {Butler}, \&
  {Vogt}}]{Wright2004}
{Wright}, J.~T., {Marcy}, G.~W., {Butler}, R.~P., \& {Vogt}, S.~S. 2004, \apjs,
  152, 261, \dodoi{10.1086/386283}

\bibitem[{{Wright} {et~al.}(2011){Wright}, {Drake}, {Mamajek}, \&
  {Henry}}]{Wright2011}
{Wright}, N.~J., {Drake}, J.~J., {Mamajek}, E.~E., \& {Henry}, G.~W. 2011,
  \apj, 743, 48, \dodoi{10.1088/0004-637X/743/1/48}

\bibitem[{{Zechmeister} {et~al.}(2018){Zechmeister}, {Reiners}, {Amado},
  {Azzaro}, {Bauer}, {B{\'e}jar}, {Caballero}, {Guenther}, {Hagen}, {Jeffers},
  {Kaminski}, {K{\"u}rster}, {Launhardt}, {Montes}, {Morales}, {Quirrenbach},
  {Reffert}, {Ribas}, {Seifert}, {Tal-Or}, \& {Wolthoff}}]{Zechmeister2018}
{Zechmeister}, M., {Reiners}, A., {Amado}, P.~J., {et~al.} 2018, \aap, 609,
  A12, \dodoi{10.1051/0004-6361/201731483}

\end{thebibliography}
\end{document}